\documentclass{iopart}
\usepackage{iopams}

\begin{document}
\newcommand{\ExpandDerivations}{1}
% set to 1 to expand the derivations as an aid to referees.
% set to 0 for the journal version

\title[Einstein-Schr\"{o}dinger theory in the presence of zero-point fluctuations]
{Einstein-Schr\"{o}dinger theory in the presence of zero-point fluctuations
}

\author{J. A. Shifflett}

\address{
Department of Physics,
Washington University, St.~Louis, Missouri 63130
}

\begin{abstract}
The Einstein-Schr\"{o}dinger theory is modified by
adding a cosmological constant contribution caused by zero-point fluctuations.
This cosmological constant which multiplies the symmetric
metric is assumed to be nearly cancelled
by Schr\"{o}dinger's ``bare'' cosmological constant which multiplies the
nonsymmetric fundamental tensor, such that the total ``physical'' cosmological
constant matches measurement.
We first derive the field equations of the theory from a Lagrangian density.
We show that the divergence of the Einstein equations vanishes using
the Christoffel connection formed from the symmetric metric,
allowing additional fields to be included
in the same manner as with ordinary general relativity.
We show that the field equations match the ordinary electro-vac
Einstein and Maxwell equations except for additional terms which are
$<\!10^{-16}$ of the usual terms for worst-case field
strengths and rates-of-change accessible to measurement.
We also show that the theory avoids ghosts in an unusual way.
We show that the Einstein-Infeld-Hoffmann (EIH) equations of motion
for this theory match the equations of motion for
Einstein-Maxwell theory to Newtonian/Coulombian order,
which proves the existence of a Lorentz force.
We derive an exact electric monopole solution,
and show that it matches the Reissner-Nordstr\"{o}m solution except for
additional terms which are $\sim\!10^{-66}$ of the usual terms for
worst-case radii accessible to measurement.
%We also derive an exact electromagnetic plane-wave
%solution which is identical to a solution of
%Einstein-Maxwell theory.
Finally, we show that the theory becomes exactly electro-vac Einstein-Maxwell theory
in the limit as the cosmological constant from zero-point fluctuations goes to infinity.
\end{abstract}

\newcommand{\rmt}{\sqrt{2}\,i}
\newcommand{\TA}{\xi}
\newcommand{\K}{N}
\newcommand{\hR}{\mathcal R}
\newcommand{\hhR}{\hat\mathcal R}
\newcommand{\tR}{\tilde\hR}
\newcommand{\sR}{{\tilde R}}
\newcommand{\tG}{{\tilde G}}
\newcommand{\tB}{{\tilde \Pi}}
\newcommand{\ca}{c_1}
\newcommand{\cb}{c_2}
\newcommand{\cc}{c_3}
\newcommand{\cUps}{\check\Upsilon}
\newcommand{\bUps}{\bar\Upsilon}
\newcommand{\nGam}{\widehat\Gamma}
\newcommand{\sGam}{{\tilde\Gamma}}
\newcommand{\tGam}{{\tilde\Gamma}}
\newcommand{\bGam}{\acute\Gamma}
\newcommand{\tpsi}{\tilde\psi}
\newcommand{\Nbar}{W}
\newcommand{\tT}{\tilde T}
\newcommand{\tS}{\tilde S}
\newcommand{\tf}{\tilde f}
\newcommand{\hf}{\hat f}
\newcommand{\hj}{\hat j}
\newcommand{\tC}{\acute C}
\newcommand{\tdual}{\tilde\vartheta}
\newcommand{\dual}{\vartheta}
\newcommand{\dualmag}{\vartheta}
\newcommand{\Fdash}{\raise1pt\hbox{\rlap\textendash} F}
\newcommand{\Idash}{\raise1pt\hbox{\rlap\textendash} I}
\newcommand{\cosht}{\check{c}}
\newcommand{\sinht}{\check{s}}
\newcommand{\tanht}{\check{u}}
\newcommand{\rmg}{\sqrt{-g}}
\newcommand{\rmN}{\sqrt{\!-N}}
\newcommand{\ff}{\,\ell}
\newcommand{\extra}{V}
\newcommand{\Lambdap}{\Lambda}
\newcommand{\Aphi}{A}
\newcommand{\Aphimag}{A}
\newcommand{\dyadic}{{\rm tensor}}
\newcommand{\ord}{{\mathcal O}}

% Einstein Maxwell spacetimes (04.40.Nr)
% cosmological constant (98.80.Es)
% dark matter (95.35.+d)
% exact solutions of general relativity (04.20.Jb)
% unified field theories and models (12.10.-g)
% relativity and gravitation (95.30.Sf) - probably covered by (04.40.Nr)
% alternative theories of gravity (04.50.+h) - it isn't about gravity
% No more than 4 of these are supposed to be used

% PACS codes
%\pacs{04.40.Nr,98.80.Es,04.20.Jb,12.10.-g}% PACS codes
%\keywords{Einstein-Schrodinger Theory, Hermitian Theory of Relativity,
%Schrodinger Affine Field Theory, Einstein-Straus Theory,
%Cosmological Constant, Zero-Point Fluctuations}
%Use showkeys class option
\ead{shifflet@hbar.wustl.edu}
%\bigskip
\section{\label{Introduction}Introduction}
The Einstein-Schr\"{o}dinger theory without a cosmological constant
was originally proposed by Einstein and
Straus\cite{EinsteinStraus,Einstein3,EinsteinBianchi,EinsteinKaufman,EinsteinMOR}.
Schr\"{o}dinger generalized the theory to include a cosmological constant,
and showed that the theory can be derived from a very simple Lagrangian density
if this cosmological constant is assumed to be
non-zero\cite{SchrodingerI,SchrodingerIII,SchrodingerSTS}.
This more general theory is usually called Schr\"{o}dinger's Affine Field Theory
or the Einstein-Schr\"{o}dinger Theory.
The theory is a generalization of
ordinary general relativity which allows a non-symmetric fundamental tensor and
connection. Einstein and Schr\"{o}dinger suspected that the fundamental tensor
might contain both the metric and the electromagnetic field,
but this was never demonstrated.

In this paper we show that
a well motivated modification of the Einstein-Schr\"{o}dinger theory does indeed
closely approximate ordinary electro-vac Einstein-Maxwell theory, the modification
being the addition of a cosmological constant caused by zero-point
fluctuations\cite{Sahni,Zeldovich,Peskin}.
This cosmological constant which multiplies the symmetric metric $g_{\nu\mu}$
is assumed to be nearly cancelled by Schr\"{o}dinger's ``bare'' cosmological
constant which multiplies the nonsymmetric fundamental tensor $N_{\nu\mu}$,
resulting in a total ``physical" cosmological constant which is consistent with measurement.
This is essentially the same as vacuum energy renormalization in quantum field
theory, and it can be viewed as a kind of zeroth order quantization effect.
Note that we are not attempting to quantize the Einstein-Schr\"{o}dinger theory.
We are considering the classical Einstein-Schr\"{o}dinger theory, but with a
modification to account for a quantum mechanical effect.

For electro-vac Einstein-Maxwell theory, the
Einstein-Infeld-Hoffmann (EIH)
method\cite{EinsteinInfeld,Wallace}
allows the equations-of-motion for both charged and neutral particles
to be derived directly from the field equations.
%As proven by this method, and other more powerful
%methods developed later\cite{JohnsonI},
%The EIH method is valuable because it does not require any additional assumptions,
%such as the postulate that neutral particles follow geodesics,
%or the ad-hoc inclusion of matter terms in the Lagrangian density.
When the EIH method was originally
applied to the Einstein-Schr\"{o}dinger theory, no Lorentz force was found
between charged particles\cite{Callaway,Infeld}. This is the primary reason
that the Einstein-Schr\"{o}dinger theory was abandoned by most researchers
long ago. It is significant that no cosmological constant was assumed in
\cite{Callaway,Infeld}. Quantum field theory predicts that zero-point
fluctuations should cause a very large cosmological constant. Also, recent
evidence suggests an ordinary cosmological constant as the most likely reason
that the expansion of the universe is accelerating\cite{Perlmutter,Astier}.
%With this knowledge, it is important to examine the equation-of-motion
%problem for the Einstein-Schr\"{o}dinger theory with a cosmological constant.
It is shown in \S\ref{EIHEquationsOfMotion} that when a cosmological constant is assumed,
and when $g_{\mu\nu}$ and $F_{\mu\nu}$ are defined differently than
in \cite{Callaway,Infeld}, the EIH method definitely predicts a Lorentz force.
This can be confirmed by
including charged matter terms in the Lagrangian density as in \cite{sShifflett},
in which case the exact Lorentz-force equation can be derived from the theory.

In much previous work on the Einstein-Schr\"{o}dinger theory,
the electromagnetic field is assumed to be the dual of $N_{[\nu\mu]}$ or
$N^{*[\alpha\rho]}=\varepsilon^{\alpha\rho\nu\mu}N_{[\nu\mu]}/2$,
as originally proposed by Einstein\cite{EinsteinMOR}.
However, all efforts have failed to connect the resulting current source
$(\varepsilon^{\rho\alpha\nu\mu}N_{[\nu\mu]})_{;\alpha}/2$ to a real charge current.
%With this definition, an Ampere's law results with a non-vanishing current source,
%$(\varepsilon^{\rho\alpha\nu\mu}N_{[\nu\mu]})_{;\alpha}/2$. However,
%all efforts have failed to connect this apparent current with a real charge current.
With this definition, a Lorentz-like force can be
demonstrated\cite{Treder57,JohnsonII,Narlikar} without a cosmological
constant. However, the solutions\cite{Wyman,Tiwari} that must be used for test
particles have bad asymptotic behavior, such as a radial electrostatic field
which is independent of radius at large distances.
Our electric monopole solution in \S\ref{Monopole} has no such problem.

Recent work\cite{Moffat78,Damour92,Damour93}
shows that the original Einstein-Schr\"{o}dinger theory
has problems with negative energy ``ghosts''.
%and a strong coupling of matter fields to the curvature tensor.
As will be seen in \S\ref{AntisymmetricPart}, this problem is avoided in
the present theory in an unusual way. In most of the work referenced above,
the electromagnetic field is assumed to be an ``added on'' field,
unrelated to $N_{[\nu\mu]}$.
%existing in addition to the electromagnetic field.
Because this approach is not taken in the present theory,
%matches electro-vac Einstein-Maxwell theory to a very close approximation.
problems\cite{Will,Zhou,Gabriel} caused by the coupling of $N_{\nu\mu}$ to the
electromagnetic field do not apply here.

In some previous work, the equations of the Einstein-Schr\"{o}dinger theory
are significantly modified\cite{Moffat95,Clayton}. In some
theories\cite{Kursunoglu,Bonnor,Borchsenius,Moffat74},
$N_{[\tau\rho]}$ is interpreted as the electromagnetic field, and a Lorentz
force is derived, but only because a term
$\rmN \K^{\dashv[\rho\tau]}N_{[\tau\rho]}$ is appended onto the
Einstein-Schr\"{o}dinger Lagrangian density.
None of these theories have been shown to approximate Einstein-Maxwell
theory close enough to agree with experiment, and the modifications assumed
in these theories have no clear physical motivation.
Neither criticism applies to the present theory.

Recently there has been much interest in Born-Infeld
electrodynamics\cite{Born,Deser} because it appears to result from
string theory. The similarity of the Einstein-Schr\"{o}dinger
theory to Born-Infeld electrodynamics is noted by \cite{Gibbons} who suggest
a connection to string theory. While a possible connection
between the Einstein-Schr\"{o}dinger theory and string theory
is beyond the scope of
this paper, it is nevertheless an additional reason to investigate the theory.

This paper is organized as follows.
In \S\ref{LagrangianDensity}-\ref{Derivation} we first derive the field equations
of this modified Einstein-Schr\"{o}dinger theory from a Lagrangian density.
We show that the divergence of the Einstein equations vanishes using
the Christoffel connection formed from the symmetric metric,
allowing additional (non-electromagnetic) fields
to be included in the same manner as with ordinary general relativity.
In \S\ref{SymmetricPart}-\ref{AntisymmetricPart} we show that the field equations match the
ordinary electro-vac Einstein and Maxwell equations except for additional
terms which are $<\!10^{-16}$ of the usual terms for worst-case field
strengths and rates-of-change accessible to measurement.
We also show that the theory avoids ghosts in an unusual way. In \S\ref{EIHEquationsOfMotion} we
derive the Einstein-Infeld-Hoffmann (EIH) equations of motion for this
theory and show that they match the equations of motion for Einstein-Maxwell theory
to Newtonian/Coulombian order, which proves the existence of a Lorentz force.
In \S\ref{Monopole} derive an exact electric monopole solution,
and show that it matches the Reissner-Nordstr\"{o}m solution except for
additional terms which are $\sim\!10^{-66}$ of the usual terms for
worst-case radii accessible to measurement.
%Finally, in \S\ref{Planewave} we derive an exact electromagnetic plane-wave
%solution which is identical to a solution of
%Einstein-Maxwell theory.

\section{\label{LagrangianDensity}The Lagrangian Density}
Ordinary vacuum general relativity
%with cosmological constant $\Lambda_b$ and dimension ``n''
can be derived from a Palatini Lagrangian density,
\begin{eqnarray}
\label{GR}
\fl~~~~{\mathcal L}(\Gamma^{\lambda}_{\!\rho\tau},g_{\rho\tau})
=-\frac{\lower1pt\hbox{$1$}}{16\pi}\rmg\left[\,g^{\mu\nu}
R_{\nu\mu}({\Gamma})+(n\!-\!2)\Lambda_b\,\right].
\end{eqnarray}
Here and throughout this paper we are assuming that $n\!=\!4$, but the dimension ``n''
will be included in the equations to show how easily the results can be generalized
to arbitrary dimension.
The original unmodified Einstein-Schr\"{o}dinger
theory\cite{SchrodingerI,SchrodingerIII,SchrodingerSTS,EinsteinStraus,Einstein3,EinsteinBianchi,EinsteinKaufman,EinsteinMOR}
can be derived from a generalization of (\ref{GR}) formed from a connection
$\nGam^\alpha_{\nu\mu}$ and a fundamental tensor $N_{\nu\mu}$
with no symmetry properties (see \ref{AffineDerivation} for an alternative derivation),
\begin{eqnarray}
\label{Palatini}
\fl~~~~{\mathcal L}(\nGam^{\lambda}_{\!\rho\tau},N_{\rho\tau})
=-\frac{\lower1pt\hbox{$1$}}{16\pi}\rmN\left[N^{\dashv\mu\nu}
\hR_{\nu\mu}({\nGam})+(n\!-\!2)\Lambda_b\,\right].
\end{eqnarray}
%\bigbreak
Our theory includes a cosmological
constant $\Lambda_z$ caused by zero-point fluctuations,
\begin{eqnarray}
\label{JSlag1}
\fl~~~~{\mathcal L}(\nGam^{\lambda}_{\!\rho\tau},N_{\rho\tau})
&=&-\frac{\lower1pt\hbox{$1$}}{16\pi}\left[\rmN N^{\dashv\mu\nu}
\hR_{\nu\mu}({\nGam})+(n\!-\!2)\Lambda_b\rmN+(n\!-\!2)\Lambda_z\rmg\,\right],
%\fl&&\!-\frac{\lower1pt\hbox{$1$}}{16\pi},
\end{eqnarray}
where the ``bare'' $\Lambda_b$ obeys $\Lambda_b\!\approx\!-\Lambda_z$ so that the ``physical'' $\Lambda$ matches measurement,
\begin{eqnarray}
\label{Lambdadef}
\Lambda=\Lambda_b+\Lambda_z,
\end{eqnarray}
and the metric and electromagnetic potential are defined as
\begin{eqnarray}
\label{gdef}
\rmg\,g^{\mu\nu}=\rmN N^{\dashv(\mu\nu)},\\
\label{A}
\Aphi_\nu=\frac{\lower1pt\hbox{$1$}}{2(n\!-\!1)}\,{\nGam}{^\sigma_{\![\sigma\nu]}}\rmt\Lambda_b^{\!-1/2}.
\end{eqnarray}
Here and throughout this paper we use geometrized units with $c\!=\!G\!=\!1$,
the symbols
$\raise2pt\hbox{$_{(~)}$}$ and $\raise2pt\hbox{$_{[~]}$}$ around indices
indicate symmetrization and antisymmetrization, ``n'' is the dimension,
$N\!=\!det(N_{\mu\nu})$, and $N^{\dashv\sigma\nu}$ is the inverse of $N_{\nu\mu}$ so that
$N^{\dashv\sigma\nu}N_{\nu\mu}=\delta^\sigma_\mu$.
In (\ref{JSlag1}), $\hR_{\nu\mu}(\nGam)$
%is not the usual Ricci tensor but
is a form of the so-called Hermitianized Ricci tensor\cite{EinsteinStraus},
%\begin{eqnarray}
%\label{HermitianizedRicci0}
%\hR_{\nu\mu}(\nGam)
%&=&\nGam^\alpha_{\nu\mu,\alpha}
%-\nGam^\alpha_{\nu\alpha,\mu}
%+\nGam^\sigma_{\nu\mu}\nGam^\alpha_{\sigma\alpha}
%-\nGam^\sigma_{\nu\alpha}\nGam^\alpha_{\sigma\mu}
%+\nGam^\alpha_{[\nu|\alpha,|\mu]}\\
%\label{HermitianizedRicci}
%&=&\nGam^\alpha_{\nu\mu,\alpha}
%-\nGam^\alpha_{(\nu|\alpha,|\mu)}
%+\nGam^\sigma_{\nu\mu}\nGam^\alpha_{\sigma\alpha}
%-\nGam^\sigma_{\nu\alpha}\nGam^\alpha_{\sigma\mu}.
%\end{eqnarray}
\begin{eqnarray}
\label{HermitianizedRicci0}
\fl~~~~~~~~\hR_{\nu\mu}(\nGam)
%&=&\nGam^\alpha_{\nu\mu,\alpha}
%-~\nGam^\alpha_{\!(\alpha\nu),\mu}
%~+\nGam^\sigma_{\nu\mu}\nGam^\alpha_{\!(\alpha\sigma)}
%-\nGam^\sigma_{\nu\alpha}\nGam^\alpha_{\sigma\mu}
%\!-\nGam^\tau_{\![\tau\nu]}\nGam^\alpha_{\![\alpha\mu]}/(n\!-\!1)
%+\nGam^\alpha_{\!(\alpha[\nu),\mu]}\\
\label{HermitianizedRicci}
&=&\nGam^\alpha_{\nu\mu,\alpha}
-\nGam^\alpha_{\!(\alpha(\nu),\mu)}
+\nGam^\sigma_{\nu\mu}\nGam^\alpha_{\!(\alpha\sigma)}
-\nGam^\sigma_{\nu\alpha}\nGam^\alpha_{\sigma\mu}
\!-\nGam^\tau_{\![\tau\nu]}\nGam^\alpha_{\![\alpha\mu]}/(n\!-\!1).
\end{eqnarray}
This tensor reduces to the ordinary Ricci tensor for symmetric fields, where we have
$\Gamma^\alpha_{\![\nu\mu]}\!=\!0$ and
$\Gamma^\alpha_{\alpha[\nu,\mu]}\!=\!R^\alpha_{~\alpha\mu\nu}/2\!=\!0$.

It is convenient to decompose ${\nGam}{^\alpha_{\nu\mu}}$
into another connection ${\tGam}{^\alpha_{\nu\mu}}$, and $A_\sigma$ from (\ref{A}),
\begin{eqnarray}
\label{gamma_natural}
\fl~~~~~~~~~~~~~~~~~~~~~{\nGam}{^\alpha_{\nu\mu}}&=&\tGam{^\alpha_{\nu\mu}}
+(\delta^\alpha_\mu\Aphi_\nu
\!-\delta^\alpha_\nu \Aphi_\mu)\rmt\Lambda_b^{\!1/2},\\
\label{gamma_tilde}
\fl~~~~~~~~~~{\rm where}~~~\tGam{^\alpha_{\nu\mu}}
&=&{\nGam}{^\alpha_{\nu\mu}}\!+
%\frac{\lower1pt\hbox{$1$}}{(n\!-\!1)}
(\delta^\alpha_\mu{\nGam}{^\sigma_{\![\sigma\nu]}}
-\delta^\alpha_\nu{\nGam}{^\sigma_{\![\sigma\mu]}})/(n\!-\!1).
\end{eqnarray}
By contracting (\ref{gamma_tilde}) on the right and left we see that
$\tGam{^\alpha_{\nu\mu}}$ has the symmetry
\begin{eqnarray}
\label{contractionsymmetric}
\tGam^\alpha_{\nu\alpha}
%\!=\!(n\nGam^\alpha_{\nu\alpha}\!-\nGam^\alpha_{\alpha\nu})/(n-1)
\!=\!\nGam^\alpha_{\!(\nu\alpha)}
\!=\!\tGam^\alpha_{\alpha\nu},
\end{eqnarray}
so it has only $n^3\!-\!n$ independent components.
%In \ref{ExtractionofConnectionAddition} it is shown that
%\begin{eqnarray}
Using ${\mathcal R}_{\nu\mu}(\nGam)\!=\!{\mathcal R}_{\nu\mu}(\tGam)+2A_{[\nu,\mu]}\rmt\Lambda_b^{\!1/2}$
%\end{eqnarray}
from (\ref{RnGam}), the Lagrangian density (\ref{JSlag1}) can be rewritten in terms of
${\tGam}{^\alpha_{\nu\mu}}$ and $A_\sigma$,
\begin{eqnarray}
\label{JSlag3}
\fl{\mathcal L}
%(\nGam^{\lambda}_{\!\rho\tau},N_{\rho\tau})
\!=-\frac{\lower1pt\hbox{$1$}}{16\pi}\!\left[\rmN N^{\dashv\mu\nu}(\tR_{\nu\mu}
%+\ca\tGam^\alpha_{\alpha[\nu,\mu]}
\!+2\Aphi_{[\nu,\mu]}\rmt\Lambda_b^{\!1/2})\!+(n\!-\!2)\Lambda_b\rmN\!+(n\!-\!2)\Lambda_z\rmg\,\right]\!.
%\fl&&\!-\frac{\lower1pt\hbox{$1$}}{16\pi}(n\!-\!2)\Lambda_z\rmg\,.
\end{eqnarray}
Here $\tR_{\nu\mu}\!=\!\hR_{\nu\mu}(\tGam)$, and from (\ref{contractionsymmetric})
the Hermitianized Ricci tensor (\ref{HermitianizedRicci}) simplifies to
\begin{eqnarray}
\label{HermitianizedRiccit}
\tR_{\nu\mu}
&=&\tGam^\alpha_{\nu\mu,\alpha}
-\tGam^\alpha_{\alpha(\nu,\mu)}
+\tGam^\sigma_{\nu\mu}\tGam^\alpha_{\sigma\alpha}
-\tGam^\sigma_{\nu\alpha}\tGam^\alpha_{\sigma\mu}.
\end{eqnarray}
From (\ref{gamma_natural},\ref{contractionsymmetric}), $\tGam^\alpha_{\nu\mu}$ and $A_\nu$ fully
parameterize $\nGam^\alpha_{\nu\mu}$ and can be treated as independent variables.
So when we set $\delta{\mathcal L}/\delta\tGam^\alpha_{\nu\mu}\!=0$
and $\delta{\mathcal L}/\delta A_\nu\!=0$,
the same field equations must result as with
$\delta{\mathcal L}/\delta\nGam^\alpha_{\nu\mu}\!=0$.
It is simpler to calculate the field equations using
$\tGam^\alpha_{\nu\mu}$ and $A_\nu$ instead of
$\nGam^\alpha_{\nu\mu}$, so we will follow this method.

We will usually assume that $\Lambda_z$ is limited by a
cutoff frequency\cite{Sakharov,Padmanabhan,Padmanabhan2,Ashtekar}
\begin{eqnarray}
\label{cutoff}
\omega_c\!\sim\!1/l_P,
\end{eqnarray}
where
%\begin{eqnarray}
%\label{Planck}
$l_P\!=\!({\rm Planck~length})\!=\!\sqrt{\hbar G/c^3}\!=\!1.6\times 10^{-33}cm$.
%\end{eqnarray}
Then from (\ref{Lambdadef},\ref{cutoff}) and assuming all of the known
fundamental particles we have\cite{Sahni},
\begin{eqnarray}
\label{Lambdab}
\Lambda_b&\approx&-\Lambda_z\sim C_z \omega_c^4 l_P^2
\sim 10^{66}\,{\rm cm}^{-2},\\
\label{Cz}
C_z&=&\frac{\lower2pt\hbox{1}}{2\pi}\!\left({{\lower2pt\hbox{fermion}}\atop{\raise2pt\hbox{spin~states}}}
-{{\lower2pt\hbox{boson}}\atop{\raise2pt\hbox{spin~states}}}\right)
\sim \frac{\lower2pt\hbox{60}}{2\pi}
%~~~~ C_z={\rm(fermions\!-\!bosons)}/2\pi\sim 60/2\pi
\end{eqnarray}
and from astronomical measurements\cite{Spergel,Perlmutter,Astier,Freedman}
\begin{eqnarray}
\label{Lambda}
\Lambda &\approx&1.4\times 10^{-56}{\rm cm}^{-2},~~~~
\label{LambdaoverLambdab}
\Lambda/\Lambda_b\sim 10^{-122}.
\end{eqnarray}
However, it might be more correct to
fully renormalize with $\omega_c\!\rightarrow\!\infty$,
$|\Lambda_z|\!\rightarrow\!\infty$, $\Lambda_b\!\rightarrow\!\infty$
as in quantum electrodynamics.
To account for this possibility we will prove that
\begin{eqnarray}
\label{LRESlimit}
\fl~~~~~~~~~~~~{{\lower2pt\hbox{lim}}\atop{\raise2pt\hbox{$\Lambda_b\!\rightarrow\!\infty$}}}
\left(\lower2pt\hbox{$\Lambda$-renormalized}\atop{\raise2pt\hbox{Einstein-Schr\"{o}dinger theory}}\right)
\!=\!\left(\lower2pt\hbox{Einstein-Maxwell}\atop{\raise2pt\hbox{theory}}\right).
\end{eqnarray}

The Hermitianized Ricci tensor (\ref{HermitianizedRicci}) has the following invariance properties
\begin{eqnarray}
\label{transpositionsymmetric}
\hR_{\nu\mu}(\nGam^T)=\hR_{\mu\nu}(\nGam),~~~~~~~~~~~~~~~~({\rm T=transpose})\\
\label{gaugesymmetric}
\hR_{\nu\mu}(\nGam^\alpha_{\rho\tau}\!+\delta^\alpha_{[\rho}\varphi_{,\tau]})=
\hR_{\nu\mu}(\nGam^\alpha_{\rho\tau})~~~{\rm for~an~arbitrary}~\varphi(x^\sigma).
\end{eqnarray}
From (\ref{transpositionsymmetric},\ref{gaugesymmetric}), the Lagrangian densities (\ref{JSlag1},\ref{JSlag3}) are invariant under charge conjugation,
\begin{eqnarray}
\label{transposition}
\fl~~Q\!\rightarrow\!-Q,~~
A_\sigma\!\rightarrow\!-A_\sigma,~~
\tGam^\alpha_{\nu\mu}\!\rightarrow\!\tGam^\alpha_{\mu\nu},~~
\nGam^\alpha_{\nu\mu}\!\rightarrow\!\nGam^\alpha_{\mu\nu},~~
N_{\nu\mu}\!\!\rightarrow\!N_{\mu\nu},~~
N^{\dashv\nu\mu}\!\rightarrow\!N^{\dashv\mu\nu},
\end{eqnarray}
and also under an electromagnetic gauge transformation
\begin{eqnarray}
\label{gauge}
\fl~~\psi\!\rightarrow\!\psi e^{i\phi},~~
A_\alpha\!\rightarrow\! A_\alpha\!-\!\frac{\hbar}{Q}\phi_{,\alpha},~~
\tGam^\alpha_{\rho\tau}\!\rightarrow\!\tGam^\alpha_{\rho\tau},~~
\nGam^\alpha_{\rho\tau}\!\rightarrow\!\nGam^\alpha_{\rho\tau}\!+\frac{2\hbar}{Q}\delta^\alpha_{[\rho}\phi_{,\tau]}\rmt\Lambda_b^{\!1/2}.
\end{eqnarray}
If $\Lambda_b\!>0,~\Lambda_z\!<0$ as in (\ref{Lambdab},\ref{Cz}) then
$\tGam^\alpha_{\nu\mu}$, $\nGam^\alpha_{\nu\mu}$, $N_{\nu\mu}$ and $N^{\dashv\nu\mu}$ are all Hermitian,
$\tR_{\nu\mu}$ and $\hR_{\nu\mu}(\nGam)$ are Hermitian from (\ref{transpositionsymmetric}),
and $g_{\nu\mu}$, $A_\sigma$ and ${\mathcal L}$ are real from (\ref{gdef},\ref{A},\ref{JSlag1},\ref{JSlag3}).
If instead $\Lambda_b\!<0,~\Lambda_z\!>0$, then all of the fields are real.
%The allowance of Hermitian fields is not a peculiarity of the present
%theory but is in fact a basic part of the original Einstein-Schr\"{o}dinger
%theory\cite{EinsteinStraus,EinsteinBianchi}.
%Finally, let us discuss the ``simplicity'' of the Lagrangian density (\ref{JSlag1}),
%which is of some relevance in any attempt to unify the laws of physics.
%As shown in equation (\ref{ordinary}) of \S\ref{Discussion}, the original Einstein-Schr\"{o}dinger theory
%can be derived using a Lagrangian density formed from the ordinary Ricci tensor $R_{\nu\mu}(\nGam)$,
%which may seem more acceptable than he Htermitianized Ricci tensor $\hR_{\nu\mu}(\nGam)$.
%We should point out that unlike the Riemann tensor, which can be derived from parallel transporting
%a vector along separate paths to the same point, the familiar Ricci tensor has no such geometrical
%significance. The importance of the Ricci tensor is that it is part of the Einstein equations
%and the Hilbert Lagrangian in ordinary general relativity with symmetric fields.
%It may seem unappealing that the Hermitianized Ricci tensor (\ref{HermitianizedRicci}) contains
%more terms than the ordinary Ricci tensor $R_{\nu\mu}(\nGam)$.

%\newpage
Note that (\ref{gdef}) defines $g^{\mu\nu}$ unambiguously because $\rmg\!=\![-det(\rmg\,g^{\mu\nu})]^{1/(n-2)}$.
In this theory the metric $g_{\mu\nu}$ is used for measuring space-time intervals, and for calculating geodesics,
and for raising and lowering of indices. The covariant derivative ``;'' is always
done using the Christoffel connection formed from $g_{\mu\nu}$,
\begin{eqnarray}
\label{Christoffel}
\Gamma^\alpha_{\nu\mu}&=&\frac{\lower1pt\hbox{$1$}}{2}\,g^{\alpha\sigma}(g_{\mu\sigma,\nu}
+g_{\sigma\nu,\mu}-g_{\nu\mu,\sigma}).
\end{eqnarray}
With the metric (\ref{gdef}), the divergence of the Einstein equations vanishes
when using (\ref{Christoffel}) for the covariant derivative.
And when $N_{\mu\nu}$ and $\nGam^\alpha_{\mu\nu}$ are symmetric,
the definition (\ref{gdef}) requires $g_{\mu\nu}\!=\!N_{\mu\nu}$,
the definition (\ref{A}) requires $A_\sigma\!=\!0$,
%$\delta{\mathcal L}/\delta\nGam^\alpha_{\nu\mu}\!=\!0$ requires
%$\nGam^\alpha_{\nu\mu}\!=\!\Gamma^\alpha_{\nu\mu}$,
and the theory reduces to ordinary general relativity without electromagnetism.

The electromagnetic field is defined in terms of the potential (\ref{A}) as usual
\begin{eqnarray}
\label{Fdef}
F_{\mu\nu}=A_{\nu,\mu}-A_{\mu,\nu}.
\end{eqnarray}
However, we will also define a lowercase $f_{\mu\nu}$
\begin{eqnarray}
\label{fdef}
\rmg\,f^{\mu\nu}=\rmN N^{\dashv[\nu\mu]}\Lambda_b^{\!1/2}/\rmt.
\end{eqnarray}
Then from (\ref{gdef}), $g^{\mu\nu}$ and $f^{\mu\nu}\rmt\Lambda_b^{\!-1/2}$ are
the symmetric and antisymmetric
parts of a total field,
\begin{eqnarray}
\label{Wdef}
\Nbar^{\mu\nu}\!=\!(\rmN/\rmg\,)N^{\dashv\nu\mu}=g^{\mu\nu}\!+\!f^{\mu\nu}\rmt\Lambda_b^{\!-1/2}.
\end{eqnarray}
We will see that the field equations require $f_{\mu\nu}\!\approx\!F_{\mu\nu}$
to a very high precision, so it is mainly just a matter of terminology which one
is called the electromagnetic field.

Note there are many possible nonsymmetric generalizations
of the Ricci tensor besides the
Hermitianized Ricci tensor
$\hR_{\nu\mu}(\nGam)$
from (\ref{HermitianizedRicci})
and the ordinary Ricci tensor $R_{\nu\mu}(\nGam)$.
For example, we could form any weighted average of
$R_{\nu\mu}(\nGam)$,
$R_{\mu\nu}(\nGam)$,
$R_{\nu\mu}(\nGam^T)$
and $R_{\mu\nu}(\nGam^T)$,
and then add any linear combination of the tensors
$R^\alpha{_{\alpha\nu\mu}}(\nGam)$,
$R^\alpha{_{\alpha\nu\mu}}(\nGam^T)$,
%$\nGam^\alpha_{\alpha[\nu,\mu]}$,
%$\nGam^\alpha_{\![\nu|\alpha,|\mu]}$,
%$\nGam^\alpha_{\![\alpha\nu],\mu}\!\!+\nGam^\alpha_{\mu\sigma}\nGam^\sigma_{\![\alpha\nu]}$,
$\nGam^\alpha_{\![\nu\mu]}\nGam^\sigma_{\![\sigma\alpha]}$,
$\nGam^\alpha_{\![\nu\sigma]}\nGam^\sigma_{\![\mu\alpha]}$,
and $\nGam^\alpha_{\![\alpha\nu]}\nGam^\sigma_{\![\sigma\mu]}$.
All of these generalized Ricci tensors would be linear in $\nGam^\alpha_{\nu\mu,\sigma}$, quadratic in $\nGam^\alpha_{\nu\mu}$,
and would reduce to the ordinary Ricci tensor for symmetric fields.
Even if we limit the tensor to only four terms, there are still eight possibilities.
We assert that invariance properties like (\ref{transpositionsymmetric},\ref{gaugesymmetric})
are the most sensible way to choose among the different alternatives,
not criteria such as the number of terms in the expression.

To include additional fields, we would simply append a matter
term ${\mathcal L}_m$ onto (\ref{JSlag3}),
%Any ${\mathcal L}_m$ used with ordinary general relativity can be included,
and this term is assumed to
be formed with the metric $g_{\mu\nu}$, not with $N_{\mu\nu}$.
This ${\mathcal L}_m$ may also contain the vector $A_\mu$ from (\ref{A}),
%which is the electromagnetic potential in this theory.
and charged spin-0 or spin-1/2 fields as in \cite{sShifflett}.
An ${\mathcal L}_m$ containing all of
the additional fields of the Standard Model could even be included.
Of course we would not want to add a $\rmg\,F^{\mu\nu}\!F_{\mu\nu}$
term because this term is effectively already contained in the theory.
In this paper we will only be considering
the Lagrangian density (\ref{JSlag3}) with no ${\mathcal L}_m$ term
appended onto it, which is our equivalent of electro-vac Einstein-Maxwell theory.

Finally, let us discuss some notation issues.
We use the symbol $\Gamma^\alpha_{\!\nu\mu}$ for
the Christoffel connection (\ref{Christoffel}) whereas
Einstein used it for our $\tGam^\alpha_{\!\nu\mu}$
and Schr\"{o}dinger used it for our
$\nGam^\alpha_{\!\nu\mu}$.
We use the symbol $g_{\mu\nu}$ for the symmetric metric (\ref{gdef})
whereas Einstein and Schr\"{o}dinger both used it
for our $N_{\mu\nu}$, the nonsymmetric fundamental tensor.
Also, to represent the inverse of $N_{\alpha\mu}$ we use $N^{\dashv\sigma\alpha}$
instead of the more conventional $N^{\alpha\sigma}$,
because this latter notation would be ambiguous when using
$g^{\mu\nu}$ to raise indices.
While our notation differs from previous literature on the
Einstein-Schr\"{o}dinger theory,
this change is required by our explicit metric definition,
and it is necessary to be consistent with the much
larger body of literature on Einstein-Maxwell theory.

\section{\label{Derivation}Derivation of the Field Equations}
Here we will derive the field equations resulting from the Lagrangian density
(\ref{JSlag3}). Setting $\delta{\mathcal L}/\delta\nGam^\alpha_{\nu\mu}\!=\!0$ will give the
connection equations, and we will see that Ampere's law can be derived from these.
However, as discussed previously, the same field equations must result if we instead
use $\tGam^\alpha_{\nu\mu}$ and $\Aphi_\tau$ as the independent variables,
and since this is simpler we will follow this method. Setting $\delta{\mathcal L}/\delta\Aphi_\tau\!=\!0$
and using (\ref{fdef}) gives Ampere's law,
\begin{eqnarray}
\label{Amperepreliminary}
\fl~~~~~~~~0&=&\frac{4\pi}{\rmg}\!\left[\frac{\partial {\mathcal L}}{\partial \Aphi_\tau}
-\left(\frac{\partial {\mathcal L}}{\partial \Aphi_{\tau,\omega}}\right)\!{_{,\,\omega}}\right]\\
\label{Ampere}
\fl&=&\!\frac{\rmt\Lambda_b^{\!1/2}}{2\rmg}(\rmN N^{\dashv[\omega\tau]})_{,\,\omega}
=\frac{1}{\rmg}(\rmg f^{\omega\tau})_{,\,\omega}
={f^{\omega\tau}}_{;\,\omega}.
\end{eqnarray}
To calculate $\delta{\mathcal L}/\delta\tGam^\alpha_{\nu\mu}$ let us first define
\begin{eqnarray}
\frac{\Delta{\mathcal L}}{\Delta\tGam^\beta_{\tau\rho}}
=\frac{\partial{\mathcal L}}{\partial\tGam^\beta_{\tau\rho}}
-\left(\frac{\partial{\mathcal L}}
{\partial\tGam^\beta_{\tau\rho,\omega}}\right){_{\!,\,\omega}}~....
\end{eqnarray}
Then from (\ref{JSlag3},\ref{HermitianizedRiccit},\ref{Ampere})
we can calculate,
\begin{eqnarray}
\fl -16\pi\frac{\lower1pt\hbox{$\Delta {\mathcal L}$}}
{\Delta\tGam^\beta_{\tau\rho}}&=&
\ifnum\ExpandDerivations=1
2\rmN N^{\dashv\mu\nu}
(\delta^\sigma_\beta\delta^\tau_\nu
\delta^\rho_{[\mu|}\tGam^\alpha_{\sigma|\alpha]}
+\tGam^\sigma_{\nu [\mu|}
\delta^\alpha_\beta\delta^\tau_\sigma\delta^\rho_{|\alpha]})\nonumber\\
\fl &&-2(\rmN N^{\dashv\mu\nu}\delta^\alpha_\beta\delta^\tau_\nu
\delta^\rho_{[\mu}\delta^\omega_{\alpha]}){_{\!,\,\omega}}
-(\rmN N^{\dashv\mu\nu}\delta^\alpha_\beta\delta^\tau_\alpha\delta^\rho_{[\nu}\delta^\omega_{\mu]})_{,\omega}\nonumber\\
\fl &=&
\fi
\label{semivarder}
-(\rmN N^{\dashv\rho\tau})_{\!,\,\beta}
\!-\tGam^\rho_{\beta\mu}\rmN N^{\dashv\mu\tau}
\!-\tGam^\tau_{\nu\beta}\rmN N^{\dashv\rho\nu}
\!+\tGam^\alpha_{\beta\alpha}\rmN N^{\dashv\rho\tau}\nonumber\\
\fl &&+\delta^\rho_\beta((\rmN N^{\dashv\,\omega\tau})_{\!,\,\omega}
+\tGam^\tau_{\nu\mu}\rmN N^{\dashv\mu\nu}),\\
\fl -16\pi\frac{\lower1pt\hbox{$\Delta {\mathcal L}$}}
{\Delta\tGam^\alpha_{\alpha\rho}}
&=&2(\rmN N^{\dashv[\rho\omega]})_{\!,\,\omega}=0,\\
\fl -16\pi\frac{\lower1pt\hbox{$\Delta {\mathcal L}$}}
{\Delta\tGam^\alpha_{\tau\alpha}}
&=&(n\!-\!1)((\rmN N^{\dashv\,\omega\tau})_{\!,\,\omega}
+\tGam^\tau_{\nu\mu}
\rmN N^{\dashv\mu\nu}).
\end{eqnarray}
In these last two equations, the index contractions occur after
the derivatives. At this point we must be careful.
Because ${\tGam^\alpha_{\nu\mu}}$ has the symmetry
(\ref{contractionsymmetric}), it has
only $n^3\!-n$ independent components, so there can only be $n^3\!-n$
independent field equations associated with it. It is shown in
\ref{VariationalDerivative} that instead of just setting (\ref{semivarder}) to zero,
the field equations associated with such a field are given by the expression,
\begin{eqnarray}
\fl 0&=&16\pi\left[\frac{\Delta{\mathcal L}}{\Delta\!\tGam^\beta_{\tau\rho}}
\!-\!\frac{\delta^\tau_\beta}{(n\!-\!1)}
\frac{\Delta {\mathcal L}}{\Delta\!\tGam^\alpha_{\alpha\rho}}
\!-\!\frac{\delta^\rho_\beta}{(n\!-\!1)}\frac{\Delta
{\mathcal L}}{\Delta\!\tGam^\alpha_{\tau\alpha}}\right]\\
\iffalse
\fl&=&(\rmN N^{\dashv\rho\tau})_{\!,\,\beta}
+\tGam^\rho_{\beta\mu}\rmN N^{\dashv\mu\tau}
+\tGam^\tau_{\nu\beta}
\rmN N^{\dashv\rho\nu}-\tGam^\alpha_{\beta\alpha}
\rmN N^{\dashv\rho\tau}\nonumber\\
\fl&&\!\!-\delta^\tau_\beta(\rmN N^{\dashv[\rho\omega]})_{,\omega}
\!+\!\frac{1}{(n\!-\!1)}((n\!-\!2)\delta^\tau_\beta(\rmN N^{\dashv[\rho\omega]})_{,\omega}
\!+\!\delta^\rho_\beta(\rmN N^{\dashv[\tau\omega]})_{,\omega})\nonumber\\
\fl&=&(\rmN N^{\dashv\rho\tau})_{\!,\,\beta}
+\tGam^\tau_{\nu\beta}\rmN N^{\dashv\rho\nu}
+\tGam^\rho_{\beta\mu}\rmN N^{\dashv\mu\tau}
-\tGam^\alpha_{\beta\alpha}\rmN N^{\dashv\rho\tau}\nonumber\\
\fl&&~~~~~~~~~~~~~~~~~~~~~~~~~~~~~~~
-\frac{1}{(n\!-\!1)}(\delta^\tau_\beta(\rmN N^{\dashv[\rho\omega]})_{,\omega}
-\delta^\rho_\beta(\rmN N^{\dashv[\tau\omega]})_{,\omega})\nonumber\\
\fi
\label{Lcontravariant}
\fl &=&(\rmN N^{\dashv\rho\tau})_{\!,\,\beta}
+\tGam^\tau_{\sigma\beta}\rmN N^{\dashv\rho\sigma}
+\tGam^\rho_{\beta\sigma}\rmN N^{\dashv\sigma\tau}
-\tGam^\alpha_{\beta\alpha}\rmN N^{\dashv\rho\tau}.
\end{eqnarray}
These are the connection equations, like $(\rmg g^{\rho\tau}){_{;\beta}}\!=\!0$
in the symmetric case.

%Antisymmetrizing (\ref{Lcontravariant}) and contracting gives Ampere's law
%(\ref{Ampere}).
From the definition of matrix inverse
$N^{\dashv\rho\tau}\!=(1/N)\partial N/\partial N_{\tau\rho}$,~
$N^{\dashv\rho\tau}N_{\tau\mu}\!=\nobreak\delta^\rho_\mu$ we get the identity
\begin{eqnarray}
\label{sqrtdetcomma}
\fl~~~~~~~~(\!\rmN\,)_{,\beta}
=\frac{\partial\rmN}
{\partial N_{\tau\rho}}N_{\tau\rho,\beta}
=\frac{\rmN}{2}N^{\dashv\rho\tau}N_{\tau\rho,\beta}
\label{sqrtdetcomma2}
=-\frac{\rmN}{2}
{N^{\dashv\rho\tau}}_{,\beta} N_{\tau\rho}.
\end{eqnarray}
Contracting (\ref{Lcontravariant}) with $N_{\tau\rho}$
using (\ref{sqrtdetcomma},\ref{contractionsymmetric}) and dividing by $(n\!-\!2)$ gives
\begin{eqnarray}
\label{der0}
\fl~~~~~~~~(\rmN\,)_{,\,\beta}-\tGam^\alpha_{\alpha\beta}\rmN=0.
\end{eqnarray}
This shows that the tensor
$\tGam^\alpha_{\alpha[\nu,\mu]}\!=\!\tilde R^\alpha{_{\!\alpha\mu\nu}}/2$ vanishes,
\begin{eqnarray}
\label{funnytensor}
\fl~~~~~~~~\tGam^\alpha_{\alpha[\nu,\mu]}
=(ln\rmN\,)_{,[\nu,\mu]}=0.
\end{eqnarray}
%Because of (\ref{funnytensor},\ref{Ampere}), the $\tGam^\alpha_{\!\alpha[\nu,\mu]}$
%term in (\ref{HermitianizedRiccit},\ref{JSlag3}) acts like a total divergence and does not affect the field equations.
From (\ref{Lcontravariant},\ref{der0}) we get the contravariant connection equations,
\begin{eqnarray}
\label{contravariant}
\fl~~~~~~~~{N^{\dashv\rho\tau}}_{,\beta}
\!+\!\tGam^\tau_{\sigma\beta}N^{\dashv\rho\sigma}
\!+\!\tGam^\rho_{\beta\sigma}N^{\dashv\sigma\tau}=0.
\end{eqnarray}
Multiplying this by
$-N_{\nu\rho}N_{\tau\mu}$ gives the covariant connection equations,
\begin{eqnarray}
\label{JSconnection0}
\fl~~~~~~~~N_{\nu\mu,\beta}\!-\!\tGam^\alpha_{\nu\beta}N_{\alpha\mu}
\!-\!\tGam^\alpha_{\beta\mu}N_{\nu\alpha}=0.
\end{eqnarray}

Setting $\delta{\mathcal L}/\delta N_{\nu\mu}\!=\!0$
will give the Einstein equations.
% However, the field equations must be the same
%if we instead use $g^{\nu\mu}$ and $f^{\nu\mu}$ as the independent variables.
However, the same field
equations must result if we instead use $\rmN N^{\dashv\mu\nu}$ as the independent
variable, and since this is simpler we will follow this method. Before calculating the
field equations, we need some preliminary results.
From (\ref{gdef}) we get,
\begin{eqnarray}
\label{gcontravariantderiv}
\fl~~~~~\frac{\partial\left(\rmg g^{\rho\tau}\right)}{\partial(\rmN N^{\dashv\mu\nu})}
&=&\delta^{(\rho}_\mu\delta^{\tau)}_\nu,\\
\label{gcovariantderiv}
\fl~~~~~\frac{\partial(g_{\tau\sigma}/\rmg)}{\partial(\rmN N^{\dashv\mu\nu})}
&=&-\frac{g_{\tau(\nu}g_{\mu)\sigma}}{\rmg\rmg}
~~~~~~\left({\rm because}~~
\frac{\partial\left(\rmg g^{\rho\tau}g_{\tau\sigma}/\rmg\right)}
{\partial(\rmN N^{\dashv\mu\nu})}=0\right).
\end{eqnarray}
%For an arbitrary matrix $M^{\nu\mu}$, inverse $M^\dashv_{\mu\nu}$, and scalar
%``$s$'', we have the identities
Using (\ref{gdef}) and the identities
$det(sM^{})\!=s^n det(M^{})$,~$det(M^{-1}_{})\!=1/det(M^{})$ gives
\begin{eqnarray}
%\fl~~~~~\rmg\,g^{\rho\tau}&=&\rmN N^{\dashv(\rho\tau)},\\
%\fl~~~~~\rmg\,f^{\rho\tau}&=&-\rmN N^{\dashv[\rho\tau]},\\
\label{rmN2}
\fl~~~~~\rmN&=&(-det(\rmN N^{\dashv\cdot\cdot}))^{1/(n-2)},\\
\label{rmg2}
\fl~~~~~\rmg&=&(-det(\rmg\,g^{..}))^{1/(n-2)}
=(-det(\rmN N^{\dashv(\cdot\cdot)}))^{1/(n-2)}.
\end{eqnarray}
Using (\ref{rmN2},\ref{rmg2}) and the identity
$\partial(det(M^{\cdot\cdot}))/\partial M^{\mu\nu}\!=M^{-1}_{\nu\mu}det(M^{\cdot\cdot})$
%with $M^{\nu\mu}=\rmN N^{\dashv\mu\nu}$
gives
\begin{eqnarray}
\label{rmNderiv}
\fl~~~~~\frac{\partial\rmN}{\partial(\!\rmN N^{\dashv\mu\nu})}
\!&=&\!\frac{(-det(\rmN N^{\dashv\cdot\cdot}))^{1/(n-2)-1+1}}{(n\!-\!2)}\frac{N_{\nu\mu}}{\rmN}
=\frac{N_{\nu\mu}}{(n\!-\!2)},\\
\label{rmgderiv}
\fl~~~~~\frac{\partial\rmg}{\partial(\!\rmN N^{\dashv\mu\nu})}
\!&=&\!\frac{(-det(\rmg\,g^{..}))^{1/(n-2)-1+1}}{(n\!-\!2)}\frac{g_{\nu\mu}}{\rmg}
=\frac{g_{\nu\mu}}{(n\!-\!2)}.
\end{eqnarray}
Note that from (\ref{gcontravariantderiv},\ref{gcovariantderiv},\ref{rmgderiv}), if there
was a matter term ${\mathcal L}_m$ in (\ref{JSlag3})
%Note that if (\ref{JSlag3}) contained a matter term ${\mathcal L}_m$
which depended only on $g_{\nu\mu}$ and not on $N_{\nu\mu}$,
then $\partial{\mathcal L}_m/\partial(\rmN N^{\dashv\mu\nu})\!=\partial{\mathcal L}_m/\partial(\rmg\,g^{\mu\nu})$.
Setting $\delta{\mathcal L}/\delta(\rmN N^{\dashv\mu\nu})\!=\!0$ and using (\ref{rmNderiv},\ref{rmgderiv}) gives,
\begin{eqnarray}
\fl~~~~~~~~0&=&-16\pi\left[\frac{\partial{\mathcal L}}{\partial(\rmN N^{\dashv\mu\nu})}
-\left(\frac{\partial{\mathcal L}}{\partial(\rmN N^{\dashv\mu\nu})_{\!,\,\omega}}\right){_{\!,\,\omega}}\right]\\
\label{para}
\fl~~~~~~~~&=&\tR_{\nu\mu}
%\!+\!\tGam^\alpha_{\alpha[\nu,\mu]}
\!+2\Aphi_{[\nu,\mu]}\rmt\Lambda_b^{\!1/2}
\!+\Lambda_b N_{\nu\mu}
\!+\Lambda_z g_{\nu\mu}.
\end{eqnarray}
Using the definition (\ref{Fdef}), the antisymmetric part of this is
\begin{eqnarray}
\label{JSantisymmetric}
\fl~~~~~~~~N_{[\nu\mu]}=F_{\nu\mu}\rmt\Lambda_b^{\!-1/2}-\tR_{[\nu\mu]}\Lambda_b^{\!-1}.
\end{eqnarray}

Taking the symmetric part of (\ref{para}) and the curl of (\ref{JSantisymmetric}) and repeating
(\ref{JSconnection0},\ref{contractionsymmetric}) gives the field equations
in the form usually used to define the Einstein-Schr\"{o}dinger theory,
\begin{eqnarray}
\label{JSsymmetric}
\tR_{(\nu\mu)}
+\Lambda_b N_{(\nu\mu)}+\Lambda_z g_{\nu\mu}=0,\\
%=\frac{8\pi G}{c^4}\!\left(T_{\nu\mu}
%-\frac{1}{(n\!-\!2)}g_{\nu\mu}T^\alpha_\alpha\right),\\
\label{JScurl}
\tR_{[\nu\mu,\sigma]}+\Lambda_b N_{[\nu\mu,\sigma]}=0,\\
%\label{contravariantdensity}
%\fl~~~(\!\rmN \K^{\dashv\rho\tau})_{\!,\,\beta}
%\!+\!\tGam^\tau_{\sigma\beta}\rmN \K^{\dashv\rho\sigma}
%\!+\!\tGam^\rho_{\beta\sigma}\rmN \K^{\dashv\sigma\tau}
%\!-\!\tGam^\alpha_{\beta\alpha}\rmN \K^{\dashv\rho\tau}=0,\\
%\fl~~~~~~~~~~~~~~~~~~~~~~~~~~~~~~~~~~~~~~~~~~~~~~~~~~~~~~~~~~~~~~~~~
%\!=\!\frac{8\pi}{c(n\!-\!1)\!}\rmg j^{[\rho}\delta^{\tau]}_\beta,\\
\label{JSconnection}
N_{\nu\mu,\beta}\!-\!\tGam^\alpha_{\nu\beta}N_{\alpha\mu}
\!-\!\tGam^\alpha_{\beta\mu}N_{\nu\alpha}=0,\\
\label{JScontractionsymmetric}
\tGam^\alpha_{\beta\alpha}=\tGam^\alpha_{\alpha\beta}.
\end{eqnarray}
If desired we could start from these equations instead of the Lagrangian density (\ref{JSlag1}).
That is, the symmetric part of (\ref{para}) comes from (\ref{JSsymmetric}), and
the antisymmetric part of (\ref{para}) is implied by (\ref{JScurl}) for some $\Aphi_\mu$.
%according to a theorem of tensor calculus\cite{Adler}.
Also, Ampere's law (\ref{Ampere}) can be derived from (\ref{JSconnection},\ref{JScontractionsymmetric})
by employing (\ref{contravariant},\ref{sqrtdetcomma}) to get (\ref{der0},\ref{Lcontravariant}),
and then antisymmetrizing and contracting (\ref{Lcontravariant}).
%Finally, from (\ref{funnytensor}) we see that
%the $\tGam{^\alpha_{\alpha[\nu,\mu]}}$ term in (\ref{para}) could be dropped,
%and this is done in many other treatments\cite{SchrodingerI,SchrodingerSTS}.
%However, when additional fields are included in the theory as in
%\cite{sShifflett,Shifflett2,Antoci3}, (\ref{funnytensor}) is not true
%so the $\tGam{^\alpha_{\alpha[\nu,\mu]}}$ term must be retained.
%or equivalently the Hermitianized Ricci tensor (\ref{HermitianizedRiccit}) must be used.

%\newpage
The Einstein equations are obtained by combining (\ref{JSsymmetric}) with
its contraction,
\begin{eqnarray}
\label{Einsteinexact}
\tG_{\nu\mu}
+\Lambda_b\!\left(N_{(\nu\mu)}
-\frac{\lower0pt\hbox{$1$}}{2}\,g_{\nu\mu}N^\rho_\rho\right)
+\!\Lambda_z\!\left(1\!-\!\frac{n}{2}\right)g_{\nu\mu}=0.
\end{eqnarray}
where we define
\begin{eqnarray}
\label{genEinstein}
\tG_{\nu\mu}
&=&\tR_{(\nu\mu)}
-\frac{\lower1pt\hbox{$1$}}{2}g_{\nu\mu}\tR^\rho_\rho.
\end{eqnarray}
A generalized contracted Bianchi identity for this theory is derived in \cite{EinsteinBianchi}
using only the connection equations (\ref{JSconnection})
and the symmetry (\ref{JScontractionsymmetric}).
%\begin{eqnarray}
%\fl~~~~~~~~~~(\rmN N^{\dashv\nu\sigma}\tR{_{\sigma\lambda}}
%+\rmN N^{\dashv\sigma\nu}\tR_{\lambda\sigma}){_{,\nu}}
%-\rmN N^{\dashv\nu\sigma}\tR_{\sigma\nu,\lambda}=0.
%\end{eqnarray}
When expressed
in terms of our metric (\ref{gdef})
and the definitions (\ref{fdef},\ref{genEinstein}),
this identity becomes\cite{JohnsonI,Antoci3,sShifflett}
\begin{eqnarray}
\label{contractedBianchi}
\!\!\!\tG^\sigma_{\nu;\,\sigma}
=\frac{\lower1pt\hbox{$3$}}{2}f^{\sigma\rho}\tR_{[\sigma\rho,\nu]}\rmt\Lambda_b^{\!-1/2}.
\end{eqnarray}
Another useful identity is derived in \ref{UsefulIdentity}
using only the definitions (\ref{gdef},\ref{fdef}) of $g_{\nu\mu}$ and $f_{\nu\mu}$,
\begin{eqnarray}
\label{usefulidentity}
\fl~~~~~~~~~~\left(\K^{(\mu}{_{\nu)}} \!-\!\frac{1}{2}\delta^\mu_\nu
\K^\rho_\rho\right)\!{_{;\,\mu}}
=\left(\frac{3}{2}f^{\sigma\rho}N_{[\sigma\rho,\nu]}
+f^{\sigma\rho}{_{;\sigma}}N_{[\rho\nu]}\right)\rmt\Lambda_b^{\!-1/2}.
\end{eqnarray}
Using (\ref{contractedBianchi},\ref{usefulidentity},\ref{Ampere},\ref{JScurl})
we see that the divergence of the Einstein equations (\ref{Einsteinexact}) vanishes
\begin{eqnarray}
\fl\left[\tG^\mu_\nu
\!+\!\Lambda_b\!\left(N^{(\mu}{_{\nu)}}\!-\!\frac{1}{2}\delta^\mu_\nu N^\rho_\rho\right)
\!+\!\Lambda_z\!\left(1\!-\!\frac{n}{2}\right)\delta^\mu_\nu\right]\!\!{_{;\,\mu}}\nonumber\\
\label{zerodivergence}
~~~~~~=\!\frac{3}{2}f^{\sigma\rho}\tR_{[\sigma\rho,\nu]}\rmt\Lambda_b^{\!-1/2}
\!+\!\Lambda_b\frac{3}{2}f^{\sigma\rho}N_{[\sigma\rho,\nu]}\rmt\Lambda_b^{\!-1/2}\!=0.
\end{eqnarray}
This is why the metric (\ref{gdef}) was assumed, because the
covariant derivative in the equations above is done using the Christoffel connection
(\ref{Christoffel}) formed from this metric, and the result (\ref{zerodivergence})
would not occur for any other metric.
It is also why neutral matter terms formed with $g_{\mu\nu}$
can be appended onto the Lagrangian density (\ref{JSlag3}), because
it means that such terms will create divergenceless energy-momentum terms
in (\ref{Einsteinexact}).
Now, when charged matter terms containing $A_\mu$ are appended onto (\ref{JSlag3}),
a few other equations besides (\ref{Einsteinexact}) acquire additional terms.
It is shown in \cite{sShifflett} that the divergence of the Einstein equations does not
vanish in this case, but instead gives the Lorentz force equation,
just as in Einstein-Maxwell theory with sources.

To show that the field equations (\ref{JSsymmetric}-\ref{JScontractionsymmetric})
closely approximate electro-vac Einstein-Maxwell theory we will need to make some approximations.
The definitions (\ref{gdef},\ref{fdef}) of $g_{\nu\mu}$ and $f_{\nu\mu}$
can be inverted to give $N_{\nu\mu}$ in terms of $g_{\nu\mu}$ and $f_{\nu\mu}$.
An expansion in powers of $\Lambda_b^{\!-1}$ is derived in \ref{ApproximateFandg} and
confirmed with tetrad methods in \cite{Shifflett2},
\begin{eqnarray}
\label{approximateNbar}
\fl~~~~~ N_{(\nu\mu)}&\!\!=\!& g_{\nu\mu}-2\!\left({f_\nu}^\sigma f_{\sigma\mu}
-\frac{1}{2(n\!-\!2)}g_{\nu\mu}f^{\rho\sigma}\!f_{\sigma\rho}\right)\!\Lambda_b^{\!-1}
+(f^4)\Lambda_b^{\!-2}\dots\\
\label{approximateNhat}
\fl~~~~~ N_{[\nu\mu]}&\!\!=\!& f_{\nu\mu}\rmt\Lambda_b^{\!-1/2}
+(f^3)\Lambda_b^{\!-3/2}\dots.
\end{eqnarray}
%where
%\begin{eqnarray}
%\label{ff}
%~\ff=f^{\rho\sigma}f_{\sigma\rho}.
%\end{eqnarray}
The connection equations (\ref{JSconnection}) can be solved similar
to the way that $g_{\mu\nu;\alpha}\!=\nobreak\!0$ is solved to get the
Christoffel connection\cite{Tonnelat}.
%Also, the exact algebraic solution $\tGam^\alpha_{\nu\mu}(N_{\sigma\rho})$
%to (\ref{covariant}) is very complicated
%in the general non-symmetric case.
An expansion in powers of $\Lambda_b^{\!-1}$ is derived in
\cite{sShifflett},
%\ref{ApproximateGamma},
confirmed by tetrad methods in \cite{Shifflett2}, and is also stated without derivation in \cite{Antoci3},
\begin{eqnarray}
\label{gammadecomposition}
\fl\tGam^\alpha_{\nu\mu}&\!\!=\!&\Gamma^\alpha_{\nu\mu}
+\Upsilon^\alpha_{\nu\mu},\\
\fl \Upsilon^\alpha_{\!(\nu\mu)}
\label{upsilonsymmetric}
&\!\!=\!&\!-2\left(f^\tau{_{\!\!(\nu}}f_{\mu)}{\!^\alpha}{_{\!;\tau}}
\!+f^{\alpha\tau}\!f_{\tau(\nu;\mu)}
\!+\!\frac{1}{4(n\!-\!2)}((f^{\rho\sigma}\!f_{\sigma\rho})_,{^\alpha}g_{\nu\mu}
\!-2(f^{\rho\sigma}\!f_{\sigma\rho})_{,(\nu}\delta^\alpha_{\mu)})\right)\!\Lambda_b^{\!-1}\nonumber\\
\fl &&+(f^{4\prime})\Lambda_b^{\!-2}\dots,\\
\label{upsilonantisymmetric}
\fl \Upsilon^\alpha_{\![\nu\mu]}
&\!\!=\!&\frac{1}{2}(f_{\nu\mu;}{^\alpha}+f^\alpha{_{\mu;\nu}}
-f^\alpha{_{\nu;\mu}})\rmt\Lambda_b^{\!-1/2}+(f^{3\prime})\Lambda_b^{\!-3/2}\dots.
%\label{upsiloncontracted}
%\fl \Upsilon^\alpha_{\alpha\nu}&\!\!=\!&\frac{-1}{2(n\!-\!2)}(f^{\rho\sigma}\!f_{\sigma\rho})_{,\nu}+(f^4).
\end{eqnarray}
Here $\Gamma^\alpha_{\nu\mu}$ is the Christoffel connection (\ref{Christoffel}).
%formed from the metric (\ref{gdef}).
%The tensor $\tGam^\alpha_{\![\nu\mu]}\!=\!\Upsilon^\alpha_{\![\nu\mu]}$
%is often called the ``torsion''.
In (\ref{approximateNbar}-\ref{upsilonantisymmetric})
the notation $(f^3)$ and $(f^4)$ refers to terms like
$f_{\nu\alpha}f^\alpha{_\sigma}f^\sigma{_\mu}$ and
$f_{\nu\alpha}f^\alpha{_\sigma}f^\sigma{_\rho}f^\rho{_\mu}$,
and the notation $(f^{3\prime})$ and $(f^{4\prime})$ refers to terms like
or $f^\alpha{_\tau}f^\tau{_\sigma}f^\sigma{_{[\nu;\mu]}}$ and
$f^\alpha{_\tau}f^\tau{_\sigma}f^\sigma{_\rho}f^\rho{_{(\nu;\mu)}}$.
Let us consider worst-case values of these higher order terms relative to the leading order terms.
From \S\ref{Monopole} we know there is an exact electric monopole solution
for this theory which approximates a $f^1{_0}\!\sim\!Q/r^2$ field.
%For a solar mass extremal charged black hole with $r\!=\!Q\!=\!M_\odot=1.48\times 10^5cm$ we have
%\begin{eqnarray}
%\label{BHskew}
%\frac{|f^1{_0}|_{BH}^2}{\Lambda_b}\sim
%\frac{(Q/r^2)^2}{\Lambda_b}\sim 10^{-76}
%\end{eqnarray}
In geometrized units an elementary charge has
\begin{eqnarray}
\label{redef}
\fl~~~~~~~~~Q_e=e\sqrt{\frac{G}{c^4}}=\sqrt{\frac{e^2}{\hbar c}\frac{G\hbar}{c^3}}=\sqrt{\alpha}\,l_P=1.38\times 10^{-34}cm
\end{eqnarray}
where $\alpha =e^2/\hbar c$ is the fine structure constant
and $l_P\!=\!\sqrt{G\hbar/c^3}$ is Planck's constant.
If we assume that charged particles retain $f^1{_0}\!\sim\!Q/r^2$
down to the smallest radii probed by high energy particle physics
experiments ($10^{-17}{\rm cm}$) we have,
\begin{eqnarray}
\label{highenergyskew}
|f^1{_0}|^2/\Lambda_b\sim (Q_e/(10^{-17})^2)^2/\Lambda_b\sim 10^{-66}.
\end{eqnarray}
Here $|f^1{_0}|$ is assumed to be in some
standard spherical or cartesian coordinate system. If an equation has a tensor term which can
be neglected in one coordinate system, it can be neglected in any coordinate system,
so it is only necessary to prove it in one coordinate system.
The fields at $10^{-17}{\rm cm}$ from an elementary charge
%The fields at this radius
would be larger than near any macroscopic charged
object and would also be larger than the strongest plane-wave fields.
Therefore the higher order terms in (\ref{approximateNbar}-\ref{upsilonantisymmetric})
must be $<\!10^{-66}$ of the leading order terms, so they will be completely negligible for most purposes.
%The ``smallness'' of $N_{[\mu\nu]}$ is discussed in
%\S\ref{SymmetricPart}, first paragraph, and in \ref{Monopole}, \S 4, last paragraph.
Approximate field equations can be obtained by substituting
(\ref{approximateNbar}-\ref{upsilonantisymmetric})
into (\ref{para}), and we will do this in
\S\ref{SymmetricPart}-\S\ref{AntisymmetricPart}.
This gives a set of field equations where $g_{\nu\mu}$ and $f_{\nu\mu}$
are the unknowns instead of $N_{\nu\mu}$. It matters little
whether we solve for $g_{\nu\mu}$ and $f_{\nu\mu}$ or for $N_{\nu\mu}$
because they are just related algebraically via (\ref{gdef},\ref{fdef}). The
advantage of writing equations in terms of $g_{\nu\mu}$ and $f_{\nu\mu}$
is that the close approximation of (\ref{Einsteinexact},\ref{Ampere}) to the
ordinary Einstein and Maxwell equations will become apparent.

%\pagebreak
\section{\label{SymmetricPart}The Symmetric Part of the Field Equations}
Here we show that the symmetric part of the field equations
contains a close approximation to the ordinary Einstein equations
of electro-vac Einstein-Maxwell theory.
Subsitituting (\ref{approximateNbar},\ref{Lambdadef})
into (\ref{Einsteinexact})
\ifnum\ExpandDerivations=1
%\bigskip
\begin{eqnarray}
\fl N_{(\nu\mu)}
-\frac{\lower0pt\hbox{$1$}}{2}\,g_{\nu\mu}N^\rho_\rho
\!&=&g_{\nu\mu}-2\left({f_\nu}^\sigma f_{\sigma\mu}
-\frac{1}{2(n\!-\!2)}g_{\nu\mu}f^{\rho\sigma}\!f_{\sigma\rho}\right)\!\Lambda_b^{\!-1}\nonumber\\
\fl&&-\frac{1}{2}g_{\nu\mu}n+g_{\nu\mu}\!\left(f^{\rho\sigma}\!f_{\sigma\rho}
-\frac{1}{2(n\!-\!2)}nf^{\rho\sigma}\!f_{\sigma\rho}\right)\!\Lambda_b^{\!-1}
+(f^4)\Lambda_b^{\!-2}\dots\nonumber\\
%\!&=&g_{\nu\mu}\left(1-\frac{n}{2}\right)
%-2{f_\nu}^\sigma f_{\sigma\mu}\Lambda_b^{\!-1}\nonumber\\
%\fl&&+g_{\nu\mu}\!\left(\frac{1}{(n\!-\!2)}+\!1
%\!-\frac{n}{2(n\!-\!2)}\right)f^{\rho\sigma}\!f_{\sigma\rho}\Lambda_b^{\!-1}
%+(f^4)\Lambda_b^{\!-2}\dots\nonumber\\
\!&=&-2\left({f_\nu}^\sigma f_{\sigma\mu}
-\frac{1}{4}g_{\nu\mu}f^{\rho\sigma}\!f_{\sigma\rho}\right)\!\Lambda_b^{\!-1}
-\left(\frac{n}{2}-1\right)g_{\nu\mu}
+(f^4)\Lambda_b^{\!-2}\dots\nonumber
\end{eqnarray}
\fi
gives approximate Einstein equations,
\begin{eqnarray}
\label{Einstein3}
\fl~~~~~~\tG_{\nu\mu}&=&
2\left({f_\nu}^\sigma f_{\sigma\mu}
\!-\!\frac{1}{4}g_{\nu\mu}f^{\rho\sigma}\!f_{\sigma\rho}\right)
+\Lambda\left(\frac{n}{2}-1\right)g_{\nu\mu}
+(f^4)\Lambda_b^{\!-1}\dots.
\end{eqnarray}
By substituting
(\ref{gammadecomposition}-\ref{upsilonantisymmetric},\ref{Ampere})
into (\ref{Einstein3},\ref{Ricciadditionsymmetric},\ref{genEinstein})
\ifnum\ExpandDerivations=1
with $\ff=f^{\rho\sigma}\!f_{\sigma\rho}$,
\begin{eqnarray}
\fl R_{\nu\mu}\!
&=& \tR_{(\nu\mu)}-\Upsilon^\alpha_{(\nu\mu);\,\alpha}
\!+\Upsilon^\alpha_{\alpha(\nu;\mu)}
\!+\Upsilon^\sigma_{[\nu\alpha]}\Upsilon^\alpha_{[\sigma\mu]}\dots\nonumber\\
\fl &=&\tR_{(\nu\mu)}
+2\left(f^\tau{_{(\nu}}f_{\mu)}{^\alpha}{_{;\,\tau}}
+f^{\alpha\tau}f_{\tau(\nu;\,\mu)}
+\frac{1}{4(n\!-\!2)}(\ff_,{^\alpha}g_{\nu\mu}
-2\ff_{,(\nu}\delta^\alpha_{\mu)})\right)\!{_{;\,\alpha}}\Lambda_b^{\!-1}\nonumber\\
\fl &+&\frac{1}{(n\!-\!2)}\ff_{,(\nu;\,\mu)}\Lambda_b^{\!-1}
-\frac{1}{2}\left(f_{\nu\alpha;}{^\sigma}\!+\!f^\sigma{_{\alpha;\nu}}
\!-\!f^\sigma{_{\nu;\,\alpha}}\right)
\left(f_{\sigma\mu;}{^\alpha}\!+\!f^\alpha{_{\mu;\,\sigma}}
\!-\!f^\alpha{_{\sigma;\,\mu}}
\right)\Lambda_b^{\!-1}\dots\nonumber\\
\label{Riccitensor}
\fl &=&\tR_{(\nu\mu)}
+\left(2f^\tau{_{(\nu}}f_{\mu)}{^\alpha}{_{;\tau;\alpha}}
+2f^{\alpha\tau}f_{\tau(\nu;\,\mu)}{_{;\,\alpha}}
+\frac{1}{2(n\!-\!2)}\ff_,{^\alpha}{_{;\alpha}}g_{\nu\mu}\right.\nonumber\\
\fl&&~~~~~~~~~~~~~~~~~~~~~~\left.-f^\sigma{_{\nu;\alpha}}f^\alpha{_{\mu;\sigma}}
+f^\sigma{_{\nu;\,\alpha}}f_{\sigma\mu;}{^\alpha}
+\frac{1}{2}f^\sigma{_{\alpha;\nu}}f^\alpha{_{\sigma;\,\mu}}\right)\!\Lambda_b^{\!-1}\dots,\nonumber\\
\fl~~R&=&\tR^\rho_\rho
+\left(2f^{\tau\beta}f_{\beta}{^\alpha}{_{;\tau;\alpha}}
+\frac{n}{2(n\!-\!2)}\ff_,{^\alpha}{_{;\alpha}}
-f{^{\sigma\beta}}_{;\alpha}f^\alpha{_{\beta;\sigma}}
+\frac{1}{2}f{^{\sigma\beta}}_{;\,\alpha}f_{\sigma\beta;}{^\alpha}\right)\!\Lambda_b^{\!-1}\dots\nonumber\\
\label{Ricciscalar}
\fl &=&\tR^\rho_\rho
+\left(2f^{\tau\beta}f_{\beta}{^\alpha}{_{;\tau;\alpha}}
+\frac{n}{2(n\!-\!2)}\ff_,{^\alpha}{_{;\alpha}}
+\frac{3}{2}f_{[\sigma\beta;\alpha]}f^{[\sigma\beta}{_;}{^{\alpha]}}\right)\!\Lambda_b^{\!-1}\dots,\nonumber
\end{eqnarray}
\fi
we see that the Einstein equations (\ref{Einstein3}) can be rewritten in the form
\begin{eqnarray}
\label{Einstein}
\fl~~~G_{\nu\mu}&=&
8\pi\tT_{\nu\mu}+\Lambda\left(\frac{n}{2}-1\right)g_{\nu\mu},
\end{eqnarray}
where
\begin{eqnarray}
\fl~~~G_{\nu\mu}=R_{\nu\mu}-\frac{1}{2}g_{\nu\mu}R,\\
\label{Ttilde}
\fl 8\pi\tT_{\nu\mu}
=2\left({f_\nu}^\sigma f_{\sigma\mu}
-\frac{1}{4}g_{\nu\mu}f^{\rho\sigma}\!f_{\sigma\rho}\right)\nonumber\\
\fl~~+\!\left(2f^\tau{_{(\nu}}f_{\mu)}{^\alpha}{_{;\tau;\alpha}}
+2f^{\alpha\tau}f_{\tau(\nu;\,\mu)}{_{;\alpha}}
-f^\sigma{_{\nu;\alpha}}f^\alpha{_{\mu;\sigma}}
\!+f^\sigma{_{\nu;\alpha}}f_{\sigma\mu;}{^\alpha}
+\frac{1}{2}f^\sigma{_{\alpha;\nu}}f^\alpha{_{\sigma;\,\mu}}\right.\nonumber\\
\fl~~~~~~\left.-g_{\nu\mu}f^{\tau\beta}f_{\beta}{^\alpha}{_{;\tau;\alpha}}
-\frac{1}{4}g_{\nu\mu}(f^{\rho\sigma}\!f_{\sigma\rho})_,{^\alpha}{_{;\alpha}}
-\frac{3}{4}g_{\nu\mu}f_{[\sigma\beta;\alpha]}f^{[\sigma\beta}{_;}{^{\alpha]}}+(f^4)\right)\!\Lambda_b^{\!-1}\dots.
\end{eqnarray}
In (\ref{Einstein}-\ref{Ttilde}), $G_{\nu\mu}$, $R_{\nu\mu}$, and $R$
are formed from the Christoffel connection (\ref{Christoffel}),
$\Lambda$ is the small ``physical'' cosmological constant (\ref{Lambdadef},\ref{Lambda}),
and $\tT_{\nu\mu}$ is our ``effective'' energy-momentum tensor.
The $(f^4)\Lambda_b^{\!-1}$ term is $<\!10^{-66}$ of the ordinary
electromagnetic term because of (\ref{highenergyskew}). To evaluate the relative
contribution of the remaining terms let us consider some worst-case values of
$|f^\mu{_{\sigma;\alpha}}|$ and $|f^\mu{_{\sigma;\alpha;\beta}}|$ accessible to measurement.
From \S\ref{Monopole} we know there is an exact electric monopole solution
for this theory which approximates a $f^1{_0}\!\sim\!Q/r^2$ field. If we assume
that charged particles retain $f^1{_0}\!\sim\!Q/r^2$ down to very small radii,
the values of $|f^\mu{_{\sigma;\alpha}}|$ and $|f^\mu{_{\sigma;\alpha;\beta}}|$
there would be greater than from any macroscopic monopole field.
For the smallest radii probed by high energy particle physics
experiments ($10^{-17}{\rm cm}$) we have from (\ref{Lambdab}),
\begin{eqnarray}
\label{highenergyderiv1}
|f^1{_{0;1}}/f^1{_0}|^2/\Lambda_b&\sim& 4/\Lambda_b\,(10^{-17})^2\sim 10^{-32},\\
\label{highenergyderiv2}
|f^1{_{0;1;1}}/f^1{_0}|/\Lambda_b&\sim& 6/\Lambda_b\,(10^{-17})^2\sim 10^{-32}.
\end{eqnarray}
So for electric monopole fields, the extra terms in (\ref{Ttilde})
must be $<\!10^{-32}$ of the ordinary electromagnetic term.
%From \S\ref{Planewave} we also know that there is an exact electromagnetic
%plane-wave solution to this theory which is identical to one for
%Einstein-Maxwell theory.
For an electromagnetic plane-wave in a flat background space we have,
\begin{eqnarray}
\label{planewaveA}
A_\mu&=&A\epsilon_\mu{\rm sin}(k_\alpha x^\alpha)
~~,~~\epsilon^\alpha\epsilon_\alpha=-1
~~,~~k^\alpha k_\alpha=k^\alpha\epsilon_\alpha=0,\\
\label{planewavef}
f_{\nu\mu}&=&2\Aphi_{[\mu,\nu]}
=2A\epsilon_{[\mu} k_{\nu]}{\rm cos}(k_\alpha x^\alpha).
%\fl~~~~~~\frac{8\pi G}{c^4}\,{^\Aphi}T_{\nu\mu}
%\approx-A^2\Lambda_bk_\nu k_\mu{\rm cos}^2(k_\alpha x^\alpha)
%=-\frac{A^2\Lambda_b}{2}k_\nu k_\mu(1+{\rm cos}(2k_\alpha x^\alpha)).
\end{eqnarray}
Here $A$ is the magnitude, $k^\alpha$ is the wavenumber, and
$\epsilon^\alpha$ is the polarization.
Substituting (\ref{planewaveA},\ref{planewavef}) into (\ref{Ttilde}),
it is easy to see that for flat space all of the extra terms
of (\ref{Ttilde}) vanish, and we have as usual,
\begin{eqnarray}
\label{TA}
%\fl~~~~~~ \Aphi_\mu=\Aphimag\epsilon_\mu{\rm sin}(k_\alpha x^\alpha)
%~~,~~\epsilon^\alpha\epsilon_\alpha=-1
%~~,~~k^\alpha k_\alpha=k^\alpha\epsilon_\alpha=0,\\
%\fl~~~~~~ f_{\nu\mu}=2\Aphi_{[\mu,\nu]}
%=2\Aphimag\epsilon_{[\mu} k_{\nu]}{\rm cos}(k_\alpha x^\alpha),\\
\fl~~~~~~~~8\pi\,{^\Aphi}T_{\nu\mu}
&\approx&-\Aphimag^2\Lambda_bk_\nu k_\mu{\rm cos}^2(k_\alpha x^\alpha)
=-\frac{\Aphimag^2\Lambda_b}{2}k_\nu k_\mu(1+{\rm cos}(2k_\alpha x^\alpha)).
\end{eqnarray}
Also, for the highest energy gamma rays known in nature ($10^{20}$eV) we have from (\ref{Lambdab}),
\begin{eqnarray}
\label{gammaderiv1}
|f^1{_{0;1}}/f^1{_0}|^2/\Lambda_b&\sim& (E/\hbar c)^2/\Lambda_b\sim 10^{-16},\\
\label{gammaderiv2}
|f^1{_{0;1;1}}/f^1{_0}|/\Lambda_b&\sim& (E/\hbar c)^2/\Lambda_b\sim 10^{-16}.
\end{eqnarray}
So for electromagnetic plane-wave fields, even if some of the extra terms
in (\ref{Ttilde}) were non-zero because of spatial curvatures,
they must still be $<\!10^{-16}$ of the ordinary electromagnetic term.
%Therefore from
%(\ref{highenergyderiv1},\ref{highenergyderiv2},\ref{gammaderiv1},\ref{gammaderiv2}),
%even in the most extreme worst-cases accessible to measurement,
%the extra terms in (\ref{Einstein}-\ref{Ttilde})
%are all $<\!10^{-16}$ of the ordinary electromagnetic term.
Therefore the extra terms in
(\ref{Einstein}-\ref{Ttilde}) must be $<\!10^{-16}$ of the ordinary electromagnetic term
for even the most extreme worst-case fields accessible to measurement.
And we see that (\ref{Einstein}-\ref{Ttilde})
go to the exact electro-vac Einstein equations in the limit as $\Lambda_b\!\rightarrow\!\infty$.

Now, $G_{\nu\mu}$ in (\ref{Einstein}) is the ordinary Einstein tensor, so
the ordinary contracted Bianchi identity $G^\sigma_{\nu;\,\sigma}\!=\!0$ applies.
For the Einstein equations (\ref{Einstein}) to be compatible,
the divergence of these equations should vanish identically, and therefore the
divergence of $\tT_{\nu\mu}$ from (\ref{Ttilde}) should vanish.
This should be expected to occur automatically because we have already
shown in (\ref{zerodivergence}) that the divergence of the exact Einstein
equations (\ref{Einsteinexact}) vanishes, and because the field equations are
derived from a variational principle. Regardless of whether one believes this argument,
it can be shown using (\ref{Ampere},\ref{JScurl},\ref{approximateNhat})
that it is indeed true that,
\begin{eqnarray}
\label{Tdivergence}
\tT^\sigma_{\nu;\,\sigma}={\rm at~most}~\ord(\Lambda_b^{-2}).
\end{eqnarray}
The calculation is rather lengthy so we will omit it. Finally, note that because
the divergence of the exact Einstein equations (\ref{Einsteinexact}) vanishes, our
``effective'' energy-momentum tensor $\tT_{\nu\mu}$ in (\ref{Einstein},\ref{Ttilde})
can be augmented by additional energy-momentum tensor contributions
caused by non-electromagnetic matter fields. This occurs when additional fields are
included in the Lagrangian density as in \cite{sShifflett}.

\section{\label{AntisymmetricPart}The Antisymmetric Part of the Field
Equations} Here we show that the antisymmetric part of the field
equations contain a very close approximation to the ordinary Maxwell equations
of electro-vac Einstein-Maxwell theory.
Substituting (\ref{upsilonantisymmetric},\ref{Ampere}),
into (\ref{Ricciadditionantisymmetric}) gives
\begin{eqnarray}
\fl\tR_{[\nu\mu]}&=&
\ifnum\ExpandDerivations=1
{\Upsilon}^\alpha_{\![\nu\mu];\alpha}+\ord(\Lambda_b^{\!-3/2})\dots\nonumber\\
\fl &=&\!\frac{1}{2}(f_{\nu\mu;}{^\alpha}
\!+\!f^\alpha{_{\mu;\nu}}\!-\!f^\alpha{_{\nu;\mu}}){_{;\alpha}}
\rmt\Lambda_b^{\!-1/2}\dots\nonumber\\
\fl &=&\!\left(\!\frac{3}{2}f_{[\nu\mu,\alpha];}{^\alpha}
+f^\alpha{_{\mu;\nu;\alpha}}-f^\alpha{_{\nu;\mu;\alpha}}
\right)\!\rmt\Lambda_b^{\!-1/2}\dots\nonumber\\
\fl &=&
\fi
\label{antisymmetricpreliminary}
\!\left(\!\frac{3}{2}f_{[\nu\mu,\alpha];}{^\alpha}
\!+\!2f^\alpha{_{\mu;[\nu;\alpha]}}\!-\!2f^\alpha{_{\nu;[\mu;\alpha]}}
\right)\!\rmt\Lambda_b^{\!-1/2}\dots.
\end{eqnarray}
From the covariant derivative commutation rule,
%the cyclic identity $2R_{\nu[\tau\alpha]\mu}=R_{\nu\mu\alpha\tau}$,
the definition of the Weyl tensor $C_{\nu\mu\alpha\tau}$,
and the Einstein equations $R_{\nu\mu}=-\Lambda g_{\nu\mu}+(f^2)\dots$ from
(\ref{Einstein},\ref{Ttilde}) we get
\begin{eqnarray}
\fl 2f^\alpha{_{\nu;[\mu;\alpha]}}&=&
\ifnum\ExpandDerivations=1
R^\tau{_{\nu\mu\alpha}}f^\alpha{_\tau}
+R_\tau{^\alpha}{_{\mu\alpha}}f^\tau{_\nu}
=\frac{1}{2}R_{\nu\mu\alpha\tau}f^{\alpha\tau}
+R^\tau{_\mu}f_{\tau\nu}\nonumber\\
\fl\!&=&\frac{1}{2}\left(C_{\nu\mu}{^{\alpha\tau}}
+\!\frac{4}{(n\!-\!2)}\delta^{[\alpha}_{[\nu}R^{\tau]}_{\mu]}
-\frac{2}{(n\!-\!1)(n\!-\!2)}\delta^{[\alpha}_{[\nu}\delta^{\tau]}_{\mu]}R\right)f_{\alpha\tau}
-R^\tau{_\mu}f_{\nu\tau}\nonumber\\
\fl &=&
\fi
\label{Fterm0}
\frac{1}{2}f^{\alpha\tau}C_{\alpha\tau\nu\mu}
+\frac{(n\!-\!2)\Lambda}{(n\!-\!1)}f_{\nu\mu}+(f^3)\dots.
\end{eqnarray}
Substituting (\ref{antisymmetricpreliminary},\ref{Fterm0},\ref{approximateNhat})
into the antisymmetric field equations (\ref{JSantisymmetric}) gives
\begin{eqnarray}
\label{antisymmetric0}
\fl f_{\nu\mu}
\!&=&F_{\nu\mu}+\tR_{[\nu\mu]}\rmt\Lambda_b^{\!-1/2}/2+(f^3)\Lambda_b^{\!-1}\dots\\
\label{threeparts}
%\fl~~~~f_{\nu\mu}
\fl &=&F_{\nu\mu}
% 2\Aphi_{[\mu,\nu]}
+\left(\theta_{[\tau,\alpha]}\varepsilon_{\nu\mu}{^{\tau\alpha}}
+f^{\alpha\tau}C_{\alpha\tau\nu\mu}
+\frac{2(n\!-\!2)\Lambda}{(n\!-\!1)}f_{\nu\mu}
+(f^3)\right)\!\Lambda_b^{\!-1}\dots.
\end{eqnarray}
where
\begin{eqnarray}
\label{thdef}
~~~~~~\theta_\tau &=&\frac{\lower1pt\hbox{$1$}}{4}f_{[\nu\mu,\alpha]}\varepsilon_\tau{^{\nu\mu\alpha}},~~
f_{[\nu\mu,\alpha]}=-\frac{\lower1pt\hbox{$2$}}{3}
\,\theta_\tau\varepsilon^\tau{_{\nu\mu\alpha}},\\
~~\varepsilon_{\tau\nu\mu\alpha}
&=&{(\rm Levi\!-\!Civita~tensor)},\\
%&=&{[\,\nu\mu\tau\tau]}\rmg~~~~~~,~~
%\varepsilon^{\nu\mu\tau\tau}=-[\,\nu\mu\tau\tau]/\rmg~,\\
%{[\,\nu\mu\tau\tau]}
%&=&\left\{\matrix{
%+1~{\rm for~even~permutations~of~0123}\cr
%-1~{\rm for~odd~~permutations~of~0123}\cr
%~0~{\rm for~two~equal~indices~~~~~~~~~~~}
%}\right.\\
\label{Ctilde}
~C_{\alpha\tau\nu\mu}&=&({\rm Weyl~tensor}).
%\tC_{\nu\mu\alpha\tau}&=&R_{\nu\mu\alpha\tau}
%-g_{\nu[\alpha}R_{\tau]\mu}+g_{\mu[\alpha}R_{\tau]\nu},\\
%\label{Fdef}
%~~~~F_{\nu\mu}&=&2A_{[\mu,\nu]}.
\end{eqnarray}
In (\ref{threeparts}) the
%$2\Aphi_{[\mu,\nu]}$
$F_{\nu\mu}$ term is the ordinary electromagnetic field (\ref{Fdef}).
The $\theta_{[\tau,\alpha]}\varepsilon_{\nu\mu}{^{\tau\alpha}}\Lambda_b^{\!-1}$
term is divergenceless and appears as
$2\theta_{[\mu,\nu]}\Lambda_b^{\!-1}$ in the dual of $f_{\nu\mu}$.
The $(f^3)\Lambda_b^{\!-1}$ term is $<\!10^{-66}$ of $f_{\nu\mu}$ from (\ref{highenergyskew}).
The $f_{\nu\mu}\Lambda/\Lambda_b$ term is $\sim 10^{-122}$ of
$f_{\nu\mu}$ from (\ref{LambdaoverLambdab}).
%To assess the magnitude of $C_{\nu\mu\alpha\tau}f^{\alpha\tau}/\Lambda_b$ relative to
%$f_{\nu\mu}$ we must consider worst-case values of the ratio
%$C_{\nu\mu\alpha\tau}/\Lambda_b$.
%The symbol $\tC_{\tau\nu\mu\alpha}$ is used because this
%quantity can be written in terms of the Weyl tensor and terms which can
%usually be ignored. From (\ref{Ctilde},\ref{Einstein},\ref{Ttilde}) we have
%\begin{eqnarray}
%\tC_{\nu\mu\alpha\tau}
%\!&=&C_{\nu\mu\alpha\tau}+\left(\!\frac{2}{(n\!-\!2)}-1\!\right)
%(g_{\nu[\alpha}R_{\tau]\mu}-g_{\mu[\alpha}R_{\tau]\nu})
%-\frac{2}{(n\!-\!1)(n\!-\!2)}Rg_{\nu[\alpha}g_{\tau]\mu}\\
%\label{Ctildeapprox}
%\!&=&C_{\nu\mu\alpha\tau}
%+\frac{2 (n\!-\!2)\Lambda}{(n\!-\!1)} g_{\nu[\alpha}
%g_{\tau]\mu}+(f^2).
%%+\left({{\rm order}~f^2\atop{\rm terms}}\right)\!.
%\end{eqnarray}
%, where $\Lambda_b\!\sim\!10^{66}~{\rm cm}^{-2}$ from (\ref{Lambdab}).
%Here the $(f^2)$ terms will create $(f^3)$ terms in
%(\ref{threeparts}), so they can be ignored from (\ref{highenergyskew}).
%The $\Lambda$ term can be ignored from (\ref{LambdaoverLambdab}).
%So let us focus on the Weyl tensor component of (\ref{Ctildeapprox}).
The Weyl tensor term might be expected to have the largest observable values
near the Schwarzschild radius, $r_s\!\nobreak=\nobreak\!2Gm/c^2$, of black
holes, where $C_{\nu\mu\alpha\tau}$ takes on values around $r_s/r^3$.
However, since the lightest black holes have the smallest
Schwarzschild radius, they will create the largest value of
$r_s/r_s^3=1/r_s^2$. The lightest black hole that we can expect to
observe would be of about one solar mass, where from (\ref{Lambdab}),
\begin{eqnarray}
\label{blackholecurvature}
\frac{C_{trtr}}{\Lambda_b}
&\sim&\frac{1}{\Lambda_br^2_s}
=\frac{1}{\Lambda_b}\left(\frac{c^2}{2Gm_\odot}\right)^2\!\sim\!10^{-77}.
\end{eqnarray}
So even in the most extreme worst-cases accessible to measurement,
the last three terms in (\ref{threeparts})
are all $<\!10^{-66}$ of $f_{\nu\mu}$.
And we set that (\ref{threeparts}) gives
exactly $f_{\nu\mu}=F_{\nu\mu}$ in the limit as $\Lambda_b\!\rightarrow\!\infty$.
%\newpage

Taking the divergence of (\ref{threeparts}) using (\ref{Ampere}),
%and ignoring $C_{\nu\mu\alpha\tau}f^{\alpha\tau}/\Lambda_b$
%and $f_{\nu\mu}\Lambda/\Lambda_b$
%due to (\ref{blackholecurvature},\ref{LambdaoverLambdab}),
the divergenceless term $\theta_{[\tau,\alpha]}\varepsilon_{\nu\mu}{^{\tau\alpha}}\Lambda_b^{\!-1}$
falls out and we get an extremely close approximation to Maxwell's equations,
\begin{eqnarray}
\label{Maxwell}
F_{\nu\mu;}{^\nu}&=&
\left[(f^{\tau\alpha}C_{\alpha\tau\nu\mu});{^\nu}+(f^{3\prime})\right]\Lambda_b^{\!-1}\dots,\\
\label{Faraday}
F_{[\nu\mu,\sigma]}&=&0.
\end{eqnarray}
%where we define
%\begin{eqnarray}
%\end{eqnarray}
As usual, Faraday's law (\ref{Faraday}) is just an identity which follows from
the definition (\ref{Fdef}).
% of $F_{\nu\mu}$.
The extra terms in Ampere's law (\ref{Maxwell})
are $<\nobreak\!10^{-66}$ of the primary terms because this is true for (\ref{threeparts}).
In most so-called ``exact'' equations in physics, there are really many known
corrections due to QED and other effects which are
ignored because they are too small to measure.
We should emphasize that the extra terms in (\ref{Maxwell}) are at least
50 orders of magnitude
% $10^{-50}$
smaller than known corrections to Maxwell's equations which are routinely ignored\cite{Jackson}.
%are two order higher in $f^\mu{_\sigma}$,
%than the leading order term, and from (\ref{highenergyskew}), these terms
%must be $<\!10^{-66}$ of the leading order term.
And we see that (\ref{Maxwell},\ref{Faraday})
go to the exact electro-vac Maxwell equations in the limit as $\Lambda_b\!\rightarrow\!\infty$.
Of course we are just considering the
electro-vac case, so (\ref{Maxwell}) has no source term.
As in electro-vac Einstein-Maxwell theory, the lack of a source term does not
preclude the existence of charges, as evidenced by the exact electric monopole
solution derived in \S\ref{Monopole}.
To get a source term in Ampere's law (\ref{Maxwell}), charged matter terms must
be included in the Lagrangian density as in \cite{sShifflett}.

%As noted previously, as long as one accepts the existence of singularities,
%these equations still allow charges as
%and currents of charge as collections of
%individual Green-function solutions.
%Since physicists now generally accept the existence of black holes,
%and regard electrons as point particles, this idea is not as controversial
%as it was in the early days of the Einstein-Schr\"{o}dinger theory.
%\newpage
The divergencless term $\theta_{[\tau,\alpha]}\varepsilon_{\nu\mu}{^{\tau\alpha}}$ of
(\ref{threeparts}) should also be expected to be $<\!10^{-32}$ of $f_{\nu\mu}$ from
(\ref{highenergyderiv1},\ref{highenergyderiv2},\ref{thdef}).
However, we need to consider the possibility where $\theta_\tau$ changes extremely rapidly,
%The divergenceless term $\theta_{[\tau,\alpha]}\varepsilon_{\nu\mu}{^{\tau\alpha}}\Lambda_b^{\!-1}$
%of (\ref{threeparts}) should also be expected to be negligible relative
%to $f_{\nu\mu}$ because of
%(\ref{thdef},\ref{highenergyderiv1},\ref{highenergyderiv2},\ref{gammaderiv1},\ref{gammaderiv2}).
%Nevertheless, we need to investigate if this term could have some significance,
so let us consider the ``dual'' part of (\ref{threeparts}).
Taking the curl of (\ref{threeparts}),
%and ignoring $C_{\nu\mu\alpha\tau}f^{\alpha\tau}/\Lambda_b$
%and $f_{\nu\mu}\Lambda/\Lambda_b$ due to
%(\ref{blackholecurvature},\ref{LambdaoverLambdab}),
%the $2\Aphi_{[\mu,\nu]}$
the $F_{\nu\mu}$ term falls out from (\ref{Fdef}) and we have
\begin{eqnarray}
\fl~~~ f_{[\nu\mu,\sigma]}
=\!\left(\theta_{\tau;\alpha;[\sigma}\,\varepsilon_{\nu\mu]}{^{\tau\alpha}}
+(f^{\alpha\tau}C_{\alpha\tau[\nu\mu}){_{,\sigma]}}
+\frac{2(n\!-\!2)\Lambda}{(n\!-\!1)}f_{[\nu\mu,\sigma]}
+(f^{3\prime})\right)\!\Lambda_b^{\!-1}\dots.
\end{eqnarray}
%\bigskip\\
Contracting this with $\Lambda_b\varepsilon^{\rho\sigma\nu\mu}/2$ and using (\ref{thdef}) gives,
\begin{eqnarray}
\label{rawProca}
\fl~~~ 2\Lambda_b\theta^\rho
%\!&=&\!-2\,\dual^\tau{\!_;}{^\alpha}{_{;\,\sigma}}\delta^{[\rho}_{\,\tau}\delta^{\sigma]}_\alpha\\
\!&=&\!-2\,\theta^{[\rho}{_;}{^{\sigma]}}{_{\!;\,\sigma}}
+\frac{1}{2}\varepsilon^{\rho\sigma\nu\mu}(f^{\alpha\tau}C_{\alpha\tau[\nu\mu}){_{,\sigma]}}
+\frac{4(n\!-\!2)\Lambda}{(n\!-\!1)}\,\theta^\rho
+(f^{3\prime})\dots
\end{eqnarray}
Using $\theta^\sigma{_{;\sigma}}\!=0$ from (\ref{thdef}) and the covariant derivative commutation rule,
the Einstein equations $R_{\nu\mu}\!=\!-\Lambda g_{\nu\mu}+(f^2)\dots$ from (\ref{Einstein},\ref{Ttilde})
give $\theta^\sigma{_{;\rho;\sigma}}\!=\!R_{\sigma\rho}\theta^\sigma\!=\!-\theta_\rho\Lambda +(f^{3\prime})\dots$,
and we get something similar to the Proca equation\cite{Greiner,Proca},
\begin{eqnarray}
\label{Proca}
\fl~~~~2\Lambda_b\theta_\rho=-\theta_\rho{_{;\,\sigma;}}{^\sigma}
+\frac{1}{2}\varepsilon_\rho{^{\sigma\nu\mu}}(f^{\alpha\tau}C_{\alpha\tau[\nu\mu}){_{,\sigma]}}
+\frac{(3n\!-\!7)\Lambda}{(n\!-\!1)}\,\theta_\rho
+(f^{3\prime})\dots.
\end{eqnarray}
%The last three terms of (\ref{Proca})
%must be $<\!10^{-66}$ of the other terms
%because this is true for (\ref{threeparts}).
%from (\ref{blackholecurvature},\ref{LambdaoverLambdab},\ref{highenergyskew}).
%The next higher order terms in this equation
%must be of order $f^3$, and from (\ref{highenergyskew}), these terms
%must be $<\!10^{-66}$ of the leading order terms.
Here the $\Lambda\theta_\rho$ term can certainly be ignored from (\ref{Lambda}),
and the $(f^{3\prime})$ term can probably be ignored in the weak field limit.
The Weyl tensor term can be ignored if one assumes a flat background,
although Proca waves might significantly perturb the background if they exist,
so this is a rather big assumption.
If we do ignore the last three terms, this equation has the trivial solution $\theta_\rho\approx 0$.
If $\Lambda_z\!>0,\,\Lambda_b\!<0$ as with supersymmetry,
wavelike solutions to (\ref{Proca}) cannot exist.
If $\Lambda_z\!<0,\,\Lambda_b\!>0$ as in (\ref{Lambdab},\ref{Cz}),
wavelike solutions could possibly exist,
and in a flat background space they would be of the form\cite{Greiner}
\begin{eqnarray}
\label{varthetawave1}
\fl~~~~~~~~~~\theta_\rho
\!&=&\theta\epsilon_\rho{\rm sin} (k_\alpha x^\alpha)~~
,~~k_\alpha k^\alpha\!=2\Lambda_b
~~,~~\epsilon_\alpha \epsilon^\alpha\!=-1
~~,~~k_\alpha\epsilon^\alpha\!=0,\\
\label{varthetawave2}
\fl~~~~~~~~~~\omega\!&=&\sqrt{2\Lambda_b+\textbf{k}^{\smash{2}}}
~~,~~k_\alpha\!=\left(\omega,\textbf{k}\right).
\end{eqnarray}
Here $\theta$ is the magnitude, $k^\alpha$ is the wavenumber,
$\epsilon^\alpha$ is the polarization,
and $\omega$ is the frequency.
%When interpreted as a particle, $\dual_\rho$ would be a superheavy, neutral,
%spin\nobreak-\nobreak 1 vector boson. From
%(\ref{Proca},\ref{cutoff},\ref{Lambdab})
%its mass would be
%\begin{eqnarray}
%\fl~~~~ m_\dual&=&\sqrt{2\Lambda_b}\,\frac{\hbar}{c}
%\sim\sqrt{-2\Lambda_z}\,\frac{\hbar}{c}
%\label{mx}\sim\frac{\hbar \sqrt{2C_z}k_c^2 l_P}{c}
%\sim \frac{\hbar \sqrt{2C_z} C_c^2}{c\,l_P}
%\sim 10^{19}{\rm GeV/c^2}.
%\end{eqnarray}
%Now let us assume $\Lambda_b\!>0$ and consider the energy-momentum tensor
%(\ref{Ttilde}) for a Proca plane-wave solution.
Substituting (\ref{varthetawave1},\ref{varthetawave2}) into
(\ref{threeparts},\ref{Ttilde})
\ifnum\ExpandDerivations=1
\begin{eqnarray}
\fl~~~~~~~~~f^{\nu\sigma}&=&\varepsilon^{\nu\sigma\rho\tau}\theta_{[\rho,\tau]}\Lambda_b^{\!-1}
=\theta\varepsilon^{\nu\sigma\rho\tau}\epsilon_{[\rho}k_{\tau]}
{\rm cos}(k_\alpha x^\alpha)\Lambda_b^{\!-1},\nonumber\\
~~~~\fl f^{\nu\sigma}f_{\sigma\mu}
&=&\varepsilon^{\nu\sigma\rho\tau}\theta_{[\rho,\tau]}
\varepsilon_{\sigma\mu\lambda\beta}\theta^{[\lambda}{_;}^{\beta]}\Lambda_b^{\!-2}
=6{\delta^{[\nu}_\mu}{\delta^\rho_\lambda}{\delta^{\tau]}_\beta}
\theta_{\rho,\tau}\theta^{[\lambda}{_;}^{\beta]}\Lambda_b^{\!-2}\nonumber\\
\fl &=&2(\delta^\nu_\mu
\theta_{\lambda,\beta}\theta^{[\lambda}{_;}^{\beta]}
+\theta_{\mu,\lambda}\theta^{[\lambda}{_;}^{\nu]}
+\theta_{\beta,\mu}\theta^{[\nu}{_;}^{\beta]})\Lambda_b^{\!-2}\nonumber\\
\fl &=&\theta^2(\delta^\nu_\mu\epsilon_\lambda\epsilon^\lambda k_\beta k^\beta
-\epsilon_\mu\epsilon^\nu k_\lambda k^\lambda
-\epsilon_\beta\epsilon^\beta k_\mu k^\nu)
{\rm cos}^2(k_\alpha x^\alpha)\Lambda_b^{\!-2}\nonumber\\
\fl &=&\theta^2(-2\Lambda_b\delta^\nu_\mu
- 2\Lambda_b\epsilon_\mu\epsilon^\nu
+ k_\mu k^\nu){\rm cos}^2(k_\alpha x^\alpha)\Lambda_b^{\!-2},\nonumber\\
\fl ~~~~f^{\nu\sigma}f_{\sigma\nu}
&=&-4\theta^2{\rm cos}^2(k_\alpha x^\alpha)\Lambda_b^{\!-1},\nonumber\\
\fl ~~~~8\pi\,{^\theta}T_{\nu\mu}
&\approx&2\theta^2(-2\Lambda_b g_{\nu\mu}
-2\Lambda_b\epsilon_\nu\epsilon_\mu
+k_\nu k_\mu+\Lambda_b g_{\nu\mu})
{\rm cos}^2(k_\alpha x^\alpha)\Lambda_b^{\!-2}+0+0+0\nonumber\\
\fl &+&2\theta^2( 2\Lambda_b g_{\nu\mu}
+2\Lambda_b\epsilon_\nu\epsilon_\mu -k_\nu k_\mu)
{\rm sin}^2(k_\alpha x^\alpha)\Lambda_b^{\!-2}
-2\theta^2k_\nu k_\mu {\rm sin}^2(k_\alpha x^\alpha)\Lambda_b^{\!-2}\nonumber\\
\fl &+&0+4\theta^2g_{\nu\mu}({\rm sin}^2(k_\alpha x^\alpha)
-{\rm cos}^2(k_\alpha x^\alpha))\Lambda_b^{\!-1}
-2\theta^2 g_{\nu\mu}{\rm sin}^2(k_\alpha x^\alpha)\Lambda_b^{\!-1}\nonumber
\end{eqnarray}
\fi
gives the energy-momentum tensor
%for such Proca plane-waves in a flat background space,
\begin{eqnarray}
\label{Tvartheta}
\fl~~~~~~8\pi\,{^\theta}T_{\nu\mu}
&\approx&-\frac{\lower1pt\hbox{$\theta^2$}}{\Lambda_b}\,[k_\nu
k_\mu\Lambda_b^{\!-1} -(3k_\nu k_\mu\Lambda_b^{\!-1}- 6g_{\nu\mu}
-4\,\epsilon_\nu\epsilon_\mu) {\rm cos}(2k_\alpha
x^\alpha)].
\end{eqnarray}
%This can be compared to the result for an electromagnetic plane-wave (\ref{TA}).
%where from (\ref{planewaveA}-\ref{planewavef})
%only the first term of (\ref{Ttilde}) contributes,

%From (\ref{varthetawave1},\ref{varthetawave2}), it is easy to verify that
%the divergence of (\ref{Tvartheta}) vanishes.
Here we find that $\left<{^\theta}T_{00}\right><0$,
an indication that the theory might allow negative energy waves,
often called ``ghosts''.
%A similar result has been found in the unmodified Einstein-Schr\"{o}dinger theory.
%Some work has been done to avoid this problem by appending a couple additional
%terms to the Einstein-Schr\"{o}dinger Lagrangian density\cite{Moffat95,Clayton},
%but this modification is unrelated to our theory, and is not very relevant here.
%The negative energy cannot be avoided by assuming that ``$\dualmag$'' and
%$f_{\nu\mu}$ are real because this would make
%$\left<{^\Aphi}T_{00}\right>\!<\!0$ in (\ref{TA}). Neither can it be avoided by assuming
%that $\Aphi_\nu$ is imaginary and $\dual_\rho$ is real because this would
%make $f_{\nu\mu}$, $det(N_{\mu\sigma})$ and ${\mathcal L}$ complex.
However, unlike similar theories\cite{Moffat78,Damour92,Damour93},
this theory avoids ghosts in an unusual way.
Recall that this theory is the original Einstein-Schr\"{o}dinger theory,
but with a $\Lambda_z g_{\mu\nu}$ in the field equations
to account for zero-point fluctuations,
%The whole theory is based upon the assumption that zero-point fluctuations
%really do contribute to the cosmological constant,
and $\Lambda_z\!=\!-C_z\omega_c^4l_P^2$ from (\ref{Lambdab},\ref{Cz}) is finite only because
of a cutoff frequency $\omega_c\!\sim\!1/l_P$ from (\ref{cutoff}).
From these equations
and (\ref{Proca}), Proca waves would be cut off because they would have a minimum frequency
\begin{eqnarray}
\omega_{Proca}\!=\!\sqrt{2\Lambda_b}\!=\!\sqrt{-2\Lambda_z}\!=\!\!\sqrt{2C_z}\,\omega_c^2l_P>\omega_c.
\end{eqnarray}
Whether the cutoff of zero-point fluctuations is caused by a discreteness, uncertainty or foaminess of
spacetime near the Planck length\cite{Garay,Padmanabhan,Padmanabhan2,Ashtekar,Smolin}
or by some other effect, the same $\omega_c$ which cuts off $\Lambda_z$
should also cut off Proca waves in this theory. So we should expect to observe only the
trivial solution $\dual_\rho\!\approx\!0$ to (\ref{Proca}) and no ghosts.
Comparing $\omega_{Proca}$ and $\omega_c$ from above, we see that this argument only applies if
\begin{eqnarray}
\label{limit}
\omega_c>\frac{\lower1pt\hbox{$1$}}{l_P\sqrt{2C_z}}.
\end{eqnarray}
Here $C_z$ is defined by (\ref{Cz}), and the inequality
is satisfied for this theory when $\omega_c$ and $C_z$ are chosen as in
(\ref{cutoff},\ref{Lambdab}) to be consistent
with a cosmological constant caused by zero-point fluctuations.
Since the prediction of negative energy waves would probably be inconsistent with
reality, this theory should be approached cautiously when considering it with
values of $\omega_c$ and $C_z$ which do not satisfy (\ref{limit}).

Finally, if we fully renormalize with $\omega_c\!\rightarrow\!\infty$
as in quantum electrodynamics,
then $\Lambda_b\!\rightarrow\!\infty$ and $\omega_{Proca}\!\rightarrow\!\infty$,
so the potential ghost goes away completely.
%From this perspective, our theory does not contain ghosts any more than quantum electrodynamics
%contains ghosts when the renormalization is done using Pauli-Villars fields
%(which are negative energy fields).
In the limit $\omega_c\!\rightarrow\!\infty$ our theory becomes exactly Einstein-Maxwell theory.
%Assuming such a full renormalization does not diminish the value of the theory in any way.
%It still unifies gravitation with electromagnetism, it still suggests untried
%approaches to a complete unified field theory, and it still
%offers an untried approach to the quantization of gravity.
However, the theory would still be much different than Einstein-Maxwell theory from
the standpoint of quantization.
In any attempt to quantize this theory, the cutoff frequency $\omega_c$ would need to be the same
cutoff which is taken to infinity during renormalization.
For example, Pauli-Villars masses would probably go as $M\!=\!\hbar\omega_c$
if Pauli-Villars renormalization was used.
Since $\Lambda_b$ and $\Lambda_z$ in the Lagrangian density go as $\omega_c^4$,
quantization and renormalization would certainly need to be done a bit different than usual.
Also, because $\omega_{Proca}$ goes as $\omega_c^2$, Proca waves would not
represent a ghost from the standpoint of quantization.

\section{\label{EIHEquationsOfMotion}The Einstein-Infeld-Hoffmann equations of motion}
Here we derive the Lorentz force from the theory using the
Einstein-Infeld-Hoffmann (EIH) method\cite{EinsteinInfeld}. For
Einstein-Maxwell theory, the EIH method allows the equations of
motion to be derived directly from the electro-vac field
equations. For neutral particles the method has been verified to
Post-Newtonian order\cite{EinsteinInfeld}, and in fact it was the
method first used to derive the Post-Newtonian equations of
motion\cite{EIH}. For charged particles the method has been verified to Post-Coulombian
order\cite{Wallace,WallaceThesis,Gorbatenko}, (see also \ref{EofMcomparison})
meaning that it gives the same result as the Darwin Lagrangian\cite{Jackson}.
The EIH method is valuable because it does not require any additional
assumptions, such as the postulate that neutral particles follow
geodesics, or the {\it ad hoc} inclusion of matter terms in the
Lagrangian density. When the EIH method was applied to the
original Einstein-Schr\"{o}dinger theory, no Lorentz force was
found between charged particles\cite{Callaway,Infeld}. The basic
difference between our case and \cite{Callaway,Infeld} is that our
effective energy-momentum tensor (\ref{Ttilde}) contains the
familiar term
$f_\nu{^\sigma}f_{\sigma\mu}\!-\!(1/4)g_{\nu\mu}f^{\rho\sigma}f_{\sigma\rho}$.
This term appears because we assumed $\Lambda_b\ne 0$, and because
of our metric definition (\ref{gdef}) and (\ref{approximateNbar}).
With this term, the EIH method predicts the same Lorentz force as
it does for electro-vac Einstein-Maxwell theory. Also, it happens
that the extra terms in our approximate Einstein and Maxwell
equations
(\ref{Einstein}-\ref{Ttilde},\ref{Maxwell},\ref{Faraday}) cause no
contribution beyond the Lorentz force, to Newtonian/Coulombian
order. The basic reason for the null result of
\cite{Callaway,Infeld} is that they assumed $\Lambda_b\!=\!0$ and
$g_{\mu\nu}\!=\!N_{(\mu\nu)}$, so that every term in their
effective energy-momentum tensor has ``extra
derivatives''\cite{Voros}. For the same reason that
\cite{Callaway,Infeld} found no Lorentz force, the extra
derivative terms in our effective energy-momentum tensor
(\ref{Ttilde}) cause no contribution to the equations of motion.

The exact Lorentz force equation can be derived for this theory by
including charged matter terms in the Lagrangian
density\cite{sShifflett}. Here we derive the Lorentz force using
the EIH method because it requires no assumptions about matter
terms, and also to show definitely that the well known negative
result of \cite{Callaway,Infeld} for the unmodified
Einstein-Schr\"{o}dinger theory does not apply to the present
theory. We will only cover the bare essentials of the EIH method
which are necessary to derive the Lorentz force, and the
references above should be consulted for a more complete
explanation. We will also only calculate the equations of motion
to Newtonian/Coulombian order, because this is the order where the
Lorentz force first appears.

With the EIH method, one does not just find equations of motion,
but rather one finds approximate solutions $g_{\mu\nu}$ and
$f_{\mu\nu}$ of the field equations which correspond to a system
of two or more particles. These approximate solutions will in
general contain $1/r^p$ singularities, and these are considered to
represent particles. It happens that acceptable solutions to the
field equations can only be found if the motions of these
singularities are constrained to obey certain equations of motion.
The assumption is that these approximate solutions for
$g_{\mu\nu}$ and $f_{\mu\nu}$ should approach exact solutions
asymptotically, and therefore the motions of the singularities
should approximate the motions of exact solutions. Any event
horizon or other unusual feature of exact solutions at small radii
is irrelevant because the singularities are assumed to be
separated by much larger distances, and because the method relies
greatly on surface integrals done at large distances from the
singularities. Some kind of exact Reissner-Nordstr\"{o}m-like
solution should probably exist in order for the EIH method to make
sense, and the electric monopole solution in \S\ref{Monopole}
fills this role in our case. However, exact solutions are really
only used indirectly to identify constants of integration.

The EIH method assumes the ``slow motion approximation'', meaning
that $v/c\!\ll\!1$. The fields are expanded in the
form\cite{EinsteinInfeld,Wallace,WallaceThesis,Gorbatenko},
\begin{eqnarray}
g_{\mu\nu}=\eta_{\mu\nu}+\gamma_{\mu\nu}\!-\eta_{\mu\nu}\eta^{\sigma\rho}\gamma_{\sigma\rho}/2,\\
\label{g00expansion}
\gamma_{00}={_2}\gamma_{00}\lambda^2+{_4}\gamma_{00}\lambda^4\dots\\
\label{g0kexpansion}
\gamma_{0k}={_3}\gamma_{0k}\lambda^3+{_5}\gamma_{0k}\lambda^5\dots\\
\label{gikexpansion}
\gamma_{ik}={_4}\gamma_{ik}\lambda^4\dots\\
\label{A0expansion}
A_{0}={_2}A_{0}\lambda^2+{_4}A_{0}\lambda^4\dots\\
\label{Akexpansion}
A_{k}={_3}A_{k}\lambda^3+{_5}A_{k}\lambda^5\dots\\
\label{f0kexpansion}
f_{0k}={_2}f_{0k}\lambda^2+{_4}f_{0k}\lambda^4\dots\\
\label{fikexpansion}
f_{ik}={_3}f_{ik}\lambda^3+{_5}f_{ik}\lambda^5\dots
\end{eqnarray}
where $\lambda\sim v/c$ is the expansion parameter, the order of
each term is indicated with a left subscript\cite{Callaway},
$\eta_{\mu\nu}={\rm diag}(1,-1,-1,-1)$, and Latin indices run from
1-3. The field $\gamma_{\mu\nu}$ (often called $\bar h_{\mu\nu}$
in other contexts) is used instead of $g_{\mu\nu}$ only because it
simplifies the calculations.
%\begin{eqnarray}
%\gamma_{\mu\nu}=h_{\mu\nu}\!-\eta_{\mu\nu}\eta^{\sigma\rho}h_{\sigma\rho}/2,~~~~~
%g_{\mu\nu}=\eta_{\mu\nu}+h_{\mu\nu},
%~~~~~h_{\mu\nu}=\gamma_{\mu\nu}\!-\eta_{\mu\nu}\eta^{\sigma\rho}\gamma_{\sigma\rho}/2,\\
%g_{\mu\nu}=\eta_{\mu\nu}+\gamma_{\mu\nu}\!-\eta_{\mu\nu}\eta^{\sigma\rho}\gamma_{\sigma\rho}/2,\\
%\eta_{\mu\nu}=({\rm Minkowski~metric})={\rm diag}(1,-1,-1,-1).
%\end{eqnarray}
Because $\lambda\sim v/c$, when the expansions are substituted
into the Einstein and Maxwell equations, a time derivative counts
the same as one higher order in $\lambda$. The general procedure
is to substitute the expansions, and solve the resulting field
equations order by order in $\lambda$, continuing to higher orders
until a desired level of accuracy is achieved. At each order in
$\lambda$, one of the ${_l}\gamma_{\mu\nu}$ terms and one of the
${_l}f_{\mu\nu}$ terms will be unknowns, and the equations will
involve known results from previous orders because of the
nonlinearity of the Einstein equations.

The expansions (\ref{g00expansion}-\ref{fikexpansion}) use only
alternate powers of $\lambda$ essentially because the Einstein and
Maxwell equations are second order differential
equations\cite{EIH}, although for higher powers of $\lambda$, all
terms must be included to predict
radiation\cite{Wallace,WallaceThesis,Gorbatenko}. Because
$\lambda\!\sim\!v/c$, the expansions have the magnetic components
$A_{k}$ and $f_{ik}$ due to motion at one order higher in
$\lambda$ than the electric components $A_0$ and $f_{0i}$. As in
\cite{Wallace,WallaceThesis,Gorbatenko}, $f_{0k}$ and $f_{ik}$
have even and odd powers of $\lambda$ respectively. This is the
opposite of \cite{Callaway,Infeld} because we are assuming a
direct definition of the electromagnetic field
(\ref{fdef},\ref{approximateNhat},\ref{threeparts},\ref{Fdef})
instead of the dual definition
$f^{\alpha\rho}=\varepsilon^{\alpha\rho\sigma\mu}N_{[\sigma\mu]}/2$
assumed in \cite{Callaway,Infeld}.

The field equations are assumed to be of the standard form
\begin{eqnarray}
\label{EIHG0} \fl~~~~~~~~~~~G_{\mu\nu}=8\pi T_{\mu\nu} ~~{\rm
where}~~
G_{\mu\nu}=R_{\mu\nu}-\frac{\lower2pt\hbox{$1$}}{2}g_{\mu\nu}g^{\alpha\beta}R_{\alpha\beta},
\end{eqnarray}
or equivalently
\begin{eqnarray}
\label{EIHR0} \fl~~~~~~~~~~~R_{\mu\nu}=8\pi S_{\mu\nu} ~~{\rm
where}~~
\,S_{\mu\nu}=\,T_{\mu\nu}-\frac{\lower2pt\hbox{$1$}}{2}g_{\mu\nu}g^{\alpha\beta}\,T_{\alpha\beta}.
\end{eqnarray}
However, with the EIH method we must solve a sort of
quasi-Einstein equations,
\begin{eqnarray}
\label{quasi} \fl~~~~~~~~~~~0&=&\breve G_{\mu\nu}-8\pi \breve
T_{\mu\nu},
\end{eqnarray}
where
\begin{eqnarray}
\fl~~~~~~~~~~~\breve G_{\mu\nu}&=&R_{\mu\nu}-\frac{\lower2pt\hbox{$1$}}{2}\eta_{\mu\nu}\eta^{\alpha\beta}R_{\alpha\beta},
\label{EIHT}
~~~~~~~~~~~\breve
T_{\mu\nu}=S_{\mu\nu}-\frac{\lower2pt\hbox{$1$}}{2}\eta_{\mu\nu}\eta^{\alpha\beta}S_{\alpha\beta}.
\end{eqnarray}
Here the use of $\eta_{\mu\nu}$ instead of $g_{\mu\nu}$ is not an
approximation because (\ref{EIHR0}) implies (\ref{quasi}) whether
$\breve G_{\mu\nu}$ and $\breve T_{\mu\nu}$ are defined with
$\eta_{\mu\nu}$ or $g_{\mu\nu}$. Note that the references use many
different notations in (\ref{quasi}): instead of $\breve
G_{\mu\nu}$ others use $\Pi_{\mu\nu}/2\!+\!\Lambda_{\mu\nu}$,
$\Phi_{\mu\nu}/2\!+\!\Lambda_{\mu\nu}$ or $[{\rm LS}\!:\!\mu\nu]$
and instead of $8\pi\breve T_{\mu\nu}$ others use $-2S_{\mu\nu}$,
$-\Lambda'_{\mu\nu}$, $-\Lambda_{\mu\nu}$ or $[{\rm
RS}\!:\!\mu\nu]$.

The equations of motion result as a condition that the field
equations (\ref{quasi}) have acceptable solutions. In the language
of the EIH method, acceptable solutions are those that contain
only ``pole'' terms and no ``dipole'' terms, and this can be
viewed as a requirement that the solutions should resemble
Reissner-Nordstr\"{o}m solutions asymptotically. To express the
condition of solvability we must consider the integral of the
field equations (\ref{quasi}) over 2D surfaces $S$ surrounding
each singularity,
\begin{eqnarray}\
\label{integral} {_l}C_\mu=\frac{1}{2\pi}\int^S(\,{_l}\breve
G_{\mu k}-8\pi\,{_l}\breve T_{\mu k})n_kdS.
\end{eqnarray}
Here $n_k$ is the surface normal and $l$ is the order in
$\lambda$. Assuming that the divergence of the Einstein equations
(\ref{EIHG0}) vanishes, and that (\ref{quasi}) has been solved to
all previous orders, it can be shown\cite{EinsteinInfeld} that in
the current order
\begin{eqnarray}
\label{zerodivergenceEIH} (\,{_l}\breve G_{\mu
k}-8\pi\,{_l}\breve T_{\mu k})_{|k}\!=\!0.
\end{eqnarray}
Here and throughout this section ``$|$'' represents ordinary
derivative\cite{EinsteinInfeld}. From Green's theorem,
(\ref{zerodivergenceEIH}) implies that ${_l}C_\mu$ in
(\ref{integral}) will be independent of surface size and
shape\cite{EinsteinInfeld}.
%because of Gauss's divergence theorem.
%However, $C_\mu$ will depend on the motion of any singularity
%at the center of the surface.
The condition for the existence of an acceptable solution for
${_4}\gamma_{ik}$ is simply
\begin{eqnarray}
\label{C} {_4}C_i=0,
\end{eqnarray}
and these are also our three $\ord(\lambda^4)$ equations of
motion\cite{EinsteinInfeld}. The $C_0$ component of
(\ref{integral}) causes no constraint on the
motion\cite{EinsteinInfeld}
so we only need to calculate $\breve G_{ik}$ and $\breve T_{ik}$.

At this point let us introduce a Lemma from \cite{EinsteinInfeld}
which is derived from Stokes's theorem. This Lemma states that
\begin{eqnarray}
\label{Lemma} \int^S \mathcal{F}_{(\cdots)kl|l}n_kdS=0 ~~~{\rm
if}~~~\mathcal{F}_{(\cdots)kl}=-\mathcal{F}_{(\cdots)lk},
\end{eqnarray}
where  $\mathcal{F}_{(\cdots)kl}$ is any antisymmetric function of
the coordinates, $n_k$ is the surface normal, and $S$ is any
closed 2D surface which may surround a singularity. The equation
${_4}C_i\!=\!0$ is a condition for the existence of a solution for
${_4}\gamma_{ik}$ because ${_4}\gamma_{ik}$ is found by solving
the $\ord(\lambda^4)$ field equations (\ref{quasi}), and ${_4}C_i$
is the integral (\ref{integral}) of these equations. However,
because of the Lemma (\ref{Lemma}), it happens that the
${_4}\gamma_{ik}$ terms in ${_4}\breve G_{ik}$ integrate to zero
in (\ref{integral}), so that ${_4}C_i$ is actually independent of
${_4}\gamma_{ik}$. In fact it is a general rule that $C_i$ for one
order can be calculated using only results from previous
orders\cite{EinsteinInfeld}, and this is a crucial aspect of the
EIH method. Therefore, the calculation of the $\ord(\lambda^4)$
equations of motion (\ref{C}) does not involve the calculation of
${_4}\gamma_{ik}$, and we will see below that it also does not
involve the calculation of ${_3}f_{ik}$ or ${_4}f_{0k}$.

The ${_4}\breve G_{ik}$ contribution to (\ref{integral}) is
derived in \cite{EinsteinInfeld}. For two particles with masses
$m_1$, $m_2$ and positions $\xi^i_1$, $\xi^i_2$, the
$\ord(\lambda^4)$ term from the integral over the first particle
is
\begin{eqnarray}
\label{neutral} \fl~~~~~~~~{^{\breve
G}_{\,4}}C_i&=&\frac{1}{2\pi}\int^1{_4}\breve G_{ik}n_kdS
=-4\left\{m_1\ddot{\xi}_1^i
-m_1m_2\frac{\partial}{\partial\xi_1^i}\left(\frac{1}{r}\right)\right\},
\end{eqnarray}
where
\begin{eqnarray}
\fl~~~~~~~~~~~r&=&\sqrt{\lower1pt\hbox{$(\xi_1^s-\xi_2^s)(\xi_1^s-\xi_2^s)$}}\,.
\end{eqnarray}
If there is no other contribution to (\ref{integral}), then
(\ref{C}) requires that ${^{\breve G}_{\,4}}C_i\!=\!0$ in
(\ref{neutral}), and the particle acceleration will be
proportional to a $\nabla(m_1m_2/r)$ Newtonian gravitational
force. These are the EIH equations of motion for vacuum general
relativity to $\ord(\lambda^4)$, or Newtonian order.

Because our effective energy-momentum tensor (\ref{Ttilde}) is
quadratic in $f_{\mu\nu}$, and the expansions
(\ref{g00expansion}-\ref{fikexpansion}) begin with $\lambda^2$
terms, the $\ord(\lambda^2)-\ord(\lambda^3)$ calculations leading
to (\ref{neutral}) are unaffected by the addition of
(\ref{Ttilde}) to the vacuum field equations. However, the
$8\pi\,{_4}\breve T_{ik}$ contribution to (\ref{integral}) will
add to the ${_4}\breve G_{ik}$ contribution. To calculate this
contribution, we will assume that our singularities in
$f_{\nu\mu}$ are simple moving Coulomb potentials, and that
$\theta^\rho\!=\!0$, $\Lambda\!=\!0$. Then from
(\ref{threeparts},\ref{f0kexpansion}-\ref{fikexpansion}) we see
that ${_2}F_{0k}\!=\!{_2}f_{0k}$, and from inspection of the extra
terms in our Maxwell equations (\ref{Maxwell},\ref{Faraday}) and
Proca equation (\ref{Proca}), we see that these equations are both
solved to $\ord(\lambda^3)$. Because (\ref{Ttilde}) is quadratic
in $f_{\mu\nu}$, we see from
(\ref{f0kexpansion}-\ref{fikexpansion}) that only ${_2}f_{0k}$ can
affect the $\ord(\lambda^4)$ equations of motion. Including only
${_2}f_{0k}$, our $f_{\mu\nu}$ is then a sum of two Coulomb
potentials with charges $Q_1$, $Q_2$ and positions $\xi^i_1$,
$\xi^i_2$ of the form
\begin{eqnarray}
\label{singularity}
\fl~~~~~~~~{_2}\Aphi_\mu&=&({_2}\varphi,0,0,0)~~,
~~~{_2}f_{0k}=2\,{_2}\Aphi_{[k|0]}=-\,{_2}\varphi_{|k},\\
%~~~{_4}f_{0k}={_4}\dot\Aphi_{,k},
%~~~{_3}f_{ik}=2\,{_3}\Aphi_{[k|i]},\\
\label{singularity2}
\fl~~~~~~~~~~{_2}\varphi\!&=&\!\psi^1+\psi^2~~~~~~,~~~\psi^1=Q_1/r_1~~~,~~~\psi^2=Q_2/r_2,\\
%~~~{_2}\varphi\!=\!\sum_{p=1}^N\psi^p,
%~~~{_3}\Aphi_k\!=\!\sum_{p=1}^N\psi^p\dot\xi_p^k,~~~\psi^p=Q_p/r_p,\\
\fl~~~~~~~~~~r_p\!&=&\sqrt{\lower1pt\hbox{$(x^s-\xi_p^s)(x^s-\xi_p^s)$}}~~~~~,~~~p=1...2\,.
\end{eqnarray}

Because our effective energy momentum tensor (\ref{Ttilde}) is
quadratic in both $f_{\mu\nu}$ and $g_{\mu\nu}$, and the
expansions (\ref{g00expansion}-\ref{fikexpansion}) start at
$\lambda^2$ in both of these quantitites, no
gravitational-electromagnetic interactions will occur at
$\ord(\lambda^4)$. This allows us to replace covariant derivatives
with ordinary derivatives, and $g_{\nu\mu}$ with $\eta_{\nu\mu}$
in (\ref{Ttilde}). This also allows us to replace $\breve
T_{\mu\nu}$ from (\ref{quasi},\ref{EIHT}) with (\ref{Ttilde}),
\begin{eqnarray}
\breve T_{\mu\nu}
=S_{\mu\nu}\!-\!\frac{\lower1pt\hbox{$1$}}{2}\eta_{\mu\nu}\eta^{\alpha\beta}
S_{\alpha\beta} \approx
\tS_{\mu\nu}\!-\!\frac{\lower1pt\hbox{$1$}}{2}g_{\mu\nu}g^{\alpha\beta}
\tS_{\alpha\beta} =\tT_{\mu\nu}.
\end{eqnarray}
Therefore, from (\ref{Ttilde}) we have
\begin{eqnarray}
\label{Ttemp} \fl 8\pi\breve T_{\nu\mu} =2\left({f_\nu}^\sigma
f_{\sigma\mu}
-\frac{1}{4}\eta_{\nu\mu}f^{\rho\sigma}f_{\sigma\rho}\right)\\
\fl~~~~~~+\!\left(2f^\tau{_{(\nu}}f_{\mu)}{^\alpha}{_{|\tau|\alpha}}
+2f^{\alpha\tau}f_{\tau(\nu|\mu)}{_{|\alpha}}
-f^\sigma{_{\nu|\alpha}}f^\alpha{_{\mu|\sigma}}
+f^\sigma{_{\nu|\alpha}}f_{\sigma\mu,}{^\alpha}
+\frac{1}{2}f^\sigma{_{\alpha|\nu}}f^\alpha{_{\sigma|\mu}}\right.\nonumber\\
\fl~~~~~~~~~~\left.-\eta_{\nu\mu}f^{\tau\beta}f_{\beta}{^\alpha}{_{|\tau|\alpha}}
\!-\frac{1}{4}\eta_{\nu\mu}(f^{\rho\sigma}f_{\sigma\rho})_|{^\alpha}{_{|\alpha}}
\!-\frac{3}{4}\eta_{\nu\mu}f_{[\sigma\beta|\alpha]}f^{[\sigma\beta}{_|}{^{\alpha]}}\!+\!(f^4)\right)\!\Lambda_b^{\!-1}.
\end{eqnarray}
%\bigskip\\
This can be simplified by keeping only $\ord(\lambda^4)$ terms.
The terms $2f^\tau{_{(\nu}}f_{\mu)}{^\alpha}{_{|\tau|\alpha}}$ and
$-\eta_{\nu\mu}f^{\tau\beta}f_{\beta}{^\alpha}{_{|\tau|\alpha}}$
vanish because (\ref{singularity}) satisfies Ampere's law to
$\ord(\lambda^2)$. The term
$-(3/4)\eta_{\nu\mu}f_{[\sigma\beta|\alpha]}f^{[\sigma\beta}{_|}{^{\alpha]}}$
vanishes because (\ref{singularity}) satisfies
$f_{[\sigma\beta|\alpha]}\!=\!2\Aphi_{[\beta|\sigma|\alpha]}\!=\!0$.
Also, since time derivatives count the same as a higher order in
$\lambda$, we can remove the term
$-f^\sigma{_{s|\alpha}}f^\alpha{_{m|\sigma}}
\!=\!-f^0{_{s|0}}f^0{_{m|0}}$, and we can change some of the
summations over Greek indices to summations over Latin indices.
The $(f^4)$ term will be $\ord(\lambda^8)$ so it can obviously be
eliminated. And as mentioned above, only ${_2}f_{0k}$ contributes
at $\ord(\lambda^4)$. Applying these results, and dropping the
order subscripts to reduce the clutter, the spatial part of
(\ref{Ttemp}) becomes,
\begin{eqnarray}
\fl 8\pi\,{_4}\breve T_{sm} &=&2\left({f_s}^0 f_{0m}
-\frac{1}{2}\eta_{sm}f^{r0}f_{0r}\right)\nonumber\\
\fl&&\!+\!\left(2f^{a0}f_{0(s|m)}{_{|a}}
+f^0{_{s|a}}f_{0m|}{^a}+f^0{_{a|s}}f^a{_{0|m}}
-\frac{1}{2}\eta_{sm}(f^{r0}f_{0r})_|{^a}{_{|a}}\right)\!\Lambda_b^{\!-1}\\
\label{Ttemp2} \fl&=&-2\left(f_{0s}f_{0m}
+\frac{1}{2}\eta_{sm}f_{0r}f_{0r}\right)\nonumber\\
\fl&&\!+\!\left(2f_{0a}f_{0(s|m)}{_{|a}}
-f{_{0s|a}}f_{0m|a}+f{_{0a|s}}f{_{0a|m}}
+\frac{1}{2}\eta_{sm}(f_{0r}f_{0r})_{|a|a}\right)\!\Lambda_b^{\!-1}.
\end{eqnarray}
Note that ${_2}\varphi$ from (\ref{singularity2}) obeys Gauss's
law,
\begin{eqnarray}
\label{Gauss} \varphi_{|a|a}=0.
\end{eqnarray}
Substituting (\ref{singularity}) into (\ref{Ttemp2}) and using
(\ref{Gauss}) gives
\begin{eqnarray}
\fl 8\pi\,{_4}\breve T_{sm} &=&-2\left(\varphi_{|s} \varphi_{|m}
+\frac{1}{2}\eta_{sm}\varphi_{|r}\varphi_{|r}\right)\nonumber\\
\fl&&+\left(2\varphi_{|a}\varphi_{|s|m|a}
-\varphi_{|s|a}\varphi_{|m|a} +\varphi{_{|a|s}}\varphi{_{|a|m}}
+\frac{1}{2}\eta_{sm}(\varphi_{|r}\varphi_{|r})_{|a|a}\right)\!\Lambda_b^{\!-1}\\
\ifnum\ExpandDerivations=1
\fl&=&-2\left(\varphi_{|s} \varphi_{|m}
+\frac{1}{2}\eta_{sm}\varphi_{|r}\varphi_{|r}\right)\nonumber\\
\fl&&-(\varphi_{|s}\varphi_{|a|m}+\varphi_{|r}\varphi_{|r|s}\eta_{am})_{|a}\Lambda_b^{\!-1}
+(\varphi_{|a}\varphi_{|s|m}+\varphi_{|r}\varphi_{|r|a}\eta_{sm})_{|a}\Lambda_b^{\!-1}\nonumber\\
\fi
\label{T} \fl &=&-2\left(\varphi_{|s} \varphi_{|m}
+\frac{1}{2}\eta_{sm}\varphi_{|r}\varphi_{|r}\right)
-2(\varphi_{|[s}\varphi_{|a]|m}+\varphi_{|r}\varphi_{|r|[s}\eta_{a]m})_{|a}\Lambda_b^{\!-1}.
%&=&\!(\varphi_{|[s}\varphi_{|a]|m})_{|a}\!-\!(\varphi_{|r}\varphi_{|r|[a}\eta_{s]m})_{|a}.
\end{eqnarray}
From (\ref{Lemma}) we see that the second group of terms in
(\ref{T}) integrates to zero in (\ref{integral}), and therefore it
can have no effect on the equations of motion. The first group of
terms in (\ref{T}) is what one gets with ordinary electro-vac
Einstein-Maxwell theory\cite{Wallace,WallaceThesis,Gorbatenko}, so
at this stage we have effectively proven that the theory predicts
a Lorentz force.

For completeness we will finish the derivation. First, we see from
(\ref{T},\ref{Gauss}) that ${_4}\breve T_{sm|s}\!=\!0$. This is to
be expected because of
(\ref{Tdivergence},\ref{zerodivergenceEIH}), and it means that the
$8\pi\,{_4}\breve T_{sm}$ contribution to the surface integral
(\ref{integral}) will be independent of surface size and shape.
This also means that only contributions from $1/{\rm distance}^2$
terms such as $\eta_{sm}/r^2$ or $x_sx_m/r^4$ can contribute to
(\ref{integral}). The integral over a term with any other
distance-dependence would necessarily depend on the surface
radius, and therefore we know beforehand that it must vanish or
cancel with other similar terms\cite{EinsteinInfeld}. Now,
$\varphi_{|i}\!=\!\psi^1_{|i}\!+\psi^2_{|i}$ from
(\ref{singularity2}). Because $\psi^1_{|i}$ and $\psi^2_{|i}$ both
go as $1/{\rm distance}^2$, but are in different locations, it is
clear from (\ref{T}) that contributions can only come from cross
terms between the two. Including only these terms gives,
\begin{eqnarray}
\label{EIHcrossterms} 8\pi\,{_4}\breve T^c_{sm}
=-2\left(\psi^1_{|s}\psi^2_{|m} \!+\!\psi^2_{|s}\psi^1_{|m} \!+\!
\,\eta_{sm}\psi^1_{|r}\psi^2_{|r}\right).
\end{eqnarray}
Some integrals we will need can be found in \cite{EinsteinInfeld}.
With $\psi=1/\sqrt{x^s x^s}$ we have,
\begin{eqnarray}
\label{EIHintegrals} &&\frac{1}{4\pi}\!\int^0
\psi_{|m}n_mdS=-1~~~,~~~ \frac{1}{4\pi}\!\int^0
\psi_{|a}n_mdS=-\frac{1}{3}\delta_{am}.
\end{eqnarray}
Using (\ref{EIHcrossterms},\ref{EIHintegrals},\ref{singularity2})
and integrating over the first particle we get,
\begin{eqnarray}
\fl\frac{1}{2\pi}\int^1\left[-8\pi\breve T_{sm}\,\right]n_mdS
&=&\frac{1}{2\pi}\!\int^1\!\!2\!\left(\psi^1_{|s}\psi^2_{|m}
\!+\!\psi^2_{|s}\psi^1_{|m}\!+\eta_{sm}\psi^1_{|r}\psi^2_{|r}\right)\!n_mdS\\
\label{EIHLorentz} \!&=&\!4Q_1\psi^2_{|s}(\xi_1)
\!\left(\!-\frac{1}{3}\!-\!1\!+\!\frac{1}{3}\right)
=-4Q_1\psi^2_{|s}(\xi_1).
\end{eqnarray}
Using
(\ref{C},\ref{integral},\ref{EIHLorentz},\ref{neutral},\ref{singularity2})
we get
\begin{eqnarray}
\fl~~~~~~~~0={_4}C_i&=&-4\left\{m_1\ddot{\xi}_1^i
-m_1m_2\frac{\partial}{\partial\xi_1^i}\left(\frac{1}{r}\right)\right\}
-4Q_1\psi^2_{|i}(\xi_1)\\
\ifnum\ExpandDerivations=1
\fl &=&-4\left\{m_1\ddot{\xi}_1^i
-m_1m_2\frac{\partial}{\partial\xi_1^i}\left(\frac{1}{r}\right)\right\}
-4Q_1\frac{\partial}{\partial\xi_1^i}\left(\frac{Q_2}{r}\right)\nonumber\\
\fi
\fl &=&-4\left\{m_1\ddot{\xi}_1^i
-m_1m_2\frac{\partial}{\partial\xi_1^i}\left(\frac{1}{r}\right)
+Q_1Q_2\frac{\partial}{\partial\xi_1^i}\left(\frac{1}{r}\right)\right\},
\end{eqnarray}
where
\begin{eqnarray}
r&=&\sqrt{\lower1pt\hbox{$(\xi_1^s-\xi_2^s)(\xi_1^s-\xi_2^s)$}}.
\end{eqnarray}
These are the EIH equations of motion for this theory to
$\ord(\lambda^4)$, or Newtonian/ Coulombian order. These equations
of motion clearly exhibit the Lorentz force, and in fact they
match the $\ord(\lambda^4)$ equations of motion of
Einstein-Maxwell theory.

\section{\label{Monopole}An exact electric monopole solution}
Here we derived an exact charged solution for this theory
which closely approximates the Reissner-Nordstr\"{o}m solution of Einstein-Maxwell theory.
%The derivation\cite{cShifflett} and
A MAPLE program\cite{LRESMAPLE} which
checks the solution is also available.
\ifnum\ExpandDerivations=1
It can be shown\cite{Papapetrou} that the assumption of spherical symmetry allows the
fundamental tensor to be written in the following form
\begin{eqnarray}
N_{\nu\mu}=
  \pmatrix{
 \gamma &-w&0&0\cr
w&-\alpha &0&0\cr
0&0&-\beta &r^2v\,{\rm sin}\,\theta\cr
0&0&-r^2v\,{\rm sin}\,\theta & -\beta\,{\rm sin}^2\theta}.
\end{eqnarray}
Both \cite{Papapetrou} and \cite{Takeno} assume this form with
$\beta=r^2,v=0$ to derive a solution to the original
Einstein-Schr\"{o}dinger field equations which looks similar to a
charged mass, but with some problems. Here we will derive a solution to the
modified field equations (\ref{JSsymmetric}-\ref{JScontractionsymmetric})
which is much closer to the Reissner-Nordstr\"{o}m
solution\cite{Reissner,Nordstrom} of ordinary electro-vac Einstein-Maxwell theory.
We will follow a similar procedure to
\cite{Papapetrou,Takeno} but will use
%the opposite sign convention for the Ricci tensor (\ref{Ricci}), and
coordinates $x_0,x_1,x_2,x_3\!=\!ct,r,\theta,\phi$ instead of
$x_1,x_2,x_3,x_4\!=\!r,\theta,\phi,ct$. We also use
the variables $a\nobreak=\nobreak1/\alpha,~b=\gamma\alpha,~\sinht=-w$,
which allow a simpler solution than the variables $\alpha,\gamma,w$. This gives
\begin{eqnarray}
\label{Nmatrix}
\fl~~~~N_{\nu\mu}=
 \pmatrix{
  ab&\sinht&0&0\cr
-\sinht&- 1/a&0&0\cr
0&0&- r^2 &0 \cr
0&0&0&- r^2{\rm sin}^2\theta
},\\
\fl~~~~N^{\dashv\mu\nu}=
\label{Nmatrixcontra}
 \pmatrix{
  1/ad&\sinht/d&0&0\cr
-\sinht/d & \!- ab/d&0&0\cr
0&0&\!\!\!- 1/r^2 &0 \cr
0&0&0&\!\!\!\!- 1/r^2{\rm sin}^2\theta
},\\
\label{rmN}
\fl~~~~\rmN=\sqrt{d}\,r^2{\rm sin}\,\theta,
\end{eqnarray}
where
\begin{eqnarray}
\label{ddef}
\fl~~~~d=b-\sinht^2.
\end{eqnarray}
From (\ref{Nmatrixcontra},\ref{rmN})
and the definitions (\ref{gdef},\ref{fdef}) of $g_{\nu\mu}$ and $f_{\nu\mu}$ we get
\begin{eqnarray}
\fl g^{\nu\mu}\!=\!
\frac{1}{\cosht}\!\pmatrix{
1/ad&0&0&0\cr
0&\!\!\!\!\!\!-ab/d&0&0\cr
0&0&\!\!\!\!\!-1/r^2&0\cr
0&0&0&\!\!\!\!\!\!\!-1/r^2{\rm sin}^2\theta
},~~
f^{\nu\mu}\!=\!
\frac{\Lambda_b^{\!1/2}}{\rmt\,\cosht }\pmatrix{
\,0&\!\!\!\!-\sinht/d&0&0\cr
\sinht/d&0&0&0\cr
0&0&0&0\cr
0&0&0&0\,
},\\
\label{gmatrix}
\fl g_{\nu\mu}\!=\!
 \,\cosht \pmatrix{
  ad &0&0&0\cr
0&- d/ab&0&0\cr
0&0&- r^2&0\cr
0&0&0&- r^2{\rm sin}^2\theta
},~~
\label{fmatrix}
f_{\nu\mu}\!=\!
\frac{\Lambda_b^{\!1/2}}{\rmt\,\cosht }\pmatrix{
\,0&\sinht&0&0\cr
-\sinht&0&0&0\cr
0&0&0&0\cr
0&0&0&0\,
},\\
\fl~~~~\rmg=\sqrt{b}\,r^2\sin\,\theta,
\end{eqnarray}
where
\begin{eqnarray}
\label{cdef}
\fl~~~~\cosht =\sqrt{b/d}=\rmg/{\lower1pt\hbox{$\rmN$}}\,.
\end{eqnarray}
Using prime ($'$) to represent $\partial/\partial r$, Ampere's law (\ref{Ampere})
and (\ref{Nmatrixcontra},\ref{rmN}) require that
\begin{eqnarray}
\label{Amperesolution}
\fl~~~~&&0=(\rmN N^{\dashv\,[01]})_{,1}
=\left(\frac{\sinht r^2 sin\,\theta}{\sqrt{d}}\right)'.
\end{eqnarray}
From (\ref{Amperesolution},\ref{ddef}), this means that for some constant $Q$ we have
\begin{eqnarray}
\label{Qintro}
\fl~~~~&&\frac{\sinht r^2}{\sqrt{d}}=\frac{\sinht r^2}{\sqrt{b-\sinht^2}}=\frac{Q\rmt}{\Lambda_b^{1/2}}.
\end{eqnarray}
Solving this for $\sinht^2$ gives
\begin{eqnarray}
\label{wsquared}
\fl~~~~\sinht^2=\frac{2bQ^2}{2Q^2\!-\Lambda_br^4}\,.
\end{eqnarray}
From (\ref{Qintro},\ref{wsquared}) we can derive the useful relationship
\begin{eqnarray}
\fl~~~~\sinht'&\!=\!&\frac{(\sinht^2)'}{2\sinht}
=\frac{1}{2\sinht}\!\left(\frac{2b'Q^2}{2Q^2\!-\Lambda_br^4}+\frac{8b\Lambda_br^3Q^2}{2Q^2\!-\Lambda_br^4}\left(\frac{\sinht^2}{2bQ^2}\right)\right)
%=\frac{1}{2\sinht}\!\left(b'\frac{\sinht^2}{b}
%-\frac{4b}{r}\frac{d}{\sinht^2}\frac{\sinht^4}{b^2}\right)
\label{bitch}
=\frac{\sinht}{b}\!\left(\frac{b'}{2}-\!\frac{2d}{r}\right)\!.
\end{eqnarray}
The connection equations (\ref{JSconnection})
are solved in \cite{Papapetrou,Takeno}.
In terms of our variables, the non-zero connections are
\begin{eqnarray}
\fl~~~~&&\tGam^1_{00}=\frac{a}{2}(ab)'+\frac{4a^2\sinht^2}{r}~~,~
\tGam^0_{10}=\tGam^0_{01}
=\frac{(ab)'}{2ab}+\frac{2\sinht^2}{br}
~~,~\tGam^1_{11}=\frac{-a'}{2a},\nonumber\\
\fl~~~~&&\tGam^2_{12}=\tGam^2_{21}
\label{finalconnections}
=\tGam^3_{13}=\tGam^3_{31}=\frac{\lower1pt\hbox{$1$}}{r},\\
\fl~~~~&&\tGam^1_{22}=-ar~~,~\tGam^1_{33}
=-ar\,{\rm sin}^2\theta~~,~
\tGam^3_{23}=\tGam^3_{32}={\rm cot}\,\theta~~,~
\tGam^2_{33}=-{\rm sin}\,\theta {\rm cos}\,\theta,\nonumber\\
\fl~~~~&&\tGam^2_{02}=-\tGam^2_{20}=\tGam^3_{03}
=-\tGam^3_{30}=-\frac{\lower1pt\hbox{$a\sinht$}}{r}~~,~
\tGam^1_{10}=-\tGam^1_{01}=-\frac{\lower1pt\hbox{$2a\sinht$}}{r},\nonumber\\
\label{contractedtGam}
\fl~~~~&&\tGam^\alpha_{\alpha0}=0,~~
\tGam^\alpha_{\alpha1}=\frac{\lower1pt\hbox{$b'$}}{2b}+\frac{\lower1pt\hbox{$2\sinht^2$}}{br}+\frac{\lower1pt\hbox{$2$}}{r},~~
\tGam^\alpha_{\alpha2}={\rm cot}\,\theta,~~\tGam^\alpha_{\alpha3}=0.
\end{eqnarray}
%This confirms the result $\tGam^\alpha_{\alpha[\nu,\mu]}\!=\!0$ from (\ref{funnytensor}).
%meaning that there is really no difference between the Ricci tensor and the Hermitianized Ricci tensor
%(\ref{HermitianizedRiccit}) when no charged matter terms are included in the Lagrangian density.

The Ricci tensor is also calculated in \cite{Papapetrou,Takeno}.
From (\ref{contractedtGam}) we have $\tGam^\alpha_{\alpha[\nu,\mu]}\!=0$
as expected from (\ref{funnytensor}), and this means that $\tR_{\nu\mu}\!=\!\tilde R_{\nu\mu}$.
In terms of our variables, and using our own sign convention,
the non-zero components of the Ricci tensor are
\begin{eqnarray}
\label{R00}
\fl~~~~~ -\tR_{00}&=&-\frac{aba''}{2}
-\frac{a^2b''}{2}-\frac{3aa'b'}{4}+\frac{a^2b'b'}{4b}
-\frac{a}{r}(ab'+a'b)-\frac{8a^2\sinht\sinht'}{r}\nonumber\\
\fl~~~~~ &&+\frac{a^2\sinht^2}{r}\left(\frac{3b'}{b}
\!-\!\frac{3a'}{a}\!-\!\frac{10}{r}\!+\!\frac{8\sinht^2}{br}\right),\\
\label{R11}
\fl~~~~~ -\tR_{11}&=&\frac{a''}{2a}
+\frac{b''}{2b}-\frac{b'b'}{4b^2}+\frac{3a'b'}{4ab}+\frac{a'}{ar}
+\frac{4\sinht\sinht'}{br}+\frac{\sinht^2}{br}\left(\frac{3a'}{a}
\!+\!\frac{4\sinht^2}{br}\!-\!\frac{2}{r}\right),\\
\label{R22}
\fl~~~~~ -\tR_{22}
&=&\frac{ar}{2}\left(\frac{2a'}{a}+\frac{b'}{b}\right)
+a-1+\frac{2a\sinht^2}{b},\\
\label{R33}
\fl~~~~~ -\tR_{33}
\!&=&\!-\tR_{22}\,{\rm sin}^2\theta,\\
\label{R10}
\fl~~~~~ -\tR_{[10]}
&=&2\left(\frac{a\sinht}{r}\right)'+\frac{6a\sinht}{r^2}.
~~\left\{{\cite{Papapetrou}~has~an~error~here}\right\}
\end{eqnarray}
From (\ref{Nmatrix},\ref{gmatrix},\ref{cdef},\ref{R33}),
the symmetric part of the field equations (\ref{JSsymmetric}) is
\begin{eqnarray}
\label{Rb00}
\fl~~~~~~~~~0&=&\tR_{00}+ \Lambda_bN_{00}+ \Lambda_zg_{00}
=\tR_{00}+\Lambda_bab+\Lambda_z\frac{\lower2pt\hbox{$ab$}}{\cosht }~,\\
\label{Rb11}
\fl~~~~~~~~~0&=&\tR_{11}+ \Lambda_bN_{11}+ \Lambda_zg_{11}
=\tR_{11}-\Lambda_b\frac{\lower2pt\hbox{$1$}}{a}-\Lambda_z\frac{\lower2pt\hbox{$1$}}{a\cosht}~,\\
\label{Rb22}
\fl~~~~~~~~~0&=&\tR_{22}+ \Lambda_bN_{22}+ \Lambda_zg_{22}
=\tR_{22}-\Lambda_br^2-\Lambda_z\cosht r^2,\\
\label{Rb33}
\fl~~~~~~~~~0&=&\tR_{33}+ \Lambda_bN_{33}+ \Lambda_zg_{33}
=(\tR_{22}+ \Lambda_bN_{22}+ \Lambda_zg_{22})\,{\rm sin}^2\theta.\frac{\vphantom{|}}{\vphantom{|}}
\end{eqnarray}
Forming a linear combination of (\ref{Rb11},\ref{Rb00}) and using
(\ref{R11},\ref{R00},\ref{bitch},\ref{ddef}), we find that many of
the terms cancel initially and we get,
\begin{eqnarray}
\fl~~~~~~~~~ 0&=&b\left(-\tR_{11}+\Lambda_b\frac{1}{a}
+\Lambda_z\frac{1}{a\cosht}\right)
+\frac{1}{a^2}\left(-\tR_{00}
-\Lambda_bab-\Lambda_z\frac{ab}{\cosht }\right)\\
\ifnum\ExpandDerivations=1
\fl~~~~~~~~~ &=&\frac{4\sinht\sinht'}{r}
+\frac{\sinht^2}{r}\left(\frac{4\sinht^2}{br}-\frac{2}{r}\right)
-\frac{b'}{r}
-\frac{8\sinht\sinht'}{r}
+\frac{\sinht^2}{r}\left(\frac{3b'}{b}-\frac{10}{r}
+\frac{8\sinht^2}{br}\right)\nonumber\\
\fl~~~~~~~~~ &=&-\frac{4\sinht}{r}\left[\frac{\sinht}{b}\left(\frac{b'}{2}
-\frac{2d}{r}\right)\right]
+\frac{12\sinht^2}{r}\left(\frac{\sinht^2}{br}-\frac{1}{r}\right)
-\frac{b'}{r}+\frac{3\sinht^2b'}{br}\nonumber\\
\fi
\label{1100}
\fl~~~~~~~~~ &=&-\frac{d}{br^2}\left(4\sinht^2+rb'\right).
\end{eqnarray}
From (\ref{wsquared}) this requires
\begin{eqnarray}
\label{thing}
0=\frac{8bQ^2}{2Q^2\!-\Lambda_br^4}+rb'.
\end{eqnarray}
Solving (\ref{thing}) and using (\ref{wsquared},\ref{ddef},\ref{cdef})
gives identical results to \cite{Papapetrou,Takeno},
\begin{eqnarray}
\label{b}
b&=&1-\frac{2Q^2}{\Lambda_br^4},\\
\label{w}
\sinht&=&\sqrt{\frac{2bQ^2}{2Q^2\!-\Lambda_br^4}}=\frac{\rmt Q}{\sqrt{\Lambda_b}\,r^2},\\
\label{d}
d&=&b-\sinht^2=1,\\
\label{c}
\cosht &=&\sqrt{b/d}=\sqrt{1-\frac{2Q^2}{\Lambda_br^4}}\,.
\end{eqnarray}

To find the variable ``$a$'', the 22 component of the field equations will
be used. The solution is guessed to be that of \cite{Papapetrou,Takeno}
plus an extra term $-\Lambda_z\extra/r$,
\begin{eqnarray}
\label{adef}
a=1-\frac{2M}{r}-\frac{\Lambda_br^2}{3}-\frac{\Lambda_z\extra}{r}.
\end{eqnarray}
Because ``$b$'' and ``$\sinht$'' are the same as \cite{Papapetrou,Takeno},
we just need to look at the extra terms that result from $\Lambda_z$.
Using (\ref{Rb22},\ref{R22},\ref{adef},\ref{1100},\ref{c}) gives,
\begin{eqnarray}
\fl~ 0&=&-\tR_{22}+\Lambda_br^2+\Lambda_z\cosht r^2
=\frac{ar}{2}\!\left(\frac{2a'}{a}+\!\frac{b'}{b}\right)
\!+a\!-\!1\!+\!\frac{2a\sinht^2}{b}\!+\Lambda_br^2\!+\Lambda_z\cosht r^2\\
\fl~~ &=&-\Lambda_z\!\left[r\left(\frac{\extra}{r}\right)'
+\frac{\extra b'}{2b}
+\frac{\extra}{r}+\frac{2\extra \sinht^2}{rb}-\cosht r^2\right]
%\fl~~~~~ &=&-\Lambda_z\!\left[\extra'+\frac{\extra}{2br}(b'r+4\sinht^2)
%-\cosht r^2\right]
\label{Veq}
=-\Lambda_z\left[\,\extra'-r^2\cosht\,\right].
\end{eqnarray}
This same equation is also obtained if the 11 or 00 components of the field
equations are used.
The solution for $V(r)$ can be written in terms of an elliptic integral
but we will not need to calculate it.
With (\ref{Veq}) and the definition
\begin{eqnarray}
\label{Vhatdef}
\fl~~~~\hat V&=&\frac{r\Lambda_b}{Q^2}\left(V-\frac{r^3}{3}\right)
\end{eqnarray}
we get the following results which will be used shortly,
\begin{eqnarray}
\label{Vhatidentity}
\fl~~~~\hat V'&=&\frac{\hat V}{r}+\frac{r^3\Lambda_b(\cosht-1)}{Q^2},~~~~~~~~~~~~~
\frac{Q^2}{\Lambda_br}\left(\frac{{\lower1pt\hbox{$\hat V$}}}{r^2}\right)'
=\cosht-1-\frac{Q^2\hat V}{\Lambda_br^4}.
\end{eqnarray}
\iffalse
%\ifnum\ExpandDerivations=1
The 11 component is,
\begin{eqnarray}
\fl 0&=&a\left(-\tR_{11}+\Lambda_b\frac{1}{a}+\Lambda_z\frac{1}{a\cosht}\right)\\
\fl &=&\frac{a''}{2}+\frac{ab''}{2b}-\frac{ab'b'}{4b^2}+\frac{3a'b'}{4b}+\frac{a'}{r}
+\frac{4a\sinht\sinht'}{br}+\frac{\sinht^2}{br}\left(3a'+\frac{4a\sinht^2}{br}
-\frac{2a}{r}\right)+\Lambda_b+\frac{\Lambda_z}{\cosht}\\
\fl &=&-\frac{\Lambda_z}{2}\left(\frac{V}{r}\right)''-\frac{\Lambda_zV b''}{2rb}
+\frac{\Lambda_zV b'b'}{4rb^2}-\frac{3\Lambda_zb'}{4b}\left(\frac{V}{r}\right)'
-\frac{\Lambda_z}{r}\left(\frac{V}{r}\right)'-\frac{4\Lambda_zV\sinht\sinht'}{br^2}\nonumber\\
\fl &&+\frac{\sinht^2}{br}\left(-3\Lambda_z\left(\frac{V}{r}\right)'
-\frac{4\Lambda_zV\sinht^2}{br^2}+\frac{2\Lambda_zV}{r^2}\right)+\frac{\Lambda_z}{\cosht}\\
\fl &=&-\Lambda_z\left[\frac{1}{2}\left(\frac{V'}{r}-\frac{V}{r^2}\right)'
+\frac{1}{r}\left(\frac{V'}{r}-\frac{V}{r^2}\right)-\frac{1}{\cosht}\right]\\
\fl &=&-\Lambda_z\left[\frac{1}{2r}(r^2\cosht)'-\frac{1}{\cosht}\right]\\
\fl &=&-\Lambda_z\left[\frac{1}{2r}\left(2r\cosht+\frac{r^2}{2\cosht}\left(\frac{8Q^2}{\Lambda_br^5}\right)\right)
-\frac{1}{\cosht}\right]\\
\fl &=&-\Lambda_z\left[\frac{1}{\cosht}(1+\sinht^2-\sinht^2)-\frac{1}{\cosht}\right]\\
\fl &=& 0.
\end{eqnarray}
\fi
Next we consider the antisymmetric part of the field equations (\ref{JSantisymmetric}),
where only the 10 component is non-vanishing. Using
(\ref{R10},\ref{Nmatrix},\ref{w},\ref{adef}) gives
\begin{eqnarray}
\fl~~F_{01}&=&\frac{\Lambda_b^{\!-1/2}}{\rmt}(\tR_{[01]}+\Lambda_b N_{[01]})
=\frac{\Lambda_b^{\!-1/2}}{\rmt}\left[2\left(\frac{a\sinht}{r}\right)'
+\frac{6a\sinht}{r^2}+\Lambda_b\sinht\right]\\
\fl &=&2\left(\frac{aQ}{\Lambda_br^3}\right)'+\frac{6aQ}{\Lambda_br^4}
+\frac{Q}{r^2}
\label{phipreliminary}
=\frac{Q}{r^2}\left(1+\frac{2a'}{\Lambda_br}\right)
\end{eqnarray}

Using
(\ref{gmatrix},\ref{b},\ref{w},\ref{d},\ref{c},\ref{adef},\ref{Veq},\ref{phipreliminary},\ref{Vhatdef},\ref{Vhatidentity})
we can put the solution in its final form.
\fi
The solution is
\begin{eqnarray}
\label{finalg}
\fl~~ds^2&=&\cosht a dt^2-\frac{1}{\cosht a}\,dr^2-\cosht r^2 d\theta^2-\cosht r^2 sin^2\theta d\phi^2,\\
\label{finalf}
\fl~~f^{10}&=&\frac{Q}{\cosht r^2},~~~~\rmN=r^2 sin\,\theta,~~~~\rmg=\cosht r^2 sin\,\theta,\\
\label{finalF}
\fl~~F_{01}&=&-A_0'
=\frac{Q}{r^2}\!\left[1+\frac{4M}{\Lambda_br^3}-\frac{4\Lambda}{3\Lambda_b}
+2\left(\cosht-1-\frac{Q^2{\lower1pt\hbox{$\hat V$}}}{\Lambda_br^4}\right)\!\left(1-\frac{\Lambda}{\Lambda_b}\right)\right],\\
\label{a}
\fl~~~~a&=&1-\frac{2M}{r}-\frac{\Lambda r^2}{3}
+\frac{Q^2\hat V}{r^2}\left(1-\frac{\Lambda}{\Lambda_b}\right),
\end{eqnarray}
where prime ($'$) means $\partial/\partial r$, and $\cosht$ and $\hat V$ are very close to one for ordinary radii,
\begin{eqnarray}
\label{finalc}
\fl~~~~\cosht &=&\sqrt{1-\frac{2Q^2}{\Lambda_br^4}}
=1-\frac{Q^2}{\Lambda_br^4}\cdots-\frac{(2i)!}{[i!]^2 4^i(2i\!-\!1)}\!\left(\frac{2Q^2}{\Lambda_br^4}\right)^i,\,\\
\fl~~~~\hat V&=&\frac{r\Lambda_b}{Q^2}\left(\int r^2\cosht\,dr -\frac{r^3}{3}\right)
=1+\frac{Q^2}{10\Lambda_br^4}\cdots +\frac{(2i)!}{i!(i\!+\!1)!\,4^i(4i\!+\!1)}\!\left(\frac{2Q^2}{\Lambda_br^4}\right)^i.
\end{eqnarray}

With $\Lambda_z\!=\!0,\Lambda_b\!=\!\Lambda$ we get the Papapetrou
solution\cite{Papapetrou,Takeno} of the unmodified Einstein-Schr\"{o}dinger
theory. In this case the $M/\Lambda_b r^3$ term in (\ref{finalF}) would be huge from (\ref{Lambda}),
and the $Q^2/r^2$ term in (\ref{a}) disappears, which is why the
Papapetrou solution was found to be unsatisfactory in \cite{Papapetrou}.
However, we are instead assuming $\Lambda_b\!\approx\!-\Lambda_z$ from (\ref{Lambdab},\ref{Cz}).
In this case the solution matches the Reissner-Nordstr\"{o}m solution except for terms which
are negligible for ordinary radii.
To see this, first recall that $\Lambda/\Lambda_b\!\sim\! 10^{-122}$ from (\ref{LambdaoverLambdab}),
so the $\Lambda$ terms are all completely negligible.
Ignoring the $\Lambda$ terms and keeping only the leading order terms in
(\ref{finalF},\ref{a},\ref{finalc}) gives
\begin{eqnarray}
\label{Fapprox}
F_{01}&=&
\frac{Q}{r^2}\!\left[1+\frac{4M}{\Lambda_br^3}-\frac{4Q^2}{\Lambda_br^4}\right]+\ord(\Lambda_b^{\!-2}),\\
\label{A0approx}
A_{0}
%&=&-\int\!F_{01}dr
&=&\frac{Q}{r}\!\left[1+\frac{M}{\Lambda_br^3}-\frac{4Q^2}{5\Lambda_br^4}\right]+\ord(\Lambda_b^{\!-2}),\\
\label{aapprox}
~~a&=&1-\frac{2M}{r}+\frac{Q^2}{r^2}\!\left[1+\frac{Q^2}{10\Lambda_br^4}\right]+\ord(\Lambda_b^{\!-2}),\\
\label{capprox}
~~\cosht &=&1-\frac{Q^2}{\Lambda_br^4}+\ord(\Lambda_b^{\!-2}).
\end{eqnarray}
%Also, near the event horizon of a solar mass extremal charged black hole we have from (\ref{BHskew})
%\begin{eqnarray}
%\label{ReisnerNordstromratio1}
%\frac{Q^2}{\Lambda_br^4}\sim 10^{-76},
%\end{eqnarray}
For the smallest radii probed by high-energy particle physics we get from (\ref{highenergyskew}),
\begin{eqnarray}
\label{ReisnerNordstromratio2}
\frac{Q^2}{\Lambda_br^4}\sim 10^{-66}.
\end{eqnarray}
The worst-case value of $M/\Lambda_b r^3$ might be near the Schwarzschild radius $r_s$
of black holes where $r\!=\!r_s\!=\!2M$ and $M/\Lambda_b r^3\!=\!1/2\Lambda_b r_s^2$.
This value will be largest for the lightest black holes,
and the lightest black hole that we can expect to
observe would be of about one solar mass, where we have
\begin{eqnarray}
\label{Qmratio1}
\frac{M}{\Lambda_b r^3}
&\sim&\frac{1}{2\Lambda_b r_s^2}
=\frac{1}{2\Lambda_b}\left(\frac{c^2}{2Gm_\odot}\right)^2\!\sim\!10^{-77}.
\end{eqnarray}
Also, an electron has $M=Gm_e/c^2=7\times 10^{-56}cm$, and using (\ref{Lambdab}) and
the smallest radii probed by high-energy particle physics ($10^{-17}cm$) we have
\begin{eqnarray}
\label{Qmratio2}
\frac{M}{\Lambda_b r^3}
\sim \frac{7\times 10^{-56}}{10^{66}(10^{-17})^3}\sim 10^{-70}.
\end{eqnarray}

From (\ref{ReisnerNordstromratio2},\ref{Qmratio1},\ref{Qmratio2},\ref{LambdaoverLambdab})
we see that our electric monopole solution (\ref{finalg}-\ref{a})
has a fractional difference from the Reissner-Nordstr\"{o}m solution
of at most $10^{-66}$ for worst-case radii accessible to measurement.
Clearly our solution does not have the deficiencies of the Papapetrou
solution\cite{Papapetrou,Takeno} in the original theory,
and it is almost certainly indistinguishable
from the Reissner-Nordstr\"{o}m solution experimentally.
Also, when this solution is expressed in Newman-Penrose
tetrad form, it can be shown to be of Petrov Type-D\cite{Shifflett2}.
And of course the solution reduces to the Schwarzschild solution for $Q\!=0$.
And we see that
the solution goes to the Reissner-Nordstr\"{o}m solution exactly in the limit
as $\Lambda_b\!\rightarrow\!\infty$.

%\bigskip
The only significant difference between our electric monopole solution and the
Reissner-Nordstr\"{o}m solution occurs on the Planck scale.
From (\ref{finalg},\ref{finalc}), the surface area of the solution is\cite{Chandrasekhar},
\begin{eqnarray}
\fl~~~\left({\rm surface}\atop{\rm area}\right)=\int_0^\pi\,d\theta\int_0^{2\pi}\!d\phi\sqrt{g_{\theta\theta}g_{\phi\phi}}
=4\pi r^2\cosht
=4\pi r^2\sqrt{1-\frac{2Q^2}{\Lambda_br^4}}.
\end{eqnarray}
The origin of the solution is where the surface area vanishes, so in our coordinates the origin
is not at $r\!=0$ but rather at
\begin{eqnarray}
\label{re}
\fl~~~~~~~~~~~r_0=\sqrt{Q}(2/\Lambda_b)^{1/4}.
%=\left(\frac{2\alpha l_P^2}{C_z\omega_c^4l_P^2}\right){\!\raise6pt\hbox{$^{1/4}$}}
%=\frac{1}{\omega_c}\left(\frac{2\alpha}{C_z}\right){\!\raise6pt\hbox{$^{1/4}$}}
%\sim 3.16\times 10^{-34}{\rm cm}.
\end{eqnarray}
From (\ref{redef},\ref{Lambdab}) we have $r_0\!\sim\!l_P\!\sim\!10^{-33}cm$ for an elementary charge,
and $r_0\!\ll\!2M$ for any realistic astrophysical black hole.
For $Q/M\!<\!1$ the behavior at the origin is hidden behind
an event horizon nearly identical to that of the Reissner-Nordstr\"{o}m solution.
For $Q/M\!>\!1$ where there is no event horizon, the behavior at the origin differs markedly
from the simple naked singularity of the Reissner-Nordstr\"{o}m solution.
For the Reissner-Nordstr\"{o}m solution all of the relevant fields have singularities at the origin,
$g_{00}\!\sim\!Q^2/r^2$, $A_0\!=\!Q/r$, $F_{01}\!=\!Q/r^2$,
$R_{00}\!\sim\!2Q^4/r^6$ and $R_{11}\!\sim\!2/r^2$.
For our solution the metric has a less severe singularity at the origin,
$g_{11}\!\sim\!-\sqrt{r_0}/2\sqrt{r-r_0}$ and $\rmg\!=\!0$.
Also, the fields $N_{\mu\nu}$, $N^{\dashv\nu\mu}$, $\rmN$, $A_\nu$, $\rmg f^{\nu\mu}$,
$\rmg f_{\nu\mu}$, $\rmg g^{\nu\mu}$, $\rmg g_{\nu\mu}$, and the functions ``a'' and $\hat V$
all have finite nonzero values and derivatives at the origin, because it can be shown that
$\hat V(r_0)\!=\!\sqrt{2}\left[\Gamma(1/4)\right]^2\!/6\sqrt{\pi}\!-\!2/3=1.08137$.
The fields $F_{\nu\mu}$, $\tGam^\alpha_{\mu\nu}$ and $\rmg\,\tR_{\nu\mu}$ are also finite
and nonzero at the origin, so if we use the tensor density form of the field equations
(\ref{Einstein},\ref{Ampere}), there is no ambiguity as to whether the field equations
are satisfied at this location.
%Based on this result and the nonsingular behavior
%of $N_{\mu\nu}$, $\tGam^\alpha_{\mu\nu}$ and $A_\nu$ at $r\!=\!r_e$, one can argue that
%the $r_e$ surface does not really represent a naked singularity even for $Q/M\!>\!1$
%where there is no event horizon to hide it. This would certainly not be true if the
%definition of the term naked singularity is instead based upon $g_{\mu\nu}$,
%but this definition is not necessarily appropriate
%in the context of alternative theories of gravity.
%Another feature of the $r_e$ surface is that it
%gives our charged solution a definite size, or non-pointlike character for $Q/M\!>\!1$,
%and like string theory, this might be significant in attempts to quantize this theory.
%Also note that if we require the area of the $r_e$ surface to be no
%smaller than the Planck area, this determines the smallest charge,
%\begin{eqnarray}
%\fl~~~~~~~~~~~~~~~l_P^2=4\pi r_e^2
%%&=&4\pi|Q|
%&=&4\pi\frac{1}{\omega_c^2}\sqrt{\frac{2\alpha}{C_z}}
%=4\pi\frac{e}{\omega_c^2}\sqrt{\frac{2}{C_z\hbar c}}\,,\\
%\Rightarrow~~e&=&\frac{\omega_c^2l_P^2}{4\pi}\sqrt{\frac{C_z\hbar c}{2}}
%=9.8\times 10^{-10} esu.
%\end{eqnarray}
%It could perhaps be significant that this charge is rather close to the elementary
%charge $e=4.8\times 10^{-10}esu$, given that the choice of $\omega_c\sim 1/l_P$ is a guess.
%Moreover, if the area of the $r_e$ surface is required to be quantized in
%units of $l_P^2$, charge quantization results.

Finally let us consider the result from (\ref{highenergyskew}) that
$|f^\mu{_\sigma}\Lambda_b^{\!-1/2}|\!<\!10^{-33}$ for worst-case electromagnetic fields
accessible to measurement. The ``smallness'' of this value may
seem unappealing at first, considering that $g^{\mu\nu}$
and $f^{\mu\nu}\rmt\Lambda_b^{\!-1/2}$ are part of the total field $(\rmN/\rmg)N^{\dashv\nu\mu}\!=\!g^{\mu\nu}\!+\!f^{\mu\nu}\rmt\Lambda_b^{\!-1/2}$
as in (\ref{Wdef}). However, for an elementary charge, $|f^{\mu\nu}\Lambda_b^{\!-1/2}|$ is not really
small if one compares it to $g^{\mu\nu}\!-\eta^{\mu\nu}$ instead of $g^{\mu\nu}$.
Our electric monopole solution (\ref{finalg},\ref{finalf},\ref{a}) has
$g^{00}\!\approx\! 1\!+2M/r\!+Q^2\!/r^2$ and $f^{01}\!\approx\!Q/r^2$.
So for an elementary charge, we see from (\ref{redef},\ref{Lambdab}) that
$|f^{01}\Lambda_b^{\!-1/2}|\!\sim\!Q^2\!/r^2$ for any radius.
%Also, using $M\!=\!Gm_e/c^2\!=\!7\!\times\!10^{-56}cm$ for an electron and
%$-\Lambda_b Q^2/2\!=\!e^2G/c^4$ from (\ref{cgs}), we see that
%$Q/r^2\!\sim\! 2M/r\!\sim\!\Lambda_bQ^2\!/2r^2\!\sim\!10^{-43}$
%near the classical electron radius $e^2/m_ec^2\!=\!2.82\times 10^{-13}cm$.
%From another perspective, it is almost required that $|f^\mu{_\sigma}\rmt\Lambda_b^{\!-1/2}|\!\ll\!1$
%at ordinary radii for any theory which unifies gravitation and electromagnetism
%by combining $g^{\mu\nu}$ and $f^{\mu\nu}\rmt\Lambda_b^{\!-1/2}$ as in (\ref{Wdef}).
%This is because $g^{00}\!\approx\! 1$, so a charged particle must always have
%some crossover radius at which $|f^{01}\rmt\Lambda_b^{\!-1/2}|$ becomes larger than $g^{00}$.
%Dramatic effects should be expected to occur at this radius, which would be difficult
%to relate to any known physical phenomena. Fortunately for the present theory,
%this radius is the $r_e$ radius from (\ref{re}), and it is near the
%Planck length and therefore inaccessible to measurement.

\section{\label{Conclusions}Conclusions}

%The Einstein-Schr\"{o}dinger theory is modified by
%adding a cosmological constant contribution caused by zero-point fluctuations.
%This cosmological constant
%which multiplies the symmetric metric
%is assumed to be nearly cancelled by Schr\"{o}dinger's ``bare''
%cosmological constant
% which multiplies the nonsymmetric fundamental tensor,
%such that the total ``physical'' cosmological constant is consistent with measurement.
%similar to vacuum energy renormalization in quantum field theory.
The Einstein-Schr\"{o}dinger theory was modified to account for
a quantum-mechanical effect.
The Einstein equations of this theory have zero divergence with the Christoffel
connection, allowing additional (non-electromagnetic) fields to be included in the theory.
The field equations match the ordinary electro-vac
Einstein and Maxwell equations except for additional terms which are
$<\!10^{-16}$ of the usual terms for worst-case field
strengths and rates-of-change accessible to measurement.
The theory avoids ghosts in the sense that ghosts could only exist if their frequency
exceeds the cutoff frequency $\omega_c\!\sim\!1/l_P$ of zero-point fluctuations.
The Einstein-Infeld-Hoffmann (EIH) equations of motion for this
theory match the equations of motion for Einstein-Maxwell theory
to Newtonian/Coulombian order, which proves the existence of a Lorentz force.
An exact electric monopole solution exists for this theory, and it matches
the Reissner-Nordstr\"{o}m solution except for additional terms which are
$\sim\!10^{-66}$ of the usual terms for worst-case radii accessible to measurement.
%An exact electromagnetic plane-wave solution also exists,
%and it is identical to a solution of Einstein-Maxwell theory.
The theory becomes exactly electro-vac Einstein-Maxwell theory in the limit as
$|\Lambda_z|\!\rightarrow\!\infty$, $\Lambda_b\!\rightarrow\!\infty$,
or more precisely as $\omega_c\!\rightarrow\!\infty$.
It seems unlikely that there is a test which is sensitive enough
to discriminate the theory from electro-vac Einstein-Maxwell theory.

\section*{Acknowledgements}
%I am grateful to Clifford Will for discussions
%and for helpful comments on drafts of this manuscript.
%%Thanks also to Claude Bernard for his help.
This work was supported in part by the National Science Foundation under grant PHY~03-53180.

\appendix

\section{\label{ExtractionofConnectionAddition}Extraction of a connection
addition from the Hermitianized Ricci tensor}
Substituting $\tGam^\alpha_{\nu\mu}\!=\!\Gamma^\alpha_{\nu\mu}
\!+\!\Upsilon^\alpha_{\nu\mu}$ from (\ref{gammadecomposition},\ref{Christoffel})
into (\ref{HermitianizedRiccit})
%and using the notation $\bUps^\alpha_{\nu\mu}
%=\Upsilon^\alpha_{(\nu\mu)}$,
%$\cUps^\alpha_{\nu\mu}=\Upsilon^\alpha_{[\nu\mu]}$
gives
\begin{eqnarray}
%\label{Ricciaddition}
\fl \hR_{\nu\mu}(\tGam)
\!&=&2[(\Gamma^\alpha_{\nu[\mu}
\!+\!\Upsilon^\alpha_{\nu[\mu}){_{,\alpha]}}
+(\Gamma^\sigma_{\nu[\mu}
\!+\!\Upsilon^\sigma_{\nu[\mu})(\Gamma^\alpha_{\sigma|\alpha]}
\!+\!\Upsilon^\alpha_{\sigma|\alpha]})]
+\!(\Gamma^\alpha_{\alpha[\nu}
\!+\!\Upsilon^\alpha_{\alpha[\nu})_{,\mu]}\\
\iftrue
\fl &=&R_{\nu\mu}(\Gamma)+\Upsilon^\alpha_{\nu\mu,\alpha}
-\Gamma^\sigma_{\nu\alpha}\Upsilon^\alpha_{\sigma\mu}
+\Gamma^\alpha_{\sigma\alpha}\Upsilon^\sigma_{\nu\mu}
-\Gamma^\alpha_{\sigma\mu}\Upsilon^\sigma_{\nu\alpha}\nonumber\\
\nopagebreak
\fl &&~~~~~~~~~-\Upsilon^\alpha_{\alpha(\nu,\mu)}
+\Gamma^\sigma_{\nu\mu}\Upsilon^\alpha_{\sigma\alpha}
-\Upsilon^\sigma_{\nu\alpha}\Upsilon^\alpha_{\sigma\mu}
+\Upsilon^\sigma_{\nu\mu}\Upsilon^\alpha_{\sigma\alpha}\\
\fi
%\label{Ricciaddition2}
\fl \!&=&R_{\nu\mu}(\Gamma)+\Upsilon^\alpha_{\nu\mu;\alpha}
-\Upsilon^\alpha_{\alpha(\nu;\mu)}
-\Upsilon^\sigma_{\nu\alpha}\Upsilon^\alpha_{\sigma\mu}
+\Upsilon^\sigma_{\nu\mu}\Upsilon^\alpha_{\sigma\alpha},\\
\fl\!R_{(\nu\mu)}(\tGam)
\label{Ricciadditionsymmetric}
%\fl
\!&=& R_{\nu\mu}(\Gamma)+{\Upsilon}^\alpha_{\!(\nu\mu);\alpha}
\!-\!\Upsilon^\alpha_{\alpha(\nu;\mu)}
\!-\!{\Upsilon}^\sigma_{\!(\nu\alpha)}{\Upsilon}^\alpha_{\!(\sigma\mu)}
\!-\!{\Upsilon}^\sigma_{\![\nu\alpha]}{\Upsilon}^\alpha_{\![\sigma\mu]}
\!+\!{\Upsilon}^\sigma_{\!(\nu\mu)}\Upsilon^\alpha_{\sigma\alpha},\\
\fl\!R_{[\nu\mu]}(\tGam)
\label{Ricciadditionantisymmetric}
\!&=&{\Upsilon}^\alpha_{\![\nu\mu];\alpha}
%\!-\!\Upsilon^\alpha_{\alpha[\nu;\mu]}
\!-\!{\Upsilon}^\sigma_{\!(\nu\alpha)}{\Upsilon}^\alpha_{\![\sigma\mu]}
\!-\!{\Upsilon}^\sigma_{\![\nu\alpha]}{\Upsilon}^\alpha_{\!(\sigma\mu)}
\!+\!{\Upsilon}^\sigma_{\![\nu\mu]}\Upsilon^\alpha_{\sigma\alpha}.
\end{eqnarray}
Also, substituting $\nGam^\alpha_{\!\nu\mu}\!=\!\tGam^\alpha_{\!\nu\mu}
\!+[\,\delta^\alpha_\mu \Aphi_\nu\!-\delta^\alpha_\nu \Aphi_\mu]\!\rmt\Lambda_b^{\!1/2}$
from (\ref{gamma_natural}) into (\ref{HermitianizedRicci}) and using
$\tGam^\alpha_{\nu\alpha}\!=\!\nGam^\alpha_{\!(\nu\alpha)}\!=\!\tGam^\alpha_{\alpha\nu}$ gives
\begin{eqnarray}
\fl \hR_{\nu\mu}(\nGam)
\!&=&\tGam^\alpha_{\!\nu\mu,\alpha}
\!+[\,\delta^\alpha_\mu \Aphi_\nu\!-\delta^\alpha_\nu \Aphi_\mu]_{,\alpha}\!\rmt\Lambda_b^{\!1/2}
\!-\tGam^\alpha_{\!\alpha(\nu,\mu)}\nonumber\\
\fl &&+\left(\tGam^\sigma_{\!\nu\mu}+[\,\delta^\sigma_\mu \Aphi_\nu
\!-\delta^\sigma_\nu \Aphi_\mu]\rmt\Lambda_b^{\!1/2}\right)\tGam^\alpha_{\!\sigma\alpha}\nonumber\\
\fl &&-\left(\tGam^\sigma_{\!\nu\alpha}\!+[\,\delta^\sigma_\alpha \Aphi_\nu
-\delta^\sigma_\nu \Aphi_\alpha]\rmt\Lambda_b^{\!1/2}\right)\!
\left(\tGam^\alpha_{\!\sigma\mu}\!+[\,\delta^\alpha_\mu \Aphi_\sigma-\delta^\alpha_\sigma \Aphi_\mu]\rmt\Lambda_b^{\!1/2}\right)\nonumber\\
\fl &&+2(n\!-\!1)A_\nu A_\mu\Lambda_b\\
\fl\!&=&\tGam^\alpha_{\!\nu\mu,\alpha}
\!+2\Aphi_{[\nu,\mu]}\rmt\Lambda_b^{\!1/2}
-\tGam^\alpha_{\!\alpha(\nu,\mu)}+\tGam^\sigma_{\!\nu\mu}\tGam^\alpha_{\!\sigma\alpha}
+[\,\Aphi_\nu\tGam^\alpha_{\!\mu\alpha}
-\Aphi_\mu\tGam^\alpha_{\!\nu\alpha}]\rmt\Lambda_b^{\!1/2}\nonumber\\
\fl &&-\tGam^\sigma_{\!\nu\alpha}\tGam^\alpha_{\!\sigma\mu}
-[\,\tGam^\sigma_{\!\nu\mu}\Aphi_\sigma-\tGam^\sigma_{\!\nu\sigma}\Aphi_\mu]\rmt\Lambda_b^{\!1/2}
-[\,\Aphi_\nu\tGam^\alpha_{\!\alpha\mu}
-\Aphi_\alpha\tGam^\alpha_{\!\nu\mu}]\rmt\Lambda_b^{\!1/2}\nonumber\\
\fl &&+2\Aphi_\nu\Aphi_\mu(1-n-1+1)\Lambda_b
+2(n\!-\!1)A_\nu A_\mu\Lambda_b\\
\label{RnGam}
\fl &=&\tGam^\alpha_{\!\nu\mu,\alpha}
-\tGam^\alpha_{\alpha(\nu,\mu)}+\tGam^\sigma_{\!\nu\mu}\tGam^\alpha_{\!\sigma\alpha}
-\tGam^\sigma_{\!\nu\alpha}\tGam^\alpha_{\!\sigma\mu}
\!+2\Aphi_{[\nu,\mu]}\rmt\Lambda_b^{\!1/2}\\
\fl &=&\hR_{\nu\mu}(\tGam)\!+2\Aphi_{[\nu,\mu]}\rmt\Lambda_b^{\!1/2}.
\end{eqnarray}

\section{\label{UsefulIdentity}A divergence identity}
Using only the definitions (\ref{gdef},\ref{fdef}) of $g_{\nu\mu}$
and $f_{\nu\mu}$, and the identity (\ref{sqrtdetcomma}) gives,
\begin{eqnarray}
\fl&&\left(\K^{(\mu}{_{\nu)}} \!-\!\frac{1}{2}\delta^\mu_\nu
\K^\rho_\rho\right)\!{_{;\,\mu}}
-\frac{3}{2}f^{\sigma\rho}N_{[\sigma\rho,\nu]}\rmt\Lambda_b^{\!-1/2}\\
\fl &&=\frac{1}{2}g^{\sigma\rho}
(N_{(\rho\nu);\sigma}\!+\!N_{(\nu\sigma);\rho}\!-\!N_{(\rho\sigma);\nu})
%-\frac{3}{2}f^{\sigma\rho}N_{[\sigma\rho,\nu]}\\
\!-\!\frac{1}{2}f^{\sigma\rho}(N_{[\sigma\rho];\nu}\!+\!N_{[\rho\nu];\sigma}
\!+\!N_{[\nu\sigma];\rho})\rmt\Lambda_b^{\!-1/2}\\
\fl &&=\frac{1}{2}\frac{\rmN}{\rmg} \!\left[ \K^{\dashv(\sigma\rho)}
(N_{(\rho\nu);\sigma} \!+\!N_{(\nu\sigma);\rho}
\!-\!N_{(\rho\sigma);\nu}) \!+\!\K^{\dashv[\sigma\rho]}
(N_{[\sigma\rho];\nu} \!+\!N_{[\rho\nu];\sigma}
\!+\!N_{[\nu\sigma];\rho})\right]\\
\fl &&=\frac{1}{2}\frac{\rmN}{\rmg} \!\left[\,\K^{\dashv\sigma\rho}
(N_{(\rho\nu);\sigma} \!+\!N_{(\nu\sigma);\rho}
\!-\!N_{(\rho\sigma);\nu}) \!+\!\K^{\dashv\sigma\rho}
(N_{[\rho\nu];\sigma} \!+\!N_{[\nu\sigma];\rho}
\!-\!N_{[\rho\sigma];\nu})\right]\\
\fl &&=\frac{1}{2}\frac{\rmN}{\rmg}\K^{\dashv\sigma\rho}
(N_{\rho\nu;\sigma}+N_{\nu\sigma;\rho}-N_{\rho\sigma;\nu})\\
\fl &&=\frac{1}{2}\frac{\rmN}{\rmg}\left[\K^{\dashv\sigma\rho}
(N_{\rho\nu;\sigma}+N_{\nu\sigma;\rho})
-\K^{\dashv\sigma\rho}(N_{\rho\sigma,\nu}
-\Gamma^\alpha_{\rho\nu}N_{\alpha\sigma}
-\Gamma^\alpha_{\sigma\nu}N_{\rho\alpha})\right]\\
\fl &&=-\frac{1}{2}\frac{\rmN}{\rmg}
(\K^{\dashv\sigma\rho}{_{;\sigma}}N_{\rho\nu}
+\K^{\dashv\sigma\rho}{_{;\rho}}N_{\nu\sigma})
-\frac{1}{\rmg}(\rmN\,)_{;\nu}\\
\fl &&=-\frac{1}{2}\left[
\left(\frac{\rmN}{\rmg}\K^{\dashv\sigma\rho}\right)
{_{\!\!;\sigma}}N_{\rho\nu}
+\left(\frac{\rmN}{\rmg}\K^{\dashv\sigma\rho}\right)
{_{\!\!;\rho}}N_{\nu\sigma}\right]\\
\fl &&=-\frac{1}{2}\left[
(g^{\rho\sigma}+f^{\rho\sigma}\rmt\Lambda_b^{\!-1/2}){_{;\sigma}}N_{\rho\nu}
+(g^{\rho\sigma}+f^{\rho\sigma}\rmt\Lambda_b^{\!-1/2}){_{;\rho}}N_{\nu\sigma}\right]\\
\fl&&=f^{\sigma\rho}{_{;\sigma}}N_{[\rho\nu]}\rmt\Lambda_b^{\!-1/2}.
\end{eqnarray}

\section{\label{VariationalDerivative}Variational derivatives for fields with
the symmetry $\tGam^\sigma_{\![\mu\sigma]}\!=\!0$}
The field equations associated with a field with symmetry properties must have the
same number of independent components as the field. For a field with the symmetry
$\tGam^\sigma_{\![\mu\sigma]}=0$, the field equations can be found by introducing
a Lagrange multiplier $\Omega^\mu$,
\begin{eqnarray}
0=\delta\int({\mathcal L}+\Omega^\mu\tGam^\sigma_{\![\mu\sigma]})d^n x.
\end{eqnarray}
Minimizing the integral with respect to $\Omega^\mu$ shows that the symmetry is enforced.
Using the definition,
\begin{eqnarray}
\frac{\Delta{\mathcal L}}{\Delta\tGam^\beta_{\tau\rho}}
=\frac{\partial{\mathcal L}}{\partial\tGam^\beta_{\tau\rho}}
-\left(\frac{\partial{\mathcal L}}
{\partial\tGam^\beta_{\tau\rho,\omega}}\right)\!{_{,\,\omega}}~...~,
\end{eqnarray}
and minimizing the integral with respect to $\tGam^\beta_{\!\tau\rho}$ gives
\begin{eqnarray}
\label{minimization}
0=\frac{\Delta{\mathcal L}}{\Delta\tGam^\beta_{\tau\rho}}
+\Omega^\mu\delta^\sigma_\beta\delta^\tau_{[\mu}\delta^\rho_{\sigma]}
=\frac{\Delta{\mathcal L}}{\Delta\tGam^\beta_{\tau\rho}}
+\frac{1}{2}(\Omega^\tau\delta^\rho_\beta-\delta^\tau_\beta\Omega^\rho).
\end{eqnarray}
Contracting this on the left and right gives
\begin{eqnarray}
\label{lagrangemultiplier}
\Omega^\rho=\frac{2}{(n\!-\!1)}\frac{\Delta{\mathcal L}}{\Delta\tGam^\alpha_{\alpha\rho}}
=-\frac{2}{(n\!-\!1)}\frac{\Delta{\mathcal L}}{\Delta\tGam^\alpha_{\rho\alpha}}.
\end{eqnarray}
Substituting (\ref{lagrangemultiplier}) back into (\ref{minimization}) gives
\begin{eqnarray}
\label{usefulresult}
0&=&\frac{\Delta{\mathcal L}}
{\Delta\tGam^\beta_{\tau\rho}}
-\frac{\delta^\tau_\beta}{(n\!-\!1)}
\frac{\Delta{\mathcal L}}{\Delta\tGam^\alpha_{\alpha\rho}}
-\frac{\delta^\rho_\beta}{(n\!-\!1)} \frac{\Delta{\mathcal L}}
{\Delta\tGam^\alpha_{\tau\alpha}}.
\end{eqnarray}
In (\ref{lagrangemultiplier},\ref{usefulresult}) the index contractions occur after
the derivatives. Contracting (\ref{usefulresult}) on the right and left gives the
same result, so it has the same number of independent components as
$\tGam^\alpha_{\mu\nu}$. This is a general expression for the field equations
associated with a field having the symmetry $\tGam^\sigma_{\![\mu\sigma]}=0$.

\section{\label{ApproximateFandg}Solution for $N_{\nu\mu}$ in terms
of $g_{\nu\mu}$ and $f_{\nu\mu}$}
Here we invert the definitions (\ref{gdef},\ref{fdef}) of
$g_{\nu\mu}$ and $f_{\nu\mu}$ to obtain
(\ref{approximateNbar},\ref{approximateNhat}), the approximation of
$N_{\nu\mu}$ in terms of $g_{\nu\mu}$ and $f_{\nu\mu}$.
First let us define the notation
\begin{eqnarray}
\label{hfdef}
\hf^{\nu\mu}\!=\!f^{\nu\mu}\rmt\,\Lambda_b^{\!-1/2}.
\end{eqnarray}
We assume that $|\hf^\nu{_\mu}|\!\ll\!1$ for all components of the
unitless field $\hf^\nu{_\mu}$, and find a solution
in the form of a power series expansion in $\hf^\nu{_\mu}$.
Lowering an index on the equation
$(\rmN/\rmg\,)N^{\dashv\mu\nu}\!=g^{\nu\mu}\!+\hf^{\nu\mu}$
from (\ref{gdef},\ref{fdef}) gives
\begin{eqnarray}
\label{gminusF2}
\frac{\lower2pt\hbox{$\rmN$}}{\rmg}N^{\dashv\mu}{_\alpha}
=\delta^\mu_\alpha-\hf^\mu{_\alpha}.
\end{eqnarray}
Let us consider the tensor $\hf^\mu{_\alpha}\!=\!\hf^{\mu\nu}g_{\nu\alpha}$.
Because $g_{\nu\alpha}$ is symmetric and $\hf^{\mu\nu}$ is antisymmetric, it is
clear that $\hf^\alpha{_\alpha}\!=\!0$. Also because $\hf_{\nu\sigma}\hf^\sigma{_\mu}$ is
symmetric it is clear that $\hf^\nu{_\sigma}\hf^\sigma{_\mu}\hf^\mu{_\nu}=0$.
In matrix language therefore $tr(\hf)\!=\!0,~tr(\hf^3)\!=\!0$,
and in fact $tr(\hf^p)\!=\!0$ for any odd p.
Using the well known formula $det(e^M)=exp\,(tr(M))$ and
the power series $ln(1\!-\!x)=-x-x^2/2-x^3/3-x^4/4\dots$
we then get\cite{Deif},
\begin{eqnarray}
\label{lndetspecial}
ln(det(I\!-\!\hf))&=&tr(ln(I\!-\!\hf))
=-\frac{1}{2}\hf^\rho{_\sigma}\hf^\sigma{_\rho}+(\hf^4)\dots
\end{eqnarray}
Here the notation $(\hf^4)$ refers to terms like
$\hf^\tau{_\alpha}\hf^\alpha{_\sigma}\hf^\sigma{_\rho}\hf^\rho{_\tau}$.
Taking $ln(det())$ on both sides of (\ref{gminusF2}) using the result (\ref{lndetspecial})
and the identities $det(sM^{})\!=s^n det(M^{})$ and $det(M^{-1}_{})\!=1/det(M^{})$
gives
\begin{eqnarray}
\fl~~~~~~~~~ln\!\left(\!\frac{\lower2pt\hbox{$\rmN$}}{\rmg}\right)
\!&=&\!\frac{1}{(n\!-\!2)}\,ln\!\left(\!\frac{\lower2pt\hbox{$N^{(n/2-1)}$}}{g^{(n/2-1)}}\right)
\label{lnapproxdetN}
=-\frac{1}{2(n\!-\!2)}\,\hf^\rho{_\sigma}\hf^\sigma{_\rho}
+(\hf^4)\dots
\end{eqnarray}
Taking $e^x$ on both sides of (\ref{lnapproxdetN}) and using $e^x=1+x+x^2/2\dots$ gives
\begin{eqnarray}
\label{approxdetN}
\frac{\lower2pt\hbox{$\rmN$}}{\rmg}
=1\!-\!\frac{1}{2(n\!-\!2)}\,\hf^{\rho\sigma}\!\hf_{\sigma\rho}
+(\hf^4)\dots
\end{eqnarray}
Using the power series $(1\!-\!x)^{-1}\!=\!1+x+x^2+x^3\dots$,
or multiplying (\ref{gminusF2}) term by term,
we can calculate the inverse of (\ref{gminusF2}) to get\cite{Deif}
\begin{eqnarray}
\frac{\rmg}{\lower2pt\hbox{$\rmN$}} N^\nu{_\mu}
=\delta^\nu_\mu+\hf^\nu{_\mu}+\hf^\nu{_\sigma}\hf^\sigma{_\mu}
+\hf^\nu{_\rho}\hf^\rho{_\sigma}\hf^\sigma{_\mu}+(\hf^4)\dots\\
\label{approxN}
N_{\nu\mu}
=\frac{\lower2pt\hbox{$\rmN$}}{\rmg}(g_{\nu\mu}+\hf_{\nu\mu}+\hf_{\nu\sigma}\hf^\sigma{_\mu}
+\hf_{\nu\rho}\hf^\rho{_\sigma}\hf^\sigma{_\mu}+(\hf^4)\dots).
\end{eqnarray}
Here the notation $(\hf^4)$ refers to terms like
$\hf_{\nu\alpha}\hf^\alpha{_\sigma}\hf^\sigma{_\rho}\hf^\rho{_\mu}$.
Since $\hf_{\nu\sigma}\hf^\sigma{_\mu}$ is symmetric and
$\hf_{\nu\rho}\hf^\rho{_\sigma}\hf^\sigma{_\mu}$ is antisymmetric, we obtain from
(\ref{approxN},\ref{approxdetN},\ref{hfdef}) the final result
(\ref{approximateNbar},\ref{approximateNhat}).
\iffalse
\begin{eqnarray}
N_{(\nu\mu)}&=&g_{\nu\mu}+\hf_{\nu\sigma}\h\hf^\sigma{_\mu}
-\frac{\lower2pt\hbox{$1$}}{2(n\!-\!2)}g_{\nu\mu}\h\hf^{\rho\sigma}\!\hf_{\sigma\rho}+(\hf^4),\\
\label{approxNhat}
N_{[\nu\mu]}&=&\hf_{\nu\mu}+(\hf^3).
\end{eqnarray}
\fi

\ifnum\ExpandDerivations=1
\section{\label{AffineDerivation}Derivation of the Einstein-Schr\"{o}dinger theory
from a purely affine Lagrangian density}
If the theory proposed in this paper is correct, we might expect that it can be derived
from some kind of simple principles. Here we will show that the original unmodified
Einstein-Schr\"{o}dinger theory can be derived from a Lagrangian density ${\mathcal L}(\nGam)$
which depends only on an affinity $\nGam^\alpha_{\sigma\mu}$ with no symmetry properties,
resulting in the field equations
\begin{eqnarray}
\label{action}
~~~~~~0&=&\delta S,~~~~~S=\!\int{\mathcal L}(\nGam)dx^1dx^2...dx^n,\\
\label{varder}
\Rightarrow~~
0&=&\frac{\delta {\mathcal L}}{\delta \nGam^\beta_{\tau\rho}}
=\frac{\partial{\mathcal L}}{\partial\nGam^\beta_{\tau\rho}}
-\left(\!\frac{\partial{\mathcal L}}
{\partial\nGam^\beta_{\tau\rho,\omega}}\!\right){_{\!,\,\omega}}
+\left(\!\frac{\partial{\mathcal L}}
{\partial\nGam^\beta_{\tau\rho,\omega,\nu}}\!\right)
{_{\!,\,\omega,\,\nu}}~...,
\end{eqnarray}
where the field equations require
%${\mathcal L}(\nGam)$ which transplants into itself.
\begin{eqnarray}
%&&{\mathcal L}+\nGam{^\alpha_{\alpha\beta}}{\mathcal L}dx^\beta
%={\mathcal L}(x+dx),\\
%~\Rightarrow
\label{der}
{\mathcal L}_{,\beta}-\nGam{^\alpha_{(\alpha\beta)}}{\mathcal L}=0.
\end{eqnarray}
%\pagebreak
Equation (\ref{der}) is a simple generalization of the result
${\mathcal L}_{,\beta}-\Gamma{^\alpha_{\alpha\beta}}{\mathcal L}=0$
that occurs with ordinary vacuum general relativity.
We will also show that the Einstein-Schr\"{o}dinger
theory appears to be unique in that it can be derived from a Lagrangian
density which satisfies (\ref{der}).
% and we will suggest an idea as to why such a Lagrangian density might be expected.

Suppose we view (\ref{varder},\ref{der}) as requirements for a purely classical field theory.
The task is then to solve (\ref{varder},\ref{der})
for the unknowns ${\mathcal L}(\nGam)$ and $\nGam^\alpha_{\mu\nu}$.
For an arbitrary ${\mathcal L}(\nGam)$, these equations
constitute more equations than unknowns, and no nontrivial
solution for $\nGam^\alpha_{\mu\nu}$ can be expected.
However, for the correct ${\mathcal L}(\nGam)$, (\ref{der}) is contained in
(\ref{varder}) and nontrivial solutions can be expected.
A Lagrangian density which allows a solution to (\ref{varder},\ref{der}) is
the following,
\begin{eqnarray}
\label{affinelag}
~~~~~~~{\mathcal L}(\nGam)=\frac{\Lambda_b}{16\pi}\sqrt{-det(N_{\nu\mu})},
\end{eqnarray}
where $N_{\nu\mu}$ is simply defined to be
\begin{eqnarray}
\label{N}
N_{\nu\mu}=-\hhR_{\nu\mu}/\Lambda_b=-(\tR_{\nu\mu}+2A_{[\nu,\mu]}\rmt\Lambda_b^{\!1/2})/\Lambda_b,
\end{eqnarray}
and $\hhR_{\sigma\mu}\!=\!\hR_{\sigma\mu}(\nGam)$ is
the Hermitianized Ricci tensor from (\ref{HermitianizedRicci}).
Here we have decomposed ${\nGam}{^\alpha_{\nu\mu}}$
into ${\tGam}{^\alpha_{\nu\mu}}$ and $A_\sigma$ as in (\ref{A},\ref{gamma_natural},\ref{gamma_tilde}),
and we have also used (\ref{RnGam}) and
$\tR_{\nu\mu}\!=\!\hR_{\nu\mu}(\tGam)$ from (\ref{HermitianizedRiccit}).
The connection $\tGam{^\alpha_{\nu\mu}}$ has the symmetry (\ref{JScontractionsymmetric})
so it has only $n^3\!-\!n$ independent components.
From (\ref{gamma_natural},\ref{JScontractionsymmetric}), $\tGam^\alpha_{\nu\mu}$ and $A_\nu$ fully
parameterize $\nGam^\alpha_{\nu\mu}$ and can be treated as independent variables.
From the invariance properties (\ref{transpositionsymmetric},\ref{gaugesymmetric}) of the
Hermitianized Ricci tensor (\ref{HermitianizedRicci}),
the Lagrangian density (\ref{affinelag}) is real, and it is also invariant under both charge conjugation
(\ref{transposition}) and under an electromagnetic gauge transformation (\ref{gauge}).

Now, it is simple to show that setting
$\delta{\mathcal L}/\delta A_\nu\!=0$ and $\delta{\mathcal L}/\delta\tGam^\alpha_{\nu\mu}\!=0$
gives identical equations as in \S\ref{Derivation}
except that $\Lambda_z\!=\!0$.
In addition, the definition (\ref{N}) matches (\ref{para}), so that this equation
and all of the subsequent equations in \S\ref{SymmetricPart} and \S\ref{AntisymmetricPart}
are identical except that $\Lambda_z\!=\!0$.
Therefore, the purely affine Lagrangian density (\ref{affinelag},\ref{N}) gives
the same theory as the Palatini Lagrangian density (\ref{JSlag1}) with
$\Lambda_z\!=\!0$, which is the original Einstein-Schr\"{o}dinger theory.
In particular (\ref{der0}) is valid, and this together with
(\ref{JScontractionsymmetric},\ref{affinelag}) gives (\ref{der}).

The derivation of the Einstein-Schr\"{o}dinger theory in this manner is
remarkable because the only fundamental field assumed \textit{a priori} was the
connection ${\nGam}{^\alpha_{\sigma\mu}}$.  The fundamental tensor
$N_{\sigma\mu}$, the metric $g_{\sigma\mu}$, the field $f_{\sigma\mu}$,
the electromagnetic potential $\Aphi_\sigma$, and the contraction-symmetric
connection $\tGam^\alpha_{\sigma\mu}$ all just appeared as convenient variables
to work with when solving the field equations.
We should emphasize that the same field equations also result from setting
$\delta{\mathcal L}/\delta{\nGam}{^\alpha_{\sigma\mu}}\!=\!0$.
When this is done, one obtains a rather complicated set of field equations in the unknowns
${\nGam}{^\alpha_{\sigma\mu}}$. However, when the equations are rewritten in
terms of the variable $\tGam{^\alpha_{\sigma\mu}}$ from (\ref{gamma_tilde}),
much simplification occurs and one eventually obtains the ordinary
Einstein-Schr\"{o}dinger field equations. The same thing occurs if one sets
$\delta{\mathcal L}_S/\delta{\nGam}{^\alpha_{\sigma\mu}}\!=\!0$
using Schr\"{o}dinger's\cite{SchrodingerI} Lagrangian density
${\mathcal L}_S\!=\!\sqrt{-det(- R_{\sigma\mu}(\nGam))}$,
but then the equations simplify when rewritten in terms of the variable
$\sGam{^\alpha_{\sigma\mu}}=\nGam^\alpha_{\sigma\mu}
-2\delta^\alpha_\sigma{\nGam}{^\nu_{[\nu\mu]}}/(n\!-\!1)$
instead of $\tGam{^\alpha_{\sigma\mu}}$.

It is important to note that this simple derivation only works for
Schr\"{o}dinger's generalization of Einstein's theory which includes
an intrinsic cosmological constant, because if
$\Lambda_b=0$, the definition (\ref{N}) would not make sense.
Also note that the only reason we do not set $\Lambda_b\!=\!1$ is because
we are assuming the convention that $N_{\sigma\mu}$ has values close to 1.
If we chose to we would be free to absorb $\Lambda_b$ into $N_{\sigma\mu}$
because both $\tGam^\alpha_{\sigma\mu}(N_{..})$ and
$R_{\sigma\mu}(\tGam(N_{..}))$ are independent of a constant multiplier
on $N_{\sigma\mu}$. We would also be free to absorb $\Lambda_b$ into the
definition of $A_\sigma$. Therefore, $\Lambda_b$ does not need to be
in either the field equations or the Lagrangian density. It is only there to
make the definitions of $N_{\sigma\mu}$ and $A_\sigma$ conform to conventions.
%In this case, when solving the field equations for the Schwarzschild solution,
%an arbitrary cosmological constant would appear in the solution in the same
%way that the mass does, as a constant of integration, and it would show up as
%an overall magnitude of $N_{\sigma\mu}$.
The cosmological constant term has often been referred to as an undesirable
complication, attached to otherwise elegant field equations to make them
conform to reality. From the standpoint of the derivation above, it is nothing
of the sort. Instead, $\Lambda_b$ appears as the magnitude of the fundamental
tensor $N_{\sigma\mu}$ when $N_{\sigma\mu}$ is put in more natural units.
The cosmological constant term is not an added-on appendage to this theory
but is instead an inherent part of it.

Given that the Einstein-Schr\"{o}dinger theory can be derived from
a Lagrangian density which obeys the equation
${\mathcal L}_{,\beta}-\nGam{^\alpha_{\!(\alpha\beta)}}{\mathcal L}=0$
as in (\ref{der}), we must next ask whether it is unique in this regard.
While a rigorous proof is probably not possible,
a strong argument will be presented below that the theory is unique in this
property. With no metric to use, the forms that a scalar density can take are
limited. Also, because (\ref{der}) exists for any dimension, we must only
consider forms which exist for any dimension. To discuss this topic, it is
convenient to use the fields $\tGam{^\alpha_{\sigma\mu}}, A_\sigma$
as defined by (\ref{gamma_tilde},\ref{A}) instead of
${\nGam}{^\alpha_{\sigma\mu}}$.
The simplest form to consider is
${\mathcal L}\nobreak=\nobreak\rmN$, where $N_{\sigma\mu}$
is a linear combination of the terms
$\tR_{\sigma\mu}$,
$\tR_{\mu\sigma}$,
$\tGam{^\alpha_{\alpha[\mu,\sigma]}}$,
$A_{[\sigma,\mu]}$,
$A_{\sigma,\mu}\!\nobreak-\nobreak\!\tGam^\alpha_{\sigma\mu}A_\alpha$,
$\tGam^\alpha_{[\sigma\mu]}A_\alpha$,
and $A_\sigma A_\mu$.
Many other terms can be decomposed into these, such as
$R_{\sigma\mu}(\tGam^T)\!\nobreak=\nobreak\!\tR_{\mu\sigma}
\!\nobreak+\nobreak\!2\tGam{^\alpha_{\alpha[\mu,\sigma]}}$,
$\tR^\alpha{_{\alpha\sigma\mu}}
\!\nobreak=\nobreak\!2\tGam{^\alpha_{\alpha[\mu,\sigma]}}$,
and anything dependent on ${\nGam}{^\alpha_{\sigma\mu}}$. Our
Lagrangian density (\ref{affinelag}) is a special case of this form. In
fact, it happens that (\ref{der}) is satisfied for any ${\mathcal L}=\rmN$
where $N_{\sigma\mu}\!=\!a\tR_{\sigma\mu}
\!+\!bA_{[\sigma,\mu]}
\!+\!c\tGam{^\alpha_{\alpha[\mu,\sigma]}}$
and $a\!\ne\!0,b\!\ne\!0$.
This would initially seem to indicate that the Einstein-Schr\"{o}dinger theory
is not unique, except for the surprising fact that the same field equations
result regardless of the coefficients in the linear combination.
The $\tGam{^\alpha_{\alpha[\mu,\sigma]}}$ term causes
$\delta^\rho_\beta(\rmN\,N^{\dashv[\tau\omega]})_{,\,\omega}$
terms in the $\delta{\mathcal L}/\delta\tGam{^\beta_{\tau\rho}}\!=\!0$
field equations (\ref{JSconnection}), but these are required to vanish by
the $\delta{\mathcal L}/\delta A_\tau\!=\!0$ field equations (\ref{Ampere}).
Also, (\ref{der}) requires that
$\tGam{^\alpha_{\alpha[\mu,\sigma]}}
\!=\!(ln{\mathcal L})_{,[\mu,\sigma]}\!=\!0$ from (\ref{funnytensor}),
so this term is of no consequence.
Different field equations result if any other terms are included
in $N_{\sigma\mu}$, but then (\ref{der}) is no longer satisfied.
To argue the case for uniqueness, we must next consider more complicated forms.
The most obvious generalization of a single $\rmN$ consists of linear
combinations of such terms, $\sqrt{-^1\!N}$ and $\sqrt{-^2\!N}$ etcetera.
The resulting field equations contain different $N^{\dashv\sigma\mu}$
terms, and there is just no way to contract the equations to remove these terms
as we did in (\ref{der0}). Linear combinations of terms such as
$\sqrt{-^1\!N}\sqrt{-^2\!N}/\sqrt{-^3\!N}$
have the same characteristic. Next one can include linear combinations of
terms like $\sqrt{-^1\!N}\,{^1}\!N^{\dashv\sigma\mu}\,{^2}\!N_{\mu\sigma}$.
In this case the field equations contain terms with different powers of
${^1}\!N^{\dashv\sigma\mu}$. From trying a few of these,
it seems very likely that the simplicity of (\ref{der}) demands
simplicity in the Lagrangian density, and that the only real prospect is
a single $\rmN$ as we considered originally.

At first glance, these results might seem
unimportant because the original Einstein-Schr\"{o}dinger theory does not seem to
represent anything physical. However, the theory proposed in this paper is just
the Einstein-Schr\"{o}dinger theory with a quantization effect, namely a $\Lambda_z$ term
caused by zero-point fluctuations. A spin-1/2 ${\mathcal L}_m$ term can be viewed
as another quantization effect, namely the first quantization of our charged monopole
solution. And as shown in \cite{sShifflett}, when these two terms are included
in the Lagrangian we get an extremely close approximation to one-particle
quantum electrodynamics. From this perspective, the fact that the original
Einstein-Schr\"{o}dinger theory can be derived from simple principles may be
important because this theory is the purely classical core of a theory which
represents a large part of reality. Furthermore, if one was to try to second quantize the
theory, the most obvious approach would be to use path integral methods with the action
(\ref{action},\ref{affinelag},\ref{N}). Since we are proposing that spin-1/2 particles
have their origin as singular solutions of the field equations, both the $\Lambda_z$
and the spin-1/2 part of our theory might be expected to appear as quantization
effects using the purely classical action (\ref{action},\ref{affinelag},\ref{N}),
and adding up the $e^{iS/\hbar}$ amplitudes for all ``paths'' of the field
$\nGam{^\alpha_{\mu\nu}}$. Now it is unclear whether such a quantization scheme
would work, or how practical it would be in terms of being able to do the calculations
and predict experimental results. However, it is at least theoretically possible.
%\pagebreak

%Finally let us consider why a Lagrangian density ${\mathcal L}(\nGam)$
%might be expected to obey the equation
%${\mathcal L}_{,\beta}-\nGam{^\alpha_{\!(\alpha\beta)}}{\mathcal L}=0$
%as in (\ref{der}). One may speculate that it is related to the
%universe we perceive being an extremum state of the real universe,
%which is fluctuating between every possible state of the field
%${\nGam^\alpha_{\beta\nu}}$. If a Lagrangian density satisfies (\ref{der}),
%that means its covariant derivative vanishes for the affinity
%${\nGam^\alpha_{\beta\nu}}$, which requires ${\mathcal L}$ to
%be in an extremum state in coordinate space. The field equations are
%derived by setting the variational derivative to zero, which requires
%${\nGam^\alpha_{\beta\nu}}$ to be in an extremum state in
%state-space. In the path-integral approach to quantum field theory,
%states which are close to an extremum add coherently, whereas other
%states cancel each other out. It may be that a similar mechanism
%causes us to perceive the extremum in both coordinate-space and
%state-space as ``the'' state of the universe. This is admittedly
%rather vague, but it is the best answer the author can offer.
\fi

\ifnum\ExpandDerivations=1
\section{\label{EofMcomparison}Verification that the EIH method applied to Einstein-Maxwell
theory gives the equations of motion of the Darwin Lagrangian to Post-Coulombian order}
Here we will compare the post-Coulombian equations of motion for Einstein-Maxwell theory
obtained by two authors\cite{Wallace,Gorbatenko} using the EIH method,
to the equations of motion obtained from the Darwin Lagrangian\cite{Jackson}.
%The Darwin Lagrangian can be found in many electrodynamics textbooks\cite{Jackson,Landau}.
For two particles the Darwin Lagrangian takes the form
\begin{eqnarray}
\fl L_a=\frac{m_av_a^2}{2}
+\frac{1}{8}\frac{m_av_a^4}{c^2}
-e_a\frac{e_b}{R_{ab}}
+\frac{e_a}{2c^2}\frac{e_b}{R_{ab}}\left[\mathbf{v}_a\cdot\mathbf{v}_b
+(\mathbf{v}_a\cdot\mathbf{n}_{ab})(\mathbf{v}_b\cdot\mathbf{n}_{ab})\right].
\end{eqnarray}
Here we are using the notation
\begin{eqnarray}
\fl \dot r_a^i\!=\!v_a^i,~~ \dot r_b^i\!=\!v_b^i,~~ r_{ab}^i\!=\!r_a^i\!-r_b^i,~~
v_{ab}^i\!=\!v_a^i\!-v_b^i,~~n_{ab}^i\!=\!r_{ab}^i/R_{ab},~~ R_{ab}^2\!=\!r_{ab}^ir_{ab}^i.
\end{eqnarray}
From this we get the equations of motion
\begin{eqnarray}
\fl 0&=&\frac{\partial L_a}{\partial r_a^i}-\frac{\partial}{\partial t}\left(\frac{\partial L_a}{\partial v_a^i}\right)\\
\fl &=&e_be_b\frac{r_{ab}^i}{R_{ab}^3}
+\frac{e_ae_b}{2c^2}\left(-\frac{r_{ab}^i}{R_{ab}^3}v_a^sv_b^s
-\frac{3r_{ab}^i}{R_{ab}^5}v_a^sr_{ab}^sv_b^ur_{ab}^u
+\frac{v_a^i}{R_{ab}^3}v_b^sr_{ab}^s
+\frac{v_b^i}{R_{ab}^3}v_a^sr_{ab}^s\right)\nonumber\\
\fl &&-m_a\dot v_a^i-\frac{m_a}{2c^2}(\dot v_a^iv_a^2+2v_a^iv_a^s\dot v_a^s)
-\frac{e_ae_b}{2c^2R_{ab}}\left(\dot v_b^i-v_b^i\frac{v_{ab}^ir_{ab}^s}{R_{ab}^2}\right)\nonumber\\
\fl &&-\frac{e_ae_b}{2c^2R_{ab}^3}\left(v_{ab}^iv_b^sr_{ab}^s
+r_{ab}^i\dot v_b^sr_{ab}^s
+r_{ab}^iv_b^sv_{ab}^s
-3r_{ab}^iv_b^ur_{ab}^u\frac{v_{ab}^sr_{ab}^s}{R_{ab}^2}\right)\\
\label{LL}
\fl &=&-m\dot v_a^i+e_ae_b\frac{r_{ab}^i}{R_{ab}^3}
+\frac{e_ae_b}{c^2}\left[-\frac{v_a^2}{2}-v_a^sv_b^s+\frac{v_b^2}{2}\right]\frac{r_{ab}^i}{R_{ab}^3}\nonumber\\
\fl&&+\frac{e_ae_b}{c^2}\left[-v_a^sv_a^i+v_a^sv_b^i\right]\frac{r_{ab}^s}{R_{ab}^3}
-\frac{3e_ae_b}{2c^2}v_b^uv_b^s\frac{r_{ab}^ur_{ab}^sr_{ab}^i}{R_{ab}^5}
+\frac{e_a^2e_b^2}{m_bc^2}\frac{r_{ab}^i}{R_{ab}^4}
\end{eqnarray}
Let us first compare the notation used in the various references,
\begin{eqnarray}
\fl~~~~\matrix{
Landau/Lifshitz &r_a^i&r_b^i&r_{ab}^i&R_{ab}&e_a&e_b&m_a&m_b\cr
Wallace &\eta^i&\zeta^i&\beta_i&r&e_1&e_2&m_1&m_2\cr
Gorbatenko &\xi^i&\eta^i&-R_i&R&Q&q&M&m\cr
%Bazanski &\xi^i&\eta^i&-R_i&r&e_1&e_2&m_1&m_2\cr
%Anderson &x_A^i&x_B^i&x_{AB}^i&x_{AB}&q_A&q_B&m_A&m_B\cr
Jackson &r_1^i&r_2^i&r_{12}^i&R&q_1&q_2&m_1&m_2
}
\end{eqnarray}

The Wallace\cite{Wallace} equations of motion (including radiation reaction term) are
\begin{eqnarray}
\fl m_1\ddot\eta^m\!+e_1e_2\frac{\partial}{\partial\eta^m}\!\left(\frac{1}{r}\right)
&=&e_1e_2\left[\left(\frac{1}{2}\dot\eta^s\dot\eta^s
+\dot\eta^s\dot\zeta^s\right)\frac{\partial}{\partial\eta^m}\!\left(\frac{1}{r}\right)\right.\nonumber\\
\fl &&~~~~~~~+\left.(\dot\eta^s\dot\eta^m
-\dot\eta^s\dot\zeta^m+\dot\zeta^s\dot\zeta^m)\frac{\partial}{\partial\eta^s}\!\left(\frac{1}{r}\right)
-\frac{1}{2}\frac{\partial^3r}{\partial\eta^m\eta^r\eta^s}\dot\zeta^r\dot\zeta^s\right]\nonumber\\
\fl &&~~~~~~~-\frac{e_1^2e_2^2}{m_2}\frac{1}{r}\frac{\partial}{\partial\eta^m}\!\left(\frac{1}{r}\right)
+\frac{2}{3}e_1(e_1\dot{\phantom{\eta}}\!\!\ddot\eta^m
+e_2\dot{\phantom{\zeta}}\!\!\ddot\zeta^m)
\end{eqnarray}
Using
\begin{eqnarray}
\fl~~\frac{\partial}{\partial\eta^m}\!\left(\frac{1}{r}\right)=-\frac{\beta_m}{r^3},~~~
\frac{\partial r}{\partial\eta^s}=\frac{1}{r}\beta_s,~~~
\frac{\partial^2r}{\partial\eta^r\eta^s}=-\frac{\beta_r\beta_s}{r^3}+\frac{1}{r}\delta_{sr}\\
\fl~~~\frac{\partial^3r}{\partial\eta^m\eta^r\eta^s}=-\delta_{rm}\frac{\beta_s}{r^3}
-\delta_{sm}\frac{\beta_r}{r^3}+\frac{3\beta_r\beta_s\beta_m}{r^5}-\frac{\beta_m}{r^3}\delta_{sr}\\
\fl-\frac{1}{2}\frac{\partial^3r}{\partial\eta^m\eta^r\eta^s}\dot\zeta^r\dot\zeta^s
\label{Kterm}
=\frac{\dot\zeta^m\dot\zeta^s\beta_s}{r^3}
-\frac{3\dot\zeta^r\dot\zeta^s\beta_r\beta_s\beta_m}{2r^5}
+\frac{\beta_m\dot\zeta^s\dot\zeta^s}{2r^3}
\end{eqnarray}
we get
\begin{eqnarray}
\fl m_1\ddot\eta^m\!-e_1e_2\frac{\beta_m}{r^3}
&=&e_1e_2\left[-\left(\frac{1}{2}\dot\eta^s\dot\eta^s
+\dot\eta^s\dot\zeta^s\right)\frac{\beta_m}{r^3}\right.\nonumber\\
\fl &&~~~~~~~-\left.(\dot\eta^s\dot\eta^m
-\dot\eta^s\dot\zeta^m+\dot\zeta^s\dot\zeta^m)\frac{\beta_s}{r^3}
-\frac{1}{2}\frac{\partial^3r}{\partial\eta^m\eta^r\eta^s}\dot\zeta^r\dot\zeta^s\right]\nonumber\\
\fl &&~~~~~~~+\frac{e_1^2e_2^2}{m_2}\frac{1}{r}\frac{\beta_m}{r^3}
+\frac{2}{3}e_1(e_1\dot{\phantom{\eta}}\!\!\ddot\eta^m
+e_2\dot{\phantom{\zeta}}\!\!\ddot\zeta^m)\\
&=&e_1e_2\left[-\left(\frac{1}{2}\dot\eta^s\dot\eta^s
+\dot\eta^s\dot\zeta^s-\frac{\dot\zeta^s\dot\zeta^s}{2}\right)\frac{\beta_m}{r^3}\right.\nonumber\\
\fl &&~~~~~~~-\left.(\dot\eta^s\dot\eta^m
-\dot\eta^s\dot\zeta^m)\frac{\beta_s}{r^3}
-\frac{3\dot\zeta^r\dot\zeta^s\beta_r\beta_s\beta_m}{2r^5}
\right]\nonumber\\
\fl &&~~~~~~~+\frac{e_1^2e_2^2}{m_2}\frac{\beta_m}{r^4}
+\frac{2}{3}e_1(e_1\dot{\phantom{\eta}}\!\!\ddot\eta^m
+e_2\dot{\phantom{\zeta}}\!\!\ddot\zeta^m)
\end{eqnarray}
Translating this into the Landau/Lifshitz notation we see that it agrees with (\ref{LL}),
\begin{eqnarray}
\fl m_a\dot v^m\!-e_ae_b\frac{r_{ab}^m}{R_{ab}^3}
&=&e_ae_b\left[-\left(\frac{v_a^2}{2}
+v_a^sv_b^s-\frac{v_b^2}{2}\right)\frac{r_{ab}^m}{R_{ab}^3}\right.\nonumber\\
\fl &&~~~~~~~-\left.(v_a^sv_a^m
-v_a^sv_b^m)\frac{r_{ab}^s}{R_{ab}^3}
-\frac{3v_b^rv_b^sr_{ab}^rr_{ab}^sr_{ab}^m}{2R_{ab}^5}
\right]\nonumber\\
\fl &&~~~~~~~+\frac{e_a^2e_b^2}{m_b}\frac{r_{ab}^m}{R_{ab}^4}
+\frac{2}{3}e_a(e_a\ddot v_a^m
+e_b\ddot v_b^m).
\end{eqnarray}

The Gorbatenko\cite{Gorbatenko} equations of motion (including radiation reaction term) are
\begin{eqnarray}
\label{Gorbatenko}
\fl M\ddot\xi_k&=&-\frac{qQ}{R^3}R_k+qQ\left[\frac{(\dot\xi_l\dot\eta_l)}{R^3}R_k
-\frac{(R_l\dot\xi_l)}{R^3}\dot\eta_k
+\frac{(R_l\dot\xi_l)}{R^3}\dot\xi_k
+\frac{(\dot\xi_l\dot\xi_l)}{2R^3}R_k\right.\nonumber\\
\fl &&\left.-\frac{\ddot\eta_k}{2R}
-\frac{(R_l\ddot\eta_l)}{2R^3}R_k
-\frac{3}{2}\frac{(R_l\dot\eta_l)^2}{R^5}R_k
-\frac{(\dot\eta_l\dot\eta_l)}{2R^3}R_k\right]
+\frac{2}{3}(Q\dot{\phantom{\xi}}\!\!\ddot\xi_k+q\dot{\phantom{\eta}}\!\!\ddot\eta_k)Q
\end{eqnarray}
The Coulombian order equations for the $\eta^k$ particle are the first two terms but with
$\xi^k\rightarrow\eta^k$,$M\rightarrow m$, $Q\rightarrow q$, $q\rightarrow Q$, $R_k\rightarrow -R_k$.
Using these equations we have
\begin{eqnarray}
\fl m\ddot\eta_k\approx\frac{qQ}{R^3}R_k
~~~\Rightarrow mR_l\ddot\eta_l\approx\frac{qQ}{R}
~~~\Rightarrow -\frac{(R_l\ddot\eta_l)}{2R^3}R_k\approx-\frac{qQ}{2m}\frac{R_k}{R^4}\\
\fl\Rightarrow -\frac{\ddot\eta_k}{2R}
-\frac{(R_l\ddot\eta_l)}{2R^3}R_k\approx-\frac{qQR_k}{mR^4}.
\end{eqnarray}
Substituting this last equation into (\ref{Gorbatenko}) and assuming $1/(mR)$ is ${\mathcal O}(\lambda^1)$ gives
\begin{eqnarray}
\fl M\ddot\xi_k&=&-\frac{qQ}{R^3}R_k+qQ\left[\frac{(\dot\xi_l\dot\eta_l)}{R^3}R_k
-\frac{(R_l\dot\xi_l)}{R^3}\dot\eta_k
+\frac{(R_l\dot\xi_l)}{R^3}\dot\xi_k
+\frac{(\dot\xi_l\dot\xi_l)}{2R^3}R_k\right.\nonumber\\
\fl &&~~~~~~~~~~~~~\left.-\frac{qQR_k}{mR^4}
-\frac{3}{2}\frac{(R_l\dot\eta_l)^2}{R^5}R_k
-\frac{(\dot\eta_l\dot\eta_l)}{2R^3}R_k\right]
+\frac{2}{3}(Q\dot{\phantom{\xi}}\!\!\ddot\xi_k+q\dot{\phantom{\eta}}\!\!\ddot\eta_k)Q
\end{eqnarray}
Translating this into the Landau/Lifshifz notation we see that it agrees with (\ref{LL}),
\begin{eqnarray}
\fl m_a\dot v_a^k&=&\frac{e_b e_a}{R_{ab}^3}r_{ab}^k+e_b e_a\left[-\frac{v_a^lv_b^l}{R_{ab}^3}r_{ab}^k
+\frac{r_{ab}^lv_a^l}{R_{ab}^3}v_b^k
-\frac{r_{ab}^lv_a^l}{R_{ab}^3}v_a^k
-\frac{v_a^2}{2R_{ab}^3}r_{ab}^k\right.\nonumber\\
\fl &&~~~~~~~~~~~~~\left.+\frac{e_b e_ar_{ab}^k}{m_bR_{ab}^4}
-\frac{3}{2}\frac{(r_{ab}^lv_b^l)^2}{R_{ab}^5}r_{ab}^k
+\frac{v_b^2}{2R_{ab}^3}r_{ab}^k\right]
+\frac{2}{3}(e_a\ddot v_a^k+e_b\ddot v_b^k)e_a.
\end{eqnarray}
\fi

%\newpage
%\bibliography{shifflettcqg}% Produces the bibliography via BibTeX.
%\section*{References}
%\begin{thebibliography}{68}
\Bibliography{68}
\expandafter\ifx\csname natexlab\endcsname\relax\def\natexlab#1{#1}\fi
\expandafter\ifx\csname bibnamefont\endcsname\relax
  \def\bibnamefont#1{#1}\fi
\expandafter\ifx\csname bibfnamefont\endcsname\relax
  \def\bibfnamefont#1{#1}\fi
\expandafter\ifx\csname citenamefont\endcsname\relax
  \def\citenamefont#1{#1}\fi
\expandafter\ifx\csname url\endcsname\relax
  \def\url#1{\texttt{#1}}\fi
\expandafter\ifx\csname urlprefix\endcsname\relax\def\urlprefix{URL }\fi
\providecommand{\bibinfo}[2]{#2}
\providecommand{\eprint}[2][]{#2}
\providecommand{\bitem}[2][]{\bibitem {#2}}

\bitem[{\citenamefont{Einstein and Straus}(1946)}]{EinsteinStraus}
\bibinfo{author}{\bibfnamefont{A.}~\bibnamefont{Einstein}} \bibnamefont{and}
  \bibinfo{author}{\bibfnamefont{E.~G.} \bibnamefont{Straus}},
  \bibinfo{journal}{Ann.\ Math.} \textbf{\bibinfo{volume}{47}}
  (\bibinfo{year}{1946}) \bibinfo{pages}{731}.

\bitem[{\citenamefont{Einstein}(1948)}]{Einstein3}
\bibinfo{author}{\bibfnamefont{A.}~\bibnamefont{Einstein}},
  \bibinfo{journal}{Rev.\ Mod.\ Phys.} \textbf{\bibinfo{volume}{20}}
  (\bibinfo{year}{1948}) \bibinfo{pages}{35}.

\bitem[{\citenamefont{Einstein}(1949)}]{EinsteinBianchi}
\bibinfo{author}{\bibfnamefont{A.}~\bibnamefont{Einstein}},
  \bibinfo{journal}{Can.\ J.\ Math.} \textbf{\bibinfo{volume}{2}}
  (\bibinfo{year}{1949}) \bibinfo{pages}{120}.

\bitem[{\citenamefont{Einstein and Kaufman}(1955)}]{EinsteinKaufman}
\bibinfo{author}{\bibfnamefont{A.}~\bibnamefont{Einstein}} \bibnamefont{and}
  \bibinfo{author}{\bibfnamefont{B.}~\bibnamefont{Kaufman}},
  \bibinfo{journal}{Ann.\ Math.} \textbf{\bibinfo{volume}{62}}
  (\bibinfo{year}{1955}) \bibinfo{pages}{128}.

\bitem[{\citenamefont{Einstein}(1956)}]{EinsteinMOR}
\bibinfo{author}{\bibfnamefont{A.}~\bibnamefont{Einstein}},
  \emph{\bibinfo{title}{The Meaning of Relativity, 5th ed. revised}}
  (\bibinfo{publisher}{Princeton U. Press}, \bibinfo{address}{Princeton
  NJ}, \bibinfo{year}{1956}).

\bitem[{\citenamefont{Schr\"{o}dinger}(1947)}]{SchrodingerI}
\bibinfo{author}{\bibfnamefont{E.}~\bibnamefont{Schr\"{o}dinger}},
  \bibinfo{journal}{Proc.\ Royal Irish Acad.} \textbf{\bibinfo{volume}{51A}}
  (\bibinfo{year}{1947}) \bibinfo{pages}{163}.

%\bitem[{\citenamefont{Schr\"{o}dinger}(1948)}]{SchrodingerII}
%\bibinfo{author}{\bibfnamefont{E.}~\bibnamefont{Schr\"{o}dinger}},
%  \bibinfo{journal}{Proc.\ Royal Irish Acad.} \textbf{\bibinfo{volume}{51A}}
%  (\bibinfo{year}{1948}) \bibinfo{pages}{205}.

\bitem[{\citenamefont{Schr\"{o}dinger}(1948)}]{SchrodingerIII}
\bibinfo{author}{\bibfnamefont{E.}~\bibnamefont{Schr\"{o}dinger}},
  \bibinfo{journal}{Proc.\ Royal Irish Acad.} \textbf{\bibinfo{volume}{52A}}
   (\bibinfo{year}{1948}) \bibinfo{pages}{1}.

\bitem[{\citenamefont{Schr\"{o}dinger}(1950)}]{SchrodingerSTS}
\bibinfo{author}{\bibfnamefont{E.}~\bibnamefont{Schr\"{o}dinger}},
  \emph{\bibinfo{title}{Space-Time Structure}}
  (\bibinfo{publisher}{Cambridge Press}, \bibinfo{address}{London},
  \bibinfo{year}{1950}) p~93,108,112.

%\bitem[{\citenamefont{Shifflett}(2003)}]{cShifflett}
%\bibinfo{author}{\bibfnamefont{J.~A.} \bibnamefont{Shifflett}},
%  \bibnamefont{(unpublished)} \eprint{arXiv:gr-qc/0310124}.

\bitem[{\citenamefont{Shifflett}(2003)}]{LRESMAPLE}
\bibinfo{author}{\bibfnamefont{J.~A.} \bibnamefont{Shifflett}},
\eprint{www.artsci.wustl.edu/$\sim$jashiffl/einstein-schrodinger.html/MAPLEtxt/}.

\bitem[{\citenamefont{Sahni and Starobinsky}(2000)}]{Sahni}
\bibinfo{author}{\bibfnamefont{V.}~\bibnamefont{Sahni}} \bibnamefont{and}
  \bibinfo{author}{\bibfnamefont{A.}~\bibnamefont{Starobinsky}},
  \bibinfo{journal}{Int.\ J.\ Mod.\ Phys.} \textbf{\bibinfo{volume}{D9}}
  (\bibinfo{year}{2000}) \bibinfo{pages}{373} [arXiv:astro-ph/9904398].

\bitem[{\citenamefont{Zeldovich}(1968)}]{Zeldovich}
\bibinfo{author}{\bibfnamefont{Ya.B.}~\bibnamefont{Zeldovich}},
  \bibinfo{journal}{Sov.\ Phys. - Uspekhi} \textbf{\bibinfo{volume}{11}}
  (\bibinfo{year}{1968}) \bibinfo{pages}{381}.

\bitem[{\citenamefont{Peskin}(1964)}]{Peskin}
\bibinfo{author}{\bibfnamefont{M.E.}~\bibnamefont{Peskin}},
\bibinfo{author}{\bibfnamefont{D.V.}~\bibnamefont{Schroeder}},
  \emph{\bibinfo{title}{An Introduction to Quantum Field Theory}}
  (\bibinfo{publisher}{Westview Press},
  \bibinfo{year}{1995}) p 790-791,394,402.

\bitem[{\citenamefont{H\'{e}ly}(1954)}]{Hely}
\bibinfo{author}{\bibfnamefont{J.}~\bibnamefont{H\'{e}ly}},
  \bibinfo{journal}{C.\ R.\ Acad.\ Sci.\ (Paris)}
  \textbf{\bibinfo{volume}{239}} (\bibinfo{year}{1954}) \bibinfo{pages}{385}.

\bitem[{\citenamefont{Treder}(1957)}]{Treder57}
\bibinfo{author}{\bibfnamefont{H.-J.}~\bibnamefont{Treder}},
  \bibinfo{journal}{Annalen der Physik} \textbf{\bibinfo{volume}{19}}
  (\bibinfo{year}{1957}) \bibinfo{pages}{369}.

\bitem[{\citenamefont{Johnson}(1985)}]{JohnsonI}
\bibinfo{author}{\bibfnamefont{C.~R.} \bibnamefont{Johnson}},
  \bibinfo{journal}{Phys.\ Rev.\ D} \textbf{\bibinfo{volume}{31}}
  (\bibinfo{year}{1985}) \bibinfo{pages}{1236}.

\bitem[{\citenamefont{Antoci}(1991)}]{Antoci3}
\bibinfo{author}{\bibfnamefont{S.}~\bibnamefont{Antoci}},
  \bibinfo{journal}{Gen.\ Rel.\ Grav.} \textbf{\bibinfo{volume}{23}}
  (\bibinfo{year}{1991}) \bibinfo{pages}{47} [arXiv:gr-qc/0108052].

%\bitem[{\citenamefont{Johnson and Nance}(1977)}]{Johnson77}
%\bibinfo{author}{\bibfnamefont{C.~R.} \bibnamefont{Johnson}} \bibnamefont{and}
%  \bibinfo{author}{\bibfnamefont{J.~R.} \bibnamefont{Nance}},
%  \bibinfo{journal}{Phys.\ Rev.\ D} \textbf{\bibinfo{volume}{15}}
%  (\bibinfo{year}{1977}) \bibinfo{pages}{377}.

\bitem[{\citenamefont{Johnson}(1985)}]{JohnsonII}
\bibinfo{author}{\bibfnamefont{C.~R.} \bibnamefont{Johnson}},
  \bibinfo{journal}{Phys.\ Rev.\ D} \textbf{\bibinfo{volume}{31}}
  (\bibinfo{year}{1985}) \bibinfo{pages}{1252}.

\bitem[{\citenamefont{Narlikar}(1953)
  \citenamefont{Narlikar, and Rao}}]{Narlikar}
  \bibinfo{author}{\bibfnamefont{V.~V.}~\bibnamefont{Narlikar}},
  \bibnamefont{and} \bibinfo{author}{\bibfnamefont{B.~R.}~\bibnamefont{Rao}},
  \bibinfo{journal}{Proc.\ Nat.\ Inst.\ Sc.\ India}
  \textbf{\bibinfo{volume}{21A}}
  (\bibinfo{year}{1953}) \bibinfo{pages}{409}.

\bitem[{\citenamefont{Wyman}(1950)}]{Wyman}
\bibinfo{author}{\bibfnamefont{M.}~\bibnamefont{Wyman}},
  \bibinfo{journal}{Can.\ J.\ Math.} \textbf{\bibinfo{volume}{2}}
  (\bibinfo{year}{1950}) \bibinfo{pages}{427}.

\bitem[{\citenamefont{Tiwari}(1972)}]{Tiwari}
\bibinfo{author}{\bibfnamefont{R.} \bibnamefont{Tiwari}} \bibnamefont{and}
  \bibinfo{author}{\bibfnamefont{D.~N.}~\bibnamefont{Pant}},
  \bibinfo{journal}{J.\ Phys.\ A} \textbf{\bibinfo{volume}{5}}
  (\bibinfo{year}{1972}) \bibinfo{pages}{394}.

\bitem[{\citenamefont{Shifflett}()}]{Shifflett2}
\bibinfo{author}{\bibfnamefont{J.~A.} \bibnamefont{Shifflett}},
   \bibnamefont{(unpublished)} \eprint{arXiv:gr-qc/0403052}.

\bitem[{\citenamefont{Shifflett}()}]{sShifflett}
\bibinfo{author}{\bibfnamefont{J.~A.} \bibnamefont{Shifflett}},
\bibnamefont{(unpublished)} \eprint{arXiv:gr-qc/0411016}.

\bitem[{\citenamefont{Einstein and Infeld}(1949)}]{EinsteinInfeld}
\bibinfo{author}{\bibfnamefont{A.}~\bibnamefont{Einstein}} \bibnamefont{and}
  \bibinfo{author}{\bibfnamefont{L.}~\bibnamefont{Infeld}},
  \bibinfo{journal}{Canad.\ J.\ Math.} \textbf{\bibinfo{volume}{1}}
  (\bibinfo{year}{1949}) \bibinfo{pages}{209}.

\bitem[{\citenamefont{Wallace}(1941)}]{Wallace}
\bibinfo{author}{\bibfnamefont{P.~R.} \bibnamefont{Wallace}},
  \bibinfo{journal}{Am.\ J.\ Math.} \textbf{\bibinfo{volume}{63}}
  (\bibinfo{year}{1941}) \bibinfo{pages}{729}.

\bitem[{\citenamefont{Callaway}(1953)}]{Callaway}
\bibinfo{author}{\bibfnamefont{J.}~\bibnamefont{Callaway}},
  \bibinfo{journal}{Phys.\ Rev.} \textbf{\bibinfo{volume}{92}}
  (\bibinfo{year}{1953}) \bibinfo{pages}{1567}.

\bitem[{\citenamefont{Infeld}(1950)}]{Infeld}
\bibinfo{author}{\bibfnamefont{L.}~\bibnamefont{Infeld}},
  \bibinfo{journal}{Acta Phys. Pol.} \textbf{\bibinfo{volume}{X}}
  (\bibinfo{year}{1950}) \bibinfo{pages}{284}.

%\bitem[{\citenamefont{Pryke et~al.}(2002)\citenamefont{Pryke, Halverson,
%  Leitch, Kovac, Carlstrom, and et.al.}}]{Pryke}
%\bibinfo{author}{\bibfnamefont{C.}~\bibnamefont{Pryke}}
%  \bibnamefont{and} \bibinfo{author}{\bibnamefont{et.al.}},
%  \bibinfo{journal}{Ap.\ J.} \textbf{\bibinfo{volume}{568}}
%  (\bibinfo{year}{2002}) \bibinfo{pages}{46} [arXiv:astro-ph/0104490].

\bitem[{\citenamefont{Spergel et~al.}(2003)\citenamefont{Spergel
 and et.al.}}]{Spergel}
\bibinfo{author}{\bibfnamefont{D.~N.}~\bibnamefont{Spergel}}
  \bibnamefont{and} \bibinfo{author}{\bibnamefont{et.al.}},
  \bibinfo{journal}{Ap.\ J. Suppl.} \textbf{\bibinfo{volume}{148}}
  (\bibinfo{year}{2003}) \bibinfo{pages}{175} [arXiv:astro-ph/0302209].

\bitem[{\citenamefont{Perlmutter and et.al.}(1999)}]{Perlmutter}
\bibinfo{author}{\bibfnamefont{S.}~\bibnamefont{Perlmutter}} \bibnamefont{and}
  \bibinfo{author}{\bibnamefont{et.al.}}, \bibinfo{journal}{Ap.\ J.}
  \textbf{\bibinfo{volume}{517}}
  (\bibinfo{year}{1999}) \bibinfo{pages}{565} [arXiv:astro-ph/9812133].

\bitem[{\citenamefont{Astier and et.al.}(1999)}]{Astier}
\bibinfo{author}{\bibfnamefont{P.}~\bibnamefont{Astier}} \bibnamefont{and}
  \bibinfo{author}{\bibnamefont{et.al.}}, \bibinfo{journal}{Astro.\ Astrophys.}
  (\bibinfo{year}{2005}) [arXiv:astro-ph/0510447].

\bitem[{\citenamefont{Freedman et~al.}(2001)\citenamefont{Freedman,
 and et.al.}}]{Freedman}
\bibinfo{author}{\bibfnamefont{W.~L.} \bibnamefont{Freedman}}
  \bibnamefont{and} \bibinfo{author}{\bibnamefont{et.al.}},
  \bibinfo{journal}{Ap.\ J.} \textbf{\bibinfo{volume}{553}}
  (\bibinfo{year}{2001}) \bibinfo{pages}{47} [arXiv:astro-ph/0012376].

\bitem[{\citenamefont{Moffat}(1979)}]{Moffat78}
\bibinfo{author}{\bibfnamefont{J.~W.} \bibnamefont{Moffat}},
  \bibinfo{journal}{Phys.\ Rev.\ D} \textbf{\bibinfo{volume}{19}},
  (\bibinfo{year}{1979}) \bibinfo{pages}{3554}.

\bitem[{\citenamefont{Damour et~al.}(1992)\citenamefont{Damour, Deser, and
  McCarthy}}]{Damour92}
\bibinfo{author}{\bibfnamefont{T.}~\bibnamefont{Damour}},
  \bibinfo{author}{\bibfnamefont{S.}~\bibnamefont{Deser}}, \bibnamefont{and}
  \bibinfo{author}{\bibfnamefont{J.}~\bibnamefont{McCarthy}},
  \bibinfo{journal}{Phys.\ Rev.\ D} \textbf{\bibinfo{volume}{45}}
  (\bibinfo{year}{1992}) \bibinfo{pages}{R3289}.

\bitem[{\citenamefont{Damour et~al.}(1993)\citenamefont{Damour, Deser, and
  McCarthy}}]{Damour93}
\bibinfo{author}{\bibfnamefont{T.}~\bibnamefont{Damour}},
  \bibinfo{author}{\bibfnamefont{S.}~\bibnamefont{Deser}}, \bibnamefont{and}
  \bibinfo{author}{\bibfnamefont{J.}~\bibnamefont{McCarthy}},
  \bibinfo{journal}{Phys.\ Rev.\ D} \textbf{\bibinfo{volume}{47}}
  (\bibinfo{year}{1993}) \bibinfo{pages}{1541} [arXiv:gr-qc/9207003].

%\bitem[{\citenamefont{Cornish and Moffat}(1993)}]{Cornish}
%\bibinfo{author}{\bibfnamefont{N.~J.} \bibnamefont{Cornish}} \bibnamefont{and}
%  \bibinfo{author}{\bibfnamefont{J.~W.} \bibnamefont{Moffat}},
%  \bibinfo{journal}{Phys.\ Rev.\ D} \textbf{\bibinfo{volume}{47}}
%  (\bibinfo{year}{1993}) \bibinfo{pages}{4421} [arXiv:gr-qc/9207007].

%\bitem[{\citenamefont{Kelly}(1991)}]{Kelly}
%\bibinfo{author}{\bibfnamefont{P.~F.} \bibnamefont{Kelly}},
%  \bibinfo{journal}{Class. Quantum Grav.} \textbf{\bibinfo{volume}{8}},
%  (\bibinfo{year}{1991}) \bibinfo{pages}{1217}.

%\bitem[{\citenamefont{Kelly and Mann}(1986)}]{KellyMann86}
%\bibinfo{author}{\bibfnamefont{P.~F.} \bibnamefont{Kelly}} \bibnamefont{and}
%  \bibinfo{author}{\bibfnamefont{R.~B.} \bibnamefont{Mann}},
%  \bibinfo{journal}{Class. Quantum Grav.} \textbf{\bibinfo{volume}{3}},
%  (\bibinfo{year}{1986}) \bibinfo{pages}{705}.

%\bitem[{\citenamefont{Kelly and Mann}(1987)}]{KellyMann87}
%\bibinfo{author}{\bibfnamefont{P.~F.} \bibnamefont{Kelly}} \bibnamefont{and}
%  \bibinfo{author}{\bibfnamefont{R.~B.} \bibnamefont{Mann}},
%  \bibinfo{journal}{Class. Quantum Grav.} \textbf{\bibinfo{volume}{4}},
%  (\bibinfo{year}{1987}) \bibinfo{pages}{1593}.

\bitem[{\citenamefont{Will}(1989)}]{Will}
\bibinfo{author}{\bibfnamefont{C.~M.} \bibnamefont{Will}},
  \bibinfo{journal}{Phys.\ Rev.\ Lett.} \textbf{\bibinfo{volume}{62}}
  (\bibinfo{year}{1989}) \bibinfo{pages}{369}.

\bitem[{\citenamefont{Zhou and Haughan}(1992)}]{Zhou}
\bibinfo{author}{\bibfnamefont{Z.} \bibnamefont{Zhou}} \bibnamefont{and}
  \bibinfo{author}{\bibfnamefont{M.~P.} \bibnamefont{Haugan}},
  \bibinfo{journal}{Phys.\ Rev.\ D} \textbf{\bibinfo{volume}{45}}
  (\bibinfo{year}{1992}) \bibinfo{pages}{3336}.

\bitem[{\citenamefont{Gabriel et~al.}(1991)}]{Gabriel}
\bibinfo{author}{\bibfnamefont{M.~D.} \bibnamefont{Gabriel}},
\bibinfo{author}{\bibfnamefont{M.~P.} \bibnamefont{Haugan}},
\bibinfo{author}{\bibfnamefont{R.~B.} \bibnamefont{Mann}} \bibnamefont{and}
  \bibinfo{author}{\bibfnamefont{J.~H.} \bibnamefont{Palmer}},
  \bibinfo{journal}{Phys.\ Rev. Lett.} \textbf{\bibinfo{volume}{67}}
  (\bibinfo{year}{1991}) \bibinfo{pages}{2123}.

\bitem[{\citenamefont{Moffat}(1995)}]{Moffat95}
\bibinfo{author}{\bibfnamefont{J.~W.} \bibnamefont{Moffat}},
  \bibinfo{journal}{J.\ Math.\ Phys.} \textbf{\bibinfo{volume}{36}},
  (\bibinfo{year}{1995}) \bibinfo{pages}{3722}.

\bitem[{\citenamefont{Clayton}(1996)}]{Clayton}
\bibinfo{author}{\bibfnamefont{M.~A.} \bibnamefont{Clayton}},
  \bibinfo{journal}{Class. Quantum Grav.} \textbf{\bibinfo{volume}{13}},
  (\bibinfo{year}{1996}) \bibinfo{pages}{2851}.

\bitem[{\citenamefont{Kur\c{s}uno\u{g}lu}(1952)}]{Kursunoglu}
\bibinfo{author}{\bibfnamefont{B.}~\bibnamefont{Kur\c{s}uno\u{g}lu}},
  \bibinfo{journal}{Phys.\ Rev.} \textbf{\bibinfo{volume}{88}}
  (\bibinfo{year}{1952}) \bibinfo{pages}{1369}.

\bitem[{\citenamefont{Bonnor}(1954)}]{Bonnor}
\bibinfo{author}{\bibfnamefont{W.~B.} \bibnamefont{Bonnor}},
  \bibinfo{journal}{Proc. R. Soc.} \textbf{\bibinfo{volume}{A226}}
  (\bibinfo{year}{1954}) \bibinfo{pages}{366}.

\bitem[{\citenamefont{Borchsenius}(1978)}]{Borchsenius}
\bibinfo{author}{\bibfnamefont{K.}~\bibnamefont{Borchsenius}},
  \bibinfo{journal}{Nuovo Cimento} \textbf{\bibinfo{volume}{46A}}
  (\bibinfo{year}{1978}) \bibinfo{pages}{403}.

\bitem[{\citenamefont{Moffat and Boal}(1974)}]{Moffat74}
\bibinfo{author}{\bibfnamefont{J.~W.} \bibnamefont{Moffat}} \bibnamefont{and}
  \bibinfo{author}{\bibfnamefont{D.~H.} \bibnamefont{Boal}},
  \bibinfo{journal}{Phys.\ Rev.\ D} \textbf{\bibinfo{volume}{11}}
  (\bibinfo{year}{1974}) \bibinfo{pages}{1375}.

\bitem[{\citenamefont{Born and Infeld}(1985)}]{Born}
\bibinfo{author}{\bibfnamefont{M.} \bibnamefont{Born}} \bibnamefont{and}
  \bibinfo{author}{\bibfnamefont{L.} \bibnamefont{Infeld}},
  \bibinfo{journal}{Proc.\ Roy.\ Soc.} \textbf{\bibinfo{volume}{A144}}
  (\bibinfo{year}{1934}) \bibinfo{pages}{425}.

\bitem[{\citenamefont{Deser and Gibbons}(2001)\citenamefont{Deser and
  Gibbons}}]{Deser}
\bibinfo{author}{\bibfnamefont{S.}~\bibnamefont{Deser}} \bibnamefont{and}
  \bibinfo{author}{\bibfnamefont{G.~W.}~\bibnamefont{Gibbons}},
  \bibinfo{journal}{Class.\ Quant.\ Grav.} \textbf{\bibinfo{volume}{15}}
  (\bibinfo{year}{1998}) \bibinfo{pages}{L35} [arXiv:hep-th/9803049].

%\bitem[{\citenamefont{Fradkin and Tseytlin}(1985)}]{Fradkin}
%\bibinfo{author}{\bibfnamefont{E.~S.} \bibnamefont{Fradkin}} \bibnamefont{and}
%\bibinfo{author}{\bibfnamefont{A.~A.} \bibnamefont{Tseytlin}},
%  \bibinfo{journal}{Phys.\ Lett.} \textbf{\bibinfo{volume}{B163}}
%  (\bibinfo{year}{1985}) \bibinfo{pages}{123}.

\bitem[{\citenamefont{Gibbons and Herdieiro}(2001)\citenamefont{Gibbons and
  Herdeiro}}]{Gibbons}
\bibinfo{author}{\bibfnamefont{G.~W.}~\bibnamefont{Gibbons}} \bibnamefont{and}
  \bibinfo{author}{\bibfnamefont{C.~A.~R.}~\bibnamefont{Herdeiro}},
  \bibinfo{journal}{Phys.\ Rev.\ D} \textbf{\bibinfo{volume}{63}}
  (\bibinfo{year}{2001}) \bibinfo{pages}{064006} [arXiv:hep-th/0008052].

\bitem[{\citenamefont{Sakharov}(1968)}]{Sakharov}
\bibinfo{author}{\bibfnamefont{A.~D.} \bibnamefont{Sakharov}},
  \bibinfo{journal}{Sov.\ Phys.\ Doklady} \textbf{\bibinfo{volume}{12}}
  (\bibinfo{year}{1968}) \bibinfo{pages}{1040}.

\bitem[{\citenamefont{Padmanabhan}(1985)}]{Padmanabhan}
\bibinfo{author}{\bibfnamefont{T.} \bibnamefont{Padmanabhan}},
  \bibinfo{journal}{Gen.\ Rel.\ Grav.} \textbf{\bibinfo{volume}{17}}
  (\bibinfo{year}{1985}) \bibinfo{pages}{215}.

\bitem[{\citenamefont{Padmanabhan2}(2002)}]{Padmanabhan2}
\bibinfo{author}{\bibfnamefont{T.} \bibnamefont{Padmanabhan}},
  \bibinfo{journal}{Class.\ Quantum\ Grav.} \textbf{\bibinfo{volume}{19}}
  (\bibinfo{year}{2002}) \bibinfo{pages}{3551}.

\bitem[{\citenamefont{Chandrasekhar}(1992)}]{Chandrasekhar}
\bibinfo{author}{\bibfnamefont{S.}~\bibnamefont{Chandrasekhar}},
  \emph{\bibinfo{title}{The Mathematical Theory of Black Holes}}
  (\bibinfo{publisher}{Oxford University Press},
  \bibinfo{address}{New York}, \bibinfo{year}{1992}) p 317.

\bitem[{\citenamefont{Ashtekar, Rovelli and Smolin}(2000)}]{Ashtekar}
\bibinfo{author}{\bibfnamefont{A.}~\bibnamefont{Ashtekar}} \bibnamefont{and}
\bibinfo{author}{\bibfnamefont{C.}~\bibnamefont{Rovelli}} \bibnamefont{and}
  \bibinfo{author}{\bibfnamefont{L.}~\bibnamefont{Smolin}},
  \bibinfo{journal}{Phys.\ Rev.\ Let.} \textbf{\bibinfo{volume}{69}}
  (\bibinfo{year}{1992}) \bibinfo{pages}{237}.

\bitem[{\citenamefont{Garay}(1995)}]{Garay}
\bibinfo{author}{\bibfnamefont{Luis.\ L.} \bibnamefont{Garay}},
  \bibinfo{journal}{Int. J. Mod. Phys.} \textbf{\bibinfo{volume}{A10}}
  (\bibinfo{year}{1995}) \bibinfo{pages}{145}.

\bitem[{\citenamefont{Smolin}(2004)}]{Smolin}
\bibinfo{author}{\bibfnamefont{L.} \bibnamefont{Smolin}},
  \bibinfo{journal}{Sci. Am.} \textbf{\bibinfo{volume}{290}}
  (\bibinfo{year}{2004}) \bibinfo{pages}{1}.

%\bitem[{\citenamefont{Rugh}()}]{Rugh}
%\bibinfo{author}{\bibfnamefont{S.~E.} \bibnamefont{Rugh}}
%\bibfnamefont{and} \bibinfo{author}{\bibfnamefont{H.}
% \bibnamefont{Zinkernagel}},
% \eprint{arXiv:astro-ph/0105130}.

%\bitem[{\citenamefont{Weinberg}()}]{Weinberg}
%\bibinfo{author}{\bibfnamefont{S.}~\bibnamefont{Weinberg}},
%  \eprint{arXiv:astro-ph/9610044}.

%\bitem[{\citenamefont{Ostriker and Steinhardt}(2001)}]{Ostriker}
%\bibinfo{author}{\bibfnamefont{J.~P.} \bibnamefont{Ostriker}} \bibnamefont{and}
%  \bibinfo{author}{\bibfnamefont{P.~J.} \bibnamefont{Steinhardt}},
%  \bibinfo{journal}{Sci.\ Am.} \textbf{\bibinfo{volume}{284}}
%  (\bibinfo{year}{2001}) \bibinfo{pages}{46}.

%\bitem[{\citenamefont{Guth}(1997)}]{GuthIU}
%\bibinfo{author}{\bibfnamefont{A.~H.} \bibnamefont{Guth}},
%  \emph{\bibinfo{title}{The Inflationary Universe}}
%  (\bibinfo{publisher}{Perseus}, \bibinfo{address}{Reading, Massachusetts},
%  \bibinfo{year}{1997}).

%\bitem[{\citenamefont{Padmanabhan3}(1916)}]{Padmanabhan3}
%\bibinfo{author}{\bibfnamefont{T.}~\bibnamefont{Padmanabhan}},
%  \bibinfo{journal}{Phys.\ Rept.} \textbf{\bibinfo{volume}{380}}
%  (\bibinfo{year}{2003}) \bibinfo{pages}{235} [arXiv:hep-th/0212290].

%\bitem[{\citenamefont{Guth}(2001)}]{Guth}
%\bibinfo{author}{\bibfnamefont{A.~H.} \bibnamefont{Guth}},
%  \emph{\bibinfo{title}{Inflation and the Accelerating Universe}}
%  (\bibinfo{year}{2001}), \eprint{http://www/physics.sunnysb.edu
% /itp/OWP/talks/aguth/}.

\bitem[{\citenamefont{Adler et~al.}(1975)\citenamefont{Adler, Bazin, and
  Schiffer}}]{Adler}
\bibinfo{author}{\bibfnamefont{R.}~\bibnamefont{Adler}},
  \bibinfo{author}{\bibfnamefont{M.}~\bibnamefont{Bazin}}, \bibnamefont{and}
  \bibinfo{author}{\bibfnamefont{M.}~\bibnamefont{Schiffer}},
  \emph{\bibinfo{title}{Intro. to General Relativity}}
  (\bibinfo{publisher}{McGraw-Hill}, \bibinfo{address}{NY},
  \bibinfo{year}{1975}) p~87.

\bitem[{\citenamefont{Tonnelat}(1954)}]{Tonnelat}
\bibinfo{author}{\bibfnamefont{M.~A.} \bibnamefont{Tonnelat}},
  \bibinfo{journal}{C.\ R.\ Acad.\ Sci.} \textbf{\bibinfo{volume}{239}}
  (\bibinfo{year}{1954}) \bibinfo{pages}{231}.

\bitem[{\citenamefont{Jackson}(1999)}]{Jackson}
\bibinfo{author}{\bibfnamefont{J.~D.}~\bibnamefont{Jackson}},
  \emph{\bibinfo{title}{Classical Electrodynamics}}
  (\bibinfo{publisher}{John Wiley and Sons},
  \bibinfo{year}{1999}) p 2-22,596.

\bitem[{\citenamefont{Greiner}(1997)}]{Greiner}
\bibinfo{author}{\bibfnamefont{W.}~\bibnamefont{Greiner}},
  \emph{\bibinfo{title}{Relativistic Quantum Mechanics - Wave Eqs.}}
  (\bibinfo{publisher}{Springer-Verlag},
  \bibinfo{address}{Berlin}, \bibinfo{year}{1997}) p~375.

\bitem[{\citenamefont{Proca}(1936)}]{Proca}
\bibinfo{author}{\bibfnamefont{A.}~\bibnamefont{Proca}}, \bibinfo{journal}{Le
  Journal de Physique et le Radium} \textbf{\bibinfo{volume}{7}}
  (\bibinfo{year}{1936}) \bibinfo{pages}{347}.

%\bitem[{\citenamefont{v.~Borzeszkowski and Treder}(1996)}]{Bor1}
%\bibinfo{author}{\bibfnamefont{H.-H.} \bibnamefont{v.~Borzeszkowski}}
%  \bibnamefont{and} \bibinfo{author}{\bibfnamefont{H.-J.}
%  \bibnamefont{Treder}}, \bibinfo{journal}{Gen.\ Rel.\ Grav.}
%  \textbf{\bibinfo{volume}{28}} (\bibinfo{year}{1996}) \bibinfo{pages}{1}.

%\bitem[{\citenamefont{v.~Borzeszkowski and Datta}(2002)}]{Bor4}
%\bibinfo{author}{\bibfnamefont{H.-H.} \bibnamefont{v.~Borzeszkowski}}
%  \bibnamefont{and} \bibinfo{author}{\bibfnamefont{et.~al.}},
%  \bibinfo{journal}{Foun.\ Phys.}
%  \textbf{\bibinfo{volume}{32}} (\bibinfo{year}{2002}) \bibinfo{pages}{1701}.

%\bitem[{\citenamefont{Deif}(1991)}]{Horn}
%\bibinfo{author}{\bibfnamefont{R.~A.} \bibnamefont{Horn}}
%\bibnamefont{and} \bibinfo{author}{\bibfnamefont{C.~R.} \bibnamefont{Johnson}},
%  \emph{\bibinfo{title}{Topics in Matrix Analysis}}
%  (\bibinfo{publisher}{Cambridge U. Press},
%  \bibinfo{address}{Cambridge}, \bibinfo{year}{1991}) p~430,439,474.

\bitem[{\citenamefont{Deif}(1982)}]{Deif}
\bibinfo{author}{\bibfnamefont{A.~S.}~\bibnamefont{Deif}},
  \emph{\bibinfo{title}{Advanced Matrix Theory For Scientists and Engineers}}
  (\bibinfo{publisher}{John Wiley and Sons},
  \bibinfo{address}{NY}, (\bibinfo{year}{1982}) p~20,153,183,171.

\bitem[{\citenamefont{Einstein et~al.}(1937)\citenamefont{Einstein, Infeld,
  and Hoffmann}}]{EIH}
\bibinfo{author}{\bibfnamefont{A.}~\bibnamefont{Einstein}},
  \bibinfo{author}{\bibfnamefont{L.}~\bibnamefont{Infeld}}, \bibnamefont{and}
  \bibinfo{author}{\bibfnamefont{B.}~\bibnamefont{Hoffmann}},
  \bibinfo{journal}{Ann.\ Math.} \textbf{\bibinfo{volume}{39}}
  (\bibinfo{year}{1937}) \bibinfo{pages}{65}.

\bitem[{\citenamefont{Wallace}(1941)}]{WallaceThesis}
\bibinfo{author}{\bibfnamefont{P.~R.} \bibnamefont{Wallace}},
  \emph{\bibinfo{title}{On the Relativistic Equations of Motion in Electromagnetic Theory}}
  (\bibinfo{publisher}{Doctoral Thesis, University of Toronto},
  \bibinfo{year}{1940}).

%\bitem[{\citenamefont{Infeld and Wallace}(1940)}]{InfeldWallace}
%\bibinfo{author}{\bibfnamefont{L.}~\bibnamefont{Infeld}} \bibnamefont{and}
%  \bibinfo{author}{\bibfnamefont{P.~R.} \bibnamefont{Wallace}},
%  \bibinfo{journal}{Phys.\ Rev.} \textbf{\bibinfo{volume}{57}}
%  (\bibinfo{year}{1940}) \bibinfo{pages}{797}.

\bitem[{\citenamefont{Gorbatenko}(2005)}]{Gorbatenko}
\bibinfo{author}{\bibfnamefont{M.~V.}~\bibnamefont{Gorbatenko}},
  \bibinfo{journal}{Theor.\ Math.\ Phys. (Springer-Verlag)} \textbf{\bibinfo{volume}{142}}
  (\bibinfo{year}{2005}) \bibinfo{pages}{138}.

%\bitem[{\citenamefont{Anderson}(1997)}]{Anderson}
%\bibinfo{author}{\bibfnamefont{J.~L.}~\bibnamefont{Anderson}},
%  \bibinfo{journal}{Phys.\ Rev.} \textbf{\bibinfo{volume}{56}}
%  (\bibinfo{year}{1997}) \bibinfo{pages}{4675} [arXiv:astro-ph/9709031].

\bitem[{\citenamefont{Voros}(1995)}]{Voros}
\bibinfo{author}{\bibfnamefont{J.}~\bibnamefont{Voros}},
  \bibinfo{journal}{Aust.\ J.\ Phys.} \textbf{\bibinfo{volume}{48}}
  (\bibinfo{year}{1995}) \bibinfo{pages}{45} [arXiv:gr-qc/9504047].

\bitem[{\citenamefont{Papapetrou}(1948)}]{Papapetrou}
\bibinfo{author}{\bibfnamefont{A.}~\bibnamefont{Papapetrou}},
  \bibinfo{journal}{Proc.\ Royal Irish Acad.} \textbf{\bibinfo{volume}{52A}}
  (\bibinfo{year}{1948}) \bibinfo{pages}{69}.

\bitem[{\citenamefont{Takeno et~al.}(1951)\citenamefont{Takeno, Ikeda, and
  Abe}}]{Takeno}
\bibinfo{author}{\bibfnamefont{H.}~\bibnamefont{Takeno}},
  \bibinfo{author}{\bibfnamefont{M.}~\bibnamefont{Ikeda}}, \bibnamefont{and}
  \bibinfo{author}{\bibfnamefont{S.}~\bibnamefont{Abe}},
  \bibinfo{journal}{Prog.\ Theor.\ Phys.} \textbf{\bibinfo{volume}{VI}}
  (\bibinfo{year}{1951}) \bibinfo{pages}{837}.

\bitem[{\citenamefont{Reissner}(1916)}]{Reissner}
\bibinfo{author}{\bibfnamefont{H.}~\bibnamefont{Reissner}},
  \bibinfo{journal}{Ann.\ d.\ Phys.\ (Leipzig)} \textbf{\bibinfo{volume}{50}}
  (\bibinfo{year}{1916}) \bibinfo{pages}{106}.

\bitem[{\citenamefont{Nordstrom}(1918)}]{Nordstrom}
\bibinfo{author}{\bibfnamefont{A.}~\bibnamefont{Nordstr\"{o}m}},
  \bibinfo{journal}{Proc.\ Kon.\ Ned.\ Akad.\ Wet.}
  \textbf{\bibinfo{volume}{20}} (\bibinfo{year}{1918}) \bibinfo{pages}{1238}.

\end{thebibliography}

\end{document}